\documentclass[aps,pra,twocolumn,
floatfix
]{revtex4-1}
\usepackage[svgnames]{xcolor}
\usepackage[colorlinks=true,urlcolor = teal, citecolor = blue, linkcolor= magenta, bookmarks=false]{hyperref}
\usepackage{amsfonts,amsmath,amssymb}
\usepackage{graphicx}
\usepackage{dsfont}

\usepackage{relsize}
\usepackage{color}
\usepackage{braket}

\usepackage{bm}

\DeclareFontFamily{T1}{pzc}{}
\DeclareFontShape{T1}{pzc}{m}{it}{<-> [1.2] pzcmi8t}{}
\DeclareMathAlphabet{\mathpzc}{T1}{pzc}{m}{it}

\DeclareMathOperator{\im}{Im}

\DeclareMathOperator{\Tr}{Tr}

\def\la{\label}

\def\be{\begin{equation}}
\def\ee{\end{equation}}

\newcommand{\bea}{\begin{eqnarray}}
\newcommand{\eea}{\end{eqnarray}}

\newcommand{\bei}{\begin{itemize}}
\newcommand{\eei}{\end{itemize}}

\newcommand{\bee}{\begin{enumerate}}
\newcommand{\eee}{\end{enumerate}}

\newcommand{\bean}{\begin{eqnarray*}}
\newcommand{\eean}{\end{eqnarray*}}

\def\a{\alpha}
\def\g{\gamma}

\def\e{\epsilon}
\def\k{\kappa}
\def\lam{\lambda}
\def\r{\rho}
\def\s{\sigma}
\def\t{\tau}

\def\om{\omega}

\def\ve{\varepsilon}

\def\S{\Sigma}

\def\sfa{{\mathsf a}}

\def\sfg{{\mathsf g}}
\def\sfh{{\mathsf h}}

\def\cG{\mathcal G}

\def\cO{\mathcal O}
\def\cS{\mathcal S}
\def\cT{\mathcal T}

\def\cW{\mathcal W}
\def\cZ{\mathcal Z}

\definecolor{grey}{rgb}{0.4,0.4,0.5}
\definecolor{darkgreen}{rgb}{0,0.5,0}
\definecolor{darkred}{rgb}{0.6,0.0,0}
\definecolor{lightbrown}{rgb}{1,0.9,0.8}
\definecolor{brown}{rgb}{0.6,0.3,0.3}
\definecolor{darkblue}{rgb}{0,0,0.8}
\definecolor{darkmagenta}{rgb}{0.5,0,0.5}

\def\pa {\partial}

\newcommand{\alg}[1]{\mathfrak{#1}}
\def\su{\alg{su}}
\def\sucex{{\su(2|2)}}
\def\ucex{{\alg{u}(2|2)}}

\def\OX{{\bm X}}

\def\OB{{\bm \phi}}
\def\OF{{\bm \psi}}
\def\OHu{{\bm \theta}}

\def\OA{{\bm a}}

\def\Oc{{\bm c}}
\def\Ocd{{\bm c}^\dagger}
\def\OQ{{\bm q}}
\def\OQd{{\bm q}^\dagger}

\def\On{{\bm n}}
\def\Obn{\bar{\bm n}}

\def\OS{{\bm s}}
\def\Os{{\bm s}}

\def\Oet{{\bm \eta}}

\def\OH{{\bm H}}
\def\OT{{\bm T}}
\def\OV{{\bm V}}

\def\Ob{{\bm b}}
\def\Obd{{\bm b}^\dagger}
\def\Of{{\bm f}}
\def\Ofd{{\bm f}^\dagger}

\def\Ocs{\Oc}
\def\OQs{\OQ}
\def\Obs{\Ob}
\def\Ofs{\Of}

\def\vOS{\bm s}

\def\vOet{\bm \eta}

\def\bs{{\bar{\sigma}}}

\def\bsu{\uparrow}
\def\bsd{\downarrow}

\def\ssu{{\mathsmaller{\uparrow}}}
\def\ssd{{\mathsmaller{\downarrow}}}

\def\qe{{\circ}}
\def\qd{{\bullet}}
\def\bqe{{\mathlarger{\mathlarger{\circ}}}}
\def\bqd{{\mathlarger{\mathlarger{\bullet}}}}

\def\oto{\leftrightarrow}

\def\tJ{{$t$-$J$} }

\def\ii{\mathrm i}

\def\DV{\nabla}
\def\vol{\mathcal V}

\def\ret{\mathrm{R}}
\def\adv{\mathrm{A}}

\def\Ut{\tilde{U}}
\def\Vt{\tilde{V}}
\def\lamt{\tilde{\lambda}}
\def\ft{\tilde{f}}

\def\vG{\varphi}

\def\GT{\mathcal G}

\def\gT{\mathfrak g}
\def\wT{\cW}
\def\Sg{\Sigma}
\def\SW{\Omega}

\def\sou{J}

\begin{document}

\title{The splitting of electrons and violation of the Luttinger sum rule}

\author{Eoin Quinn}
\email{E.P.Quinn@uva.nl}
\affiliation{Institute for Theoretical Physics, University of Amsterdam, Science Park 904, 1090 GL Amsterdam, The Netherlands}

\date{\today}

\begin{abstract}
We obtain a controlled description of a strongly correlated regime of electronic behaviour. 
We begin by arguing that there are two ways to characterise the electronic degree of freedom, either by the canonical fermion algebra or the graded Lie algebra $\su(2|2)$. The first underlies the Fermi liquid description of correlated matter, and we identify a novel regime governed by the latter. We exploit an exceptional central extension of $\su(2|2)$ to employ a perturbative scheme recently developed by Shastry, and obtain a series of successive approximations for the electronic Green's function. We then focus on the leading approximation, which reveals a splitting in two of the electronic dispersion. The Luttinger sum rule is violated, and a Mott metal-insulator transition is exhibited. We offer a perspective.
\end{abstract} 

\maketitle

\section{Introduction} 
A key step in characterising the behaviour of a system is the identification of the relevant degrees of freedom. This is exemplified by Landau's theory of the Fermi liquid \cite{landau1957,landau1959}, which offers a general description of metallic states in terms of weakly interacting fermions, degrees of freedom obeying the canonical anti-commutation relations
\be\la{can_f}
\{\Oc_{\s},\Ocd_{\s'}\}=\delta_{\s\s'}.
\ee
These account not just for the  long-wavelength phenomenology, but also the electronic band structure, and underlie powerful techniques such as density functional theory which provide a detailed description of a wide variety of materials \cite{gross2013density}.

Some of the most interesting materials have however resisted a description within this framework.
Chief among these are the cuprates, whose puzzling behaviour has provided the central challenge in the field of condensed matter for three decades \cite{BednorzMuller86,ANDERSON_1987,Keimer_rev}. 
Beyond having some of the highest known superconducting transition temperatures, they exhibit a Mott transition, a pseudogap regime displaying a landscape of intertwined orders \cite{Keimer_rev,Fradkin_2015}, and a strange metal regime which appears to defy a quasi-particle description \cite{marginalFL}. Other notable examples include  iron pnictides and chalcogenides \cite{Si_2016},  heavy-fermion compounds \cite{Gegenwart_2008}, and organic charge-transfer salts \cite{Powell_2011}.

 An important question  is whether canonical degrees of freedom,  bosons and fermions, are sufficient to account for such behaviour \cite{ANDERSON_1987}.
A quantum degree of freedom is specified by the algebra it obeys, which for bosons and fermions has a schematic form $[\OA,\OA]\sim1$.  
 Here we argue that strongly correlated electrons are instead governed by degrees of freedom which obey a non-canonical Lie algebra, i.e.~an algebra of the form $[\OA,\OA]\sim\OA$. The bracket again reduces the order of operators, but by one, as opposed to two in the canonical case. 
 The challenge then is to control the growth of correlations  generated by the Hamiltonian through $[\OH,\OA]$.

In one dimension it  is well understood how algebraic structures  govern the behaviour of correlated electrons, through the formalism of algebraic Bethe ansatz \cite{Faddeev_2016,EKS,Hbook,HS1}. 
This is specialised to one dimension however, owing to enhanced symmetries resulting from the constrained geometry \cite{ZAMOLODCHIKOV1979253}. 
Numerous efforts have been made to exploit Lie algebraic structures in higher dimensions \cite{Wiegmann_1988,Forster_1989,Chaichian_1991,Kochetov_1996,Coleman_2002,Anderson08,Avella_2011,Ramires}, most specifically through the formalism of Hubbard operators \cite{Hubbard4,Vedyaev_1984,RuckensteinSR,Izyumov_1990,Ovchinnikov_2004,Izyumov_2005,StanescuKotliar}, but a controlled theoretical framework has so far remained elusive. 
 A significant advancement has however recently been made by Shastry \cite{Shastry_2011,Shastry_2013},  who has developed a perturbative scheme for gaining  control over certain non-canonical degrees of freedom, assuming there exists a suitable expansion parameter.

In this work we readdress the question of how to characterise the behaviour of interacting electrons. As the electron has an inherent fermionic nature, we argue that graded Lie algebras  provide the natural language for the task. We consider the two such algebras relevant for the electronic degree of freedom: $\su(1|1)\otimes\su(1|1)$ and $\sucex$.
The first is the algebra of canonical fermions, Eq.~\eqref{can_f}, which underlies the Fermi liquid description of interacting electrons. The second is closely related to the algebra of Hubbard operators, and we will exploit it to obtain a distinct controlled description of interacting electrons. In particular, we will consider an exceptional central extension of $\sucex$, introduced by Beisert \cite{Beisert07,Beisert08}, which naturally provides a parameter for the use of Shastry's perturbative scheme.

We focus on the simplest setting where the novel features of this new controlled description can clearly be seen. 
We will not attempt to explicitly model any given system, but instead frame our discussion around two overarching themes: the Luttinger sum rule and the Mott metal-insulator transition. 

The Luttinger sum rule states that the volume of the region enclosed by the Fermi surface is directly proportional to the electron density, and independent of interactions.
It is proven to be valid for a Fermi liquid in the sense of Landau \cite{Luttinger_1960}, but there is strong evidence that it is violated in certain strongly correlated systems, such as the cuprates in the pseudogap regime \cite{Doiron_Leyraud_2007,Badoux_2016}. We explicitly demonstrate that $\su(2|2)$ degrees of freedom account for a violation of the Luttinger sum rule, and thus characterise an electronic state of matter which is not a  Fermi liquid. 

A Mott metal-insulator transition occurs when electronic correlations induce the opening of a gap within an electronic band, signifying a failure of band theory. 
This phenomenon  has played a pivotal role in the study of strongly correlated electrons, but remains incompletely understood \cite{Imada_1998,LNWrev}. 
It directly conflicts with the Luttinger sum rule, which implies that a partially filled band has a non-trivial Fermi surface and so is metallic. 
A controlled description consistent with Fermi liquid behaviour is however provided by dynamical mean-field theory  \cite{Metzner_1989,DMFT}, which is exact in the limit of infinite dimensions. 
Here the localisation of electronic quasi-particles is driven by the divergence of their effective mass, as previously described by Brinkman--Rice  \cite{PhysRevB.2.4302}. In contrast, we demonstrate that $\su(2|2)$ degrees of freedom result in a splitting in two of the electronic band, each carrying a fraction of the electron's spectral weight. These bands violate the Luttinger sum rule, and a  Mott transition naturally occurs when the  two bands separate. In the language of the seminal review \cite{Imada_1998}, this can be understood as a carrier-number-vanishing transition as opposed to a mass-diverging transition. 
 We thus offer a controlled framework for characterising Mott transitions in materials, such as the cuprates, where
the carrier number vanishes as the transition is approached \cite{Ando1,Ando2}.

The paper is structured as follows. In Sec.~\ref{sec:dof} we consider a general lattice model of interacting electrons, and demonstrate that it can be expressed through the generators of either $\su(1|1)\otimes\su(1|1)$ or $\sucex$. We interpret these as two ways to characterise the electronic degree of freedom. In Sec.~\ref{sec:GF} we derive a controlled framework for organising the growth of correlations in the $\sucex$ regime. That is, we obtain a series of successive approximations for the electronic Green's function, which mirrors the self-energy expansion for the canonical regime. In Sec.~\ref{sec:approx} we examine the leading approximation and find that it captures a splitting of the electronic band. We demonstrate that the Luttinger sum rule is violated, and we observe a Mott transition of carrier-number-vanishing type. Section~\ref{sec:disc} is a discussion, where we provide further context to our results and offer some perspectives. We conclude in Sec.~\ref{sec:conc}.

There are five appendices: \ref{app:su22} reviews the graded Lie algebra $\sucex$, \ref{app:params} provides explicit expressions for constants and parameters, \ref{app:canGF} reviews the Green's function analysis for the case of a canonical fermion,  \ref{app:sch} presents a schematic overview of the Green's function analysis for non-canonical $\sucex$, and \ref{app:2nd} contains the second order contributions to the $\sucex$ self-energy and adaptive spectral weight.


\section{Electronic degrees of freedom}\la{sec:dof}

We wish to address the question of how to characterise behaviour resulting from electronic correlations. Let us consider a  lattice with four states per site
\be\la{4states}
\ket{\bqe}=\ket{0},~~~\ket{\bsd}=\Ocd_{\ssd}\ket{0},~~~\ket{\bsu}=\Ocd_{\ssu}\ket{0},~~~\ket{\bqd}=\Ocd_{\ssd} \Ocd_{\ssu}\ket{0},
\ee 
which provides the Hilbert space for a single-orbital tight-binding model. We disregard disorder and lattice vibrations, focusing solely on electronic interactions. The simpler case of just the two states $\{\ket{\bsd},\ket{\bsu}\}$ at each site is relatively well understood in terms of the spin degree of freedom, governed by the Lie algebra $\su(2)$ \cite{Holstein_1940,Dyson_1956}. The complication in  the present case is the fermionic nature of the electron, which induces a graded structure between $\{\ket{\bsd},\ket{\bsu}\}$ and $\{\ket{\bqe},\ket{\bqd}\}$.

For concreteness we focus on a Hamiltonian  which encompasses both the Hubbard  and \tJ  models,
\be\la{eq:ham}
\OH=\sum_{\braket{i,j}}\OT_{ij} + J \sum_{\braket{i,j}} \vec{\Os}_i\cdot \vec{\Os}_j+U \sum_i  \OV^H_i -2\mu\sum_i \Oet_i^z,
\ee
on a $d$-dimensional hypercubic lattice.
The Heisenberg spin interaction is expressed through the local spin operators 
\be\la{eq:spin}
\Os^z=\frac{1}{2}(\On_{\ssu}-\On_{\ssd}),~~\Os^+=\Ocd_{\ssu} \Oc_{\ssd},~~\Os^-=\Ocd_{\ssd} \Oc_{\ssu},
\ee
which obey $ [\Os^z,\Os^\pm]=\pm\Os^\pm$ and $[\Os^+,\Os^-]=2\Os^z$, and generate $\su(2)$ rotations between  the local spin doublet $\{\ket{\bsd},\ket{\bsu}\}$. In addition it is useful to introduce the corresponding local charge operators
\be\la{eq:charge}
 \Oet^z=\frac{1}{2}(\On_{\ssu}+\On_{\ssd}-1), ~~\Oet^+ =\Ocd_{\ssd} \Ocd_{\ssu},~~ \Oet^- = \Oc_{\ssu} \Oc_{\ssd},
\ee
which obey $[\Oet^z,\Oet^\pm]=\pm\Oet^\pm$ and  $[\Oet^+,\Oet^-]=2\Oet^z$, and generate $\su(2)$ rotations between  the local charge doublet $\{\ket{\bqe},\ket{\bqd}\}$.
We choose the Hubbard interaction 
\be
\OV^H=(\On_{\ssu}-1/2)(\On_{\ssd}-1/2),
\ee
 to be of a particle-hole symmetric form, and the chemical potential $\mu$ couples to the charge density.

\begin{figure}[tb]
\centering
\includegraphics[width=0.55\columnwidth]{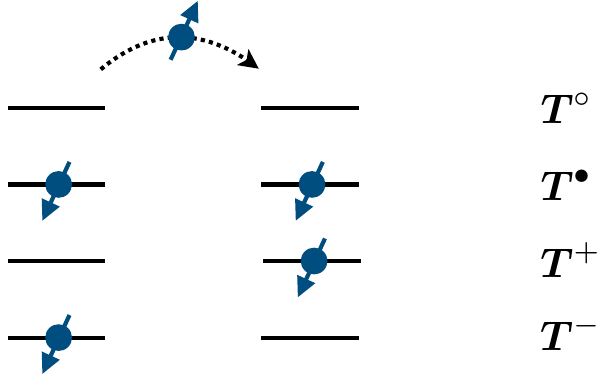}
\caption{\label{fig_chop}
Correlated hopping is when the hopping amplitude depends on how the two sites are occupied by electrons of the opposite spin. Here we illustrate the four possibilities for a hopping spin-up electron (the final two of which are hermitian conjugate). We argue that decoupling these amplitudes from the uncorrelated limit may induce a splitting of the electron.
}
\end{figure}

We take the kinetic term to be of a general correlated form 
\be\la{eq:CH}
\OT_{ij} =t(1-\lam) \OT^\qe_{ij}+t(1+\lam)\OT^\qd_{ij}+t_\pm (\OT^+_{ij}+\OT^-_{ij}),
\ee
where the three parameters $t$, $\lam$, $t_\pm$ decouple the terms 
\be
\begin{split}
\OT^\qe_{ij} &=- \sum_{\s=\ssd,\ssu} \big(\Ocd_{i\sigma} \Oc_{j\sigma} + \Ocd_{j\sigma} \Oc_{i\sigma}\big)\Obn_{i\bs}\Obn_{j\bs},\\
\OT^\qd_{ij} &=- \sum_{\s=\ssd,\ssu} \big(\Ocd_{i\sigma} \Oc_{j\sigma} + \Ocd_{j\sigma} \Oc_{i\sigma}\big)\On_{i\bs}\On_{j\bs},\\
\OT^+_{ij}&=- \sum_{\s=\ssd,\ssu} \big(\Ocd_{i\sigma} \Oc_{j\sigma}\On_{i\bs}\Obn_{j\bs} + \Ocd_{j\sigma} \Oc_{i\sigma}\Obn_{i\bs}\On_{j\bs}\big),\\
\OT^-_{ij}&=- \sum_{\s=\ssd,\ssu} \big(\Ocd_{i\sigma} \Oc_{j\sigma}\Obn_{i\bs}\On_{j\bs} + \Ocd_{j\sigma} \Oc_{i\sigma}\On_{i\bs}\Obn_{j\bs}\big),
\end{split}
\ee
with $\bs=-\s$ and $\Obn_{\s}=1-\On_{\s}$. This allows for distinct hopping amplitudes depending on the occupancy of the two sites involved by electrons of the opposite spin,  see Fig.~\ref{fig_chop}.
Correlated hopping is an important interaction in, for example, charge-transfer insulators \cite{Fujimori_1984,Zaanen_1985}, a family of materials which includes the cuprates, when described by an effective single-orbital lattice model that eliminates the low-lying ligand $p$ orbital degree of freedom \cite{Zhang_1988,Micnas89,MarsiglioHirsch,Sim_n_1993}.
In addition, it has recently been shown that correlated hopping can be induced as an effective interaction of ultracold atoms in periodically driven optical lattice setups \cite{Rapp12,Liberto2014}.
The \tJ model corresponds to an extreme form of correlated hopping $\lam=-1$, $t_\pm=0$, which disallows hopping processes involving doubly occupied sites. While the Hubbard and \tJ models are often regarded as good minimal models for characterising strong correlation effects, we will see that a rich and useful structure arises by considering this more general model which encompasses them both.

Conventional band theory is founded upon having a kinetic term that is bilinear in $\Ocs$, a feature that is lost when there is correlated hopping. We can however re-express the kinetic term through the generators of a different algebra as follows
\be\la{eq:CHQ}
\OT_{ij}=-\sum_{\s=\ssd,\ssu}\sum_{\nu=\qe,\qd} t_\nu
	\big(\OQd_{i\sigma\nu} \OQ_{j\sigma\nu} + \OQd_{j\sigma\nu} \OQ_{i\sigma\nu}\big),
\ee
which is now bilinear in 
\be\la{Q0s}
\begin{split}
\OQd_{\s\qe} &=\frac{1+\k}{2}\Oc_{\bs} -  \k  \On_{\s}\Oc_{\bs},\\
\OQd_{\s\qd} &=\bs\Big(\frac{1-\k}{2}\Ocd_{\s} +\k  \On_{\bs}\Ocd_{\s}\Big) ,
\end{split}
\ee
with hopping parameters given by
\be\la{eq:CHparam}
t_\nu =
\Big(\frac{2\nu }{1+\k^2}+\frac{\lam}{\k}\Big)t,
\quad \k=\sqrt{\frac{t-t_\pm}{t+t_\pm}},
\ee
where $\s$ takes values $-1,1$ for $\s=\bsd,\bsu$, and  $\nu$ takes values $-1,1$ for $\nu=\bqe,\bqd$ respectively. The $\OQs$ are the fermionic generators of the graded Lie algebra $\sucex$ \cite{Beisert07,Beisert08,HS1},  summarised in Appendix \ref{app:su22}. 
Their anti-commutation relations are 
\be\la{eq:su22}
\begin{split}
&  \{ \OQ_{\s\nu}, \OQd_{\s\nu}\}= \frac{1+\k^2}{4}+\k (\nu \Oet^z - \s \Os^z),\\
& \{\OQ_{\ssd\nu}, \OQd_{\ssu\nu}\}=\k{ \Os}^+, ~~~~~~~ 
	\{ \OQ_{\s\qe},\OQd_{\s\qd}\}= \k{ \Oet}^+,\\
& \{ \OQ_{\ssu\nu}, \OQd_{\ssd\nu}\}= \k{ \Os}^-, ~~~~~~~
	\{\OQ_{\s\qd}, \OQd_{\s\qe}\}= \k{\Oet}^-,\\
&  \{ \OQ_{\s\nu}, \OQ_{\s'\nu'}\}=\{ \OQd_{\s\nu}, \OQd_{\s'\nu'}\}=\frac{1-\k^2}{4}\e_{\s' \s}		\e_{\nu \nu'} ,
\end{split}
\ee
with $\e_{\ssd\ssu}=-\e_{\ssu\ssd}=\e_{\qe\qd}=-\e_{\qd\qe}=1$. They provide a non-canonical symmetry of the electronic degree of freedom, one that interplays with spin and charge.
The inversion of Eqs.~\eqref{Q0s} takes a linear form
\be\la{inv_rels}
\Ocd_{\ssd} =\OQ_{\ssu\qe}+\OQd_{\ssd\qd},\quad 
\Ocd_{\ssu} =\OQ_{\ssd\qe}-\OQd_{\ssu\qd},
\ee
and  we refer to this as a splitting of the electron, as opposed to `fractionalisation' which takes a product form.

While  graded Lie algebras are not commonly referred to by name in the physics literature, they are frequently used. Indeed, the canonical fermion algebra $\{\Oc,\Oc^\dagger\}=1$ is the graded Lie algebra $\su(1|1)$.  This is extended to  $\alg{u}(1|1)$ by adding $\On = \Oc^\dagger \Oc$, obeying $[\On ,\Oc^\dagger]=\Oc^\dagger$, $[\On ,\Oc]=-\Oc$. The canonical algebra of Eq.~\eqref{can_f} is  $\su(1|1)\otimes\su(1|1)$. This offers one way to characterise the electronic degree of freedom, which can be viewed as grouping the four electronic states as
\be
\{ \ket{\bqe};\ket{\bsd}\}\otimes \{ \ket{\bqe};\ket{\bsu}\}.
\ee
This canonical algebra underlies the Fermi liquid description of correlated matter.

The graded Lie algebra $\sucex$ offers an alternative way to characterise the electronic degree of freedom. Here it is useful to view the  four states grouped as
\be
\{ \ket{\bsd},\ket{\bsu}; \ket{\bqe},\ket{\bqd}\}.
\ee
The algebra contains $\su(2)$ spin generators $\vOS$ acting on the first pair,  $\su(2)$ charge generators $\vOet$ acting on the second pair, and fermionic generators $\OQs$ which act between the two pairs. The anti-commutation relations of the $\OQs$ are not canonical, but instead yield the generators $\vOS$ and $\vOet$ through Eqs.~\eqref{eq:su22}.
The algebra can be extended to $\ucex$ by adding 
$\OHu=\k \OV^H=\frac{\k}{3}(\vec{\Oet}\cdot\vec{\Oet}-\vec{\Os}\cdot\vec{\Os})$, 
which obeys
\be\la{VHQ0}
\begin{split}
\lbrack \OHu, \OQd_{\s\nu} \rbrack &= \frac{1+\k^2}{4 } \OQd_{\s\nu} +\frac{1-\k^2}{4 } \e_{\s\s'}\e_{\nu\nu'} \OQ_{\s'\nu'},\\
\lbrack \OHu, \OQ_{\s\nu} \rbrack &= -\frac{1+\k^2}{4 } \OQ_{\s\nu} -\frac{1-\k^2}{4 } \e_{\s\s'}\e_{\nu\nu'} \OQd_{\s'\nu'},
\end{split}
\ee
and commutes with the spin and charge generators. This linear action of $\OHu$ has the consequence that the parameter $U$ plays a role akin to an additional chemical potential for the $\OQs$ degrees of freedom, controlling their splitting. 
For $\k=1$,  the algebra $\ucex$ is closely related to the Hubbard algebra  \cite{Hubbard4}, see Appendix~\ref{app:su22}. The appearance of $\k$ in the algebra formally corresponds to an exceptional central extension \cite{Beisert07,Beisert08}. It has the role of suppressing the spin and charge generators in the anti-commutation relations Eqs.~\eqref{eq:su22} for small $\k$. We will exploit this to gain perturbative control over the growth of correlations. As $\k\to0$ the $\OQs$ collapse pairwise onto the $\Ocs$, the anti-commutation relations reduce to canonical relations of Eq.~\eqref{can_f}, the kinetic term becomes uncorrelated, and $\OHu$ vanishes.

We thus see there are two possibilities for characterising the electronic degree of freedom: $\su(1|1)\otimes\su(1|1)$ and $\sucex$. Both are graded algebras, which inherently take into account the grading of the four states of Eq.~\eqref{4states}. The graded Lie algebras have been classified \cite{kac1977lie}, and there do not appear to be other independent possibilities relevant for the single-orbital electronic problem.

Let us emphasise that we will not consider to what extent these algebras provide explicit symmetries of a system. Instead we will examine how they govern the underlying degrees of freedom, 
i.e.~how they organise correlations. There is no fine tuning in this approach. 

The canonical degree of freedom governs the Fermi liquid description of electronic matter. In the next two sections we will show 
that $\sucex$ degrees of freedom underlie a controlled description of an alternative strongly correlated regime.


\section{Green's function analysis}\la{sec:GF}

In the previous section we have identified two ways to characterise the electronic degree of freedom. We now demonstrate that they each offer a means to systematically organise the electronic correlations of an interacting system.

We focus our effort on obtaining the electronic Green's function. 
Let us first review how the imaginary-time formalism provides access to the retarded and advanced Green's functions
\be\la{GRA}
\begin{split}
G^{\ret}_{ij\s}(t) &= - i  \Theta(t) \braket{ \{\Oc_{i\s}(t), \Ocd_{j\s}(0)\} } ,\\
G^{\adv}_{ij\s}(t) &=  i  \Theta(-t) \braket{ \{\Oc_{i\s}(t), \Ocd_{j\s}(0)\} },
\end{split}
\ee
with $ \Theta$ the Heaviside function.
We start with the imaginary-time thermal Green's function 
\be\la{thGFcan}
\begin{split}
{\GT_{ij\s}}(\t)= &- \braket{\Oc_{i\s}(\t) \Ocd_{j\s}(0)}\\
=&-\frac{1}{\cZ}\Tr \Big(e^{-\beta \OH}\cT\big[\Oc_{i\s}(\t) \Ocd_{j\s}(0)\big]\Big),
\end{split}
\ee
where $\cZ=\Tr e^{-\beta \OH}$, $\beta$ is inverse temperature, $\OA(\t)= e^{\t\OH}\OA e^{-\t\OH}$, and $\cT$ is the $\t$-ordering operator which is antisymmetric under interchange of fermionic operators
\be
\cT\big[\Oc_{i\s}(\t) \Ocd_{j\s}(0)\big] = \Theta(\t) \Oc_{i\s}(\t) \Ocd_{j\s}(0) - \Theta(-\t) \Ocd_{j\s}(0) \Oc_{i\s}(\t).
\ee
Taking the $\t$-derivative yields the  equation of motion
\be\la{elEoM}
\pa_{\t} \cG_{ij\s}(\t) = - \delta(\t)\delta_{ij}-\braket{ [\OH,\Oc_{i\s}(\t)] \Ocd_{j\s}(0)}.
\ee
The advantage over the real time equation of motion is the anti-periodic boundary condition 
$\cG_{ij\s}(\beta) = - \cG_{ij\s}(0)$, which follows  from the cyclicity of the trace and antisymmetry of $\cT$. The Fourier transform
\be\la{FT}
\cG_{p\s}(i\om_n) = \frac{1}{\vol}\sum_{i,j}\int_0^{\beta}  d\t  e^{\ii \om_n\t-\ii p(i-j)} \cG_{ij\s}(\t),
\ee
is then defined at the Matsubara frequencies $\om_n=(2n+1)\frac{\pi}{\beta}$, with $n\in\mathbb Z$, and $\vol$ is the total number of lattice sites. We define $G_{p\s}(\om)$ by analytically continuing $\cG_{p\s}(\om)$ to all non-real $\om$, provided it satisfies the causality condition that it has no singularities in this region. The retarded and  advanced Green's functions are then obtained as
\be
G^{\ret}_{p\s}(\om)=G_{p\s}(\om+\ii 0^+),\quad G^{\adv}_{p\s}(\om)=G_{p\s}(\om-\ii 0^+).
\ee

It appears that the challenge of computing the Green's function revolves around solving the equation of motion, Eq.~\eqref{elEoM}. For example  if $\OH$ is bilinear in $\Ocs$, say 
 $\OH=- \sum_{i,j,\s} t_{ij} \Ocd_{i\s} \Oc_{j\s}- \mu\sum_{i,\s} \On_{i\s}$, then the equation of motion takes the form
\be\la{GF0can}
\begin{split}
\sum_k \Big[ \delta_{ik} \big( &-\pa_\t + \mu\big)+ t_{ik} \Big] \GT_{kj\s}(\t) = \delta(\t)\delta_{ij},
\end{split}
\ee
which upon Fourier transformation becomes 
\be
(i\om_n+\mu-\ve_p)\cG_{p\s}(i\om_n) = 1,
\ee
 with dispersion relation $\ve_p = -\frac{1}{\vol}\sum_{i,j} t_{ij} e^{\ii p(i-j)}$.  Inverting, and analytically continuing $\cG_{p\s}(\om)$ to all non-real $\om$, results in the non-interacting Green's function
\be
G_{p\s}(\om) = \frac{1}{\om+\mu-\ve_p}.
\ee
The Hamiltonian of  Eq.~\eqref{eq:ham} is not bilinear in $\Ocs$ however. It contains both biquadratic and bicubic terms, and these induce correlations in the system.

One way to proceed is to investigate how the growth of  correlations is controlled by Eq.~\eqref{elEoM}, with a perturbative treatment of the interactions. This leads to the canonical description of correlated electrons which underlies the Fermi liquid \cite{Abrikosov,kadanoff1962quantum}. We review this  in Appendix~\ref{app:canGF} for the case of spinless fermions. Our subsequent analysis parallels the discussion there, and the reader may find it useful to contrast the two.

We now however take an alternative route, and  consider the Green's functions  of the $\sucex$ degrees of freedom, e.g.~$\braket{\OQ_{i\s\nu}(\t) \OQd_{j\s'\nu'}(0)}$. We will use their equation of motion to 
gain control of correlations, employing the Green's function factorisation technique recently pioneered by Shastry \cite{Shastry_2011,Shastry_2013}. As the splitting of Eqs.~\eqref{inv_rels} is linear, the electronic Green's functions $\GT_{ij\s}(\t)$ are immediately reobtained through linear combinations of the $\sucex$ Green's functions. In this way we gain access to a regime of strongly correlated behaviour.

We will continue our analysis in an explicit manner. While this obscures the presentation to a certain extent, it has the benefit of avoiding ambiguity. We complement this with Appendix~\ref{app:sch} which contains a schematic summary of the derivation.

It is useful to introduce some simplifying notations. We collect the fermionic generators as
\be
\OF_i^\a = \left(\begin{array}{cccccccc}
\OQd_{i\ssu\qe}&\OQ_{i\ssd\qd}& \OQd_{i\ssd\qe}&\OQ_{i\ssu\qd}&
 \OQ_{i\ssu\qe}&\OQd_{i\ssd\qd}&\OQ_{i\ssd\qe}&\OQd_{i\ssu\qd}
 \end{array}\right),
 \ee
with greek indices, and the bosonic generators as
\be
 \OB_i^a = \left(\begin{array}{cccccc}
 \OS_i^z&\OS_i^-&\OS_i^+ &\Oet_i^z&\Oet_i^-&\Oet_i^+ 
\end{array}\right),
\ee 
with latin indices. The  $\sucex$ algebra is then compactly expressed as 
\be\la{su22alg}
\begin{split}
\{ \OF_i^\a,\OF_j^\beta\} & = \delta_{ij} \big(  f^{\a\beta}{}_I + f^{\a\beta}{}_a \OB_i^a\big),\\
 [ \OB_i^a,\OF_j^\beta]  &= \delta_{ij}f^{a \beta}{}_\g \OF_i^\g,\quad\\
 [ \OB_i^a,\OB_j^b]  &= \delta_{ij}f^{ab}{}_c \OB_i^c,
 \end{split}
\ee
 and the extension to $\ucex$ is given by
\be
 [ \OHu_i,\OF_j^\a]  = \delta_{ij}f^{\Theta \a}{}_\beta \OF_i^\beta,\quad [ \OHu_i,\OB_j^a]  =0.
\ee
Summation over repeated algebraic indices is implied, and we collect the  structure constants $f$ in Appendix~\ref{app:params}.

We now consider the Hamiltonian 
\be\la{hamTTR}
\begin{split}
\OH =& -\frac{1}{2}\sum_{i,j}  t_{ij,\a\beta}   \OF_i^\a  \OF_{j}^\beta 
	+\frac{1}{2}\sum_{i,j} V_{ij,ab}  \OB_i^a \OB_j^b\\
	 & \qquad -  \mu_{a} \sum_i \OB_i^a+\Ut\sum_i \OHu_i,
\end{split}
\ee
with hopping and interaction parameters obeying $t_{ii,\a\beta}=0$, $t_{ji,\a\beta}=t_{ij,\a\beta} $, $t_{ij,\beta\a}=-t_{ij,\a\beta} $ and $V_{ii,ab}=0$, $V_{ji,ab}=V_{ij,ab}$, $V_{ij,ba}=V_{ij,ab}$, chemical potentials $\mu_a=( h~0~0~2\mu~0~0)$, and $\Ut=U/\k$. This model is extremely general, as the sixteen generators $\{8\times \OF,6\times \OB, 1,\OHu\}$  provide a complete basis for the local operators at each site.
This reflects the wide range of applicability of our approach, though we remind that it is important for the model to have correlated hopping. We include the specific hopping and interaction parameters corresponding to the Hamiltonian of Eq.~\eqref{eq:ham} in Appendix~\ref{app:params}. 

To introduce the Green's function of the $\OQ$ it is useful  to first set a matrix structure via 
\be\la{eq:metric}
\OF_{i\a} = \OF_i^\beta K_{\beta\a}= \big(\OF_i^\a\big)^\dagger,
\ee
defining a metric $K$, presented explicitly in Appendix~\ref{app:params}. 
Our object of study is then the matrix Green's function 
\be
\GT_{ij}{}^\a_\beta (\t,\t')=  -{\braket{\OF_{i}^\a(\t) \OF_{j\beta}(\t')}}.
\ee
As highlighted above, the electronic Green's function is directly obtained from linear combinations of these, via Eqs.~\eqref{inv_rels},
\be\la{GeGQ}
\begin{split}
\GT_{ij\ssd}(\t) &= {\GT_{ij}}^1_1(\t)+{\GT_{ij}}^1_2(\t)+{\GT_{ij}}^2_1(\t)+{\GT_{ij}}^2_2(\t),\\
\GT_{ij\ssu}(\t) &= {\GT_{ij}}^3_3(\t)-{\GT_{ij}}^3_4(\t)-{\GT_{ij}}^4_3(\t)+{\GT_{ij}}^4_4(\t),
\end{split}
\ee
with $\GT_{ij}{}^\a_\beta (\t)=\GT_{ij}{}^\a_\beta (\t,0)$.
In addition, as the bosonic generators $\vOS$ and $\vOet$ are quadratic in $\Oc$, see Eqs.~\eqref{eq:spin} and \eqref{eq:charge}, we can also use Eqs.~\eqref{inv_rels} to obtain
\be\la{GtoT}
\begin{split}
\braket{\OB_i^a(\t)} &= \vG^a{}^\a_\beta \GT_{ii}{}^\beta_\a (\t,\t^+),
\end{split}
\ee
with coefficients $\vG^a{}^\a_\beta$ which are independent of $\k$, presented explicitly in Appendix~\ref{app:params}.

Although the Hamiltonian is at most bilinear in the generators of $\sucex$, correlations are nevertheless induced as a result of the non-canonical nature of the algebra. 
To handle these we incorporate sources for the $\OB$ into the imaginary-time thermal expectation value as follows
\be\la{ev_source}
 \braket{ \cO(\t_1,\t_2,\ldots)} = \frac{\Tr \Big( e^{-\beta  \OH} \cT \big[e^{\int_0^\beta d\t \cS(\t)} \cO(\t_1,\t_2,\ldots) \big]   \Big)}{\Tr \big( e^{-\beta H} \cT [e^{\int_0^\beta d\t \cS(\t)}]   \big)},
\ee
with $\cS(\t)  = \sum_i \sou_{ia}(\t) \OB^a_i(\t)$, and we consider all $\t$ to take values on the interval $(0,\beta)$.
The source term breaks translational invariance in both time and space, providing a means of organising correlations by trading bosonic correlations for their variations through 
\be
\begin{split}
\DV_i^a(\t)\braket{\cO(\t_1,\t_2,\ldots,\t_n)}= &\braket{\OB^a_i(\t)\cO(\t_1,\t_2,\ldots,\t_n)}\\
	& -\braket{\OB^a_i(\t)}\braket{\cO(\t_1,\t_2,\ldots,\t_n)},
\end{split}
\ee
where $\DV_i^a(\t) =  \frac{{\delta} }{ {\delta} \sou_{ia}(\t^+)}$ denotes the functional derivative, and $\t^+=\t+0^+$ incorporates an infinitesimal regulator which ensures a consistent ordering when $\t$ is one of the $\t_1,\t_2,\ldots,\t_n$. At the end of the computation the sources will be set to zero without difficulty, restoring translational invariance.

As for the electronic Green's function, there is again the  anti-periodic boundary condition 
\be
\GT_{ij}{}^\a_\beta(\beta,\t) = - \GT_{ij}{}^\a_\beta(0,\t).
\ee
The equation of motion 
\be
\begin{split}
\pa_{\t} \GT_{ij}{}^\a_\beta(\t,\t') =  -\delta(\t-\t')\braket{\{\OF_{i}^\a(\t), \OF_{j\beta}(\t)\}}&\\
	 + \braket{[\cS(\t), \OF^\a_i(\t)]\OF_{j\beta}(\t')}&\\
	 - \braket{ [\OH,\OF^\a_i(\t)]\OF_{j\beta}(\t') }&,
\end{split}
\ee
picks up an additional contribution from the source term, a consequence of the $\t$-ordering operator. 
The first two terms are straightforwardly evaluated  from Eqs.~\eqref{su22alg}
\be
\begin{split}
\braket{\{\OF_{i}^\a(\t), \OF_{j\beta}(\t)\}} &=\delta_{ij}\big( f^{\a\g}{}_I + f^{\a\g}{}_a \braket{\OB_i^a(\t)}
	\big)K_{\g\beta},\\
\braket{[\cS(\t), \OF^\a_i(\t)]\OF_{j\beta}(\t')} &= - f^{a\a}{}_\g \sou_{ia}(\t)  \GT_{ij}{}^\g_\beta(\t,\t').
\end{split}
\ee
\begin{widetext}\noindent
The commutator in the final term is
\be
\begin{split}
\lbrack\OH,\OF^\a_i\rbrack =
	\sum_{l}\big[  f^{\a\delta}{}_I t_{il,\delta\g} \OF_l^\g
	+  f^{\a\delta}{}_a t_{il,\delta\g} \OB_i^a \OF_l^\g\big]
	+ \sum_{l}f^{a\a}{}_\g V_{il,ab}\OB_l^b \OF_i^\g 
	- \mu_a f^{a\a}{}_\g \OF_i^\g + \Ut f^{\Theta\a}{}_\g \OF_i^\g,
\end{split}
\ee
and, recasting the bosonic correlations as variations of the sources, we obtain
\be
\begin{split}
\braket{\lbrack\OH,\OF^\a_i(\t)\rbrack \OF_{j\beta}(\t')}=& (\mu_a f^{a\a}{}_\g - \Ut f^{\Theta\a}{}_\g)  \GT_{ij}{}^\g_\beta(\t,\t')  
	 -\sum_l f^{a\a}{}_\g V_{il,ab}  \big(\braket{\OB_l^b(\t)}
		+ \DV_l^b(\t) \big) \GT_{ij}{}^\g_\beta(\t,\t')\\
	&- \sum_l f^{\a\delta}{}_I t_{il,\delta\g}   \GT_{lj}{}^\g_\beta(\t,\t')
	- \sum_l  f^{\a\delta}{}_a t_{il,\delta\g}  \big(
		\braket{\OB_i^a(\t)}+\DV_i^a(\t)\big) \GT_{lj}{}^\g_\beta(\t,\t').
\end{split}
\ee 
Collecting these expressions, the equation of motion  takes the form
\be\la{GFeqn}
\begin{split}
\sum_k \Big[ 
\delta_{ik}\Big(-\delta^\a_\g \pa_{\t} -f^{a\a}{}_\g \sou_{ia}(\t) 
 - \mu_a f^{a\a}{}_\g + \Ut f^{\Theta\a}{}_\g + \sum_l f^{a\a}{}_\g V_{il,ab} \big(\braket{\OB_l^b(\t)}
		 +  \DV_l^b(\t) \big)\Big) ~~~~~~~~~~~~~~~ &\\
+ f^{\a\delta}{}_I t_{ik,\delta\g}
	+f^{\a\delta}{}_a t_{ik,\delta\g} \big( \braket{\OB_i^a(\t)} 
	 + \DV_i^a(\t) \big)\Big] \GT_{kj}{}^\g_\beta(\t,\t')&\\
= \delta(\t-\t')\delta_{ij}\big(f^{\a\g}{}_I+ f^{\a\g}{}_a\braket{\OB^a_i(\t)}\big) K_{\g\beta}&.
\end{split}
\ee

We want to obtain solutions to this equation. Its analogue in the canonical case is Eq.~\eqref{canGFeqn}, to which it has a very similar structure. The primary complication of the non-canonical degree of freedom is the appearance of $\braket{\OB}$ on the right-hand side, which indicates that the spectral weight of the Green's function is dressed by correlations. Here it depends explicitly on $\GT$ through Eq.~\eqref{GtoT}. A technique for overcoming this difficulty has been pioneered by Shastry \cite{Shastry_2011,Shastry_2013}: the trick is to factorise $\GT$ into its numerator and denominator, and obtain a coupled controlled description of both \cite{Shastry11_Anatomy}. In practice we write \footnote{The asymmetry in this factorisation $\GT=\gT\wT$ results from considering the equation of motion $\pa_\t \GT(\t,\t')$. Alternatively we could consider $\pa_{\t'} \GT(\t,\t')$, and then factorise the Green's function as $\GT=\wT\gT$.}
\be\la{Shansatz}
\GT_{ij}{}^\a_\beta(\t,\t') = \sum_l \int_0^\beta d\t''  \gT_{il}{}^\a_\g(\t,\t'') \wT_{lj}{}^\g_\beta(\t'',\t').
\ee
The functional derivative in Eq.~\eqref{GFeqn} then gives two contributions
\be
\DV_l(\t'') \GT_{ij}{}^\a_\beta(\t,\t') = \sum_k \int_0^\beta d\t''' \Big[ \Big(\DV_l(\t'')\gT_{ik}{}^\a_\g(\t,\t''')\Big) \wT_{kj}{}^\g_\beta(\t''',\t')+  \gT_{ik}{}^\a_\g(\t,\t''') \Big(\DV_l(\t'')\wT_{kj}{}^\g_\beta(\t''',\t')\Big)\Big].
\ee
Substituting these into Eq.~\eqref{GFeqn}, and bringing the terms with $\DV\wT$ to the right-hand side,  
permits a factorisation of the equation of motion. 
Setting
\be\la{eq:cW}
\begin{split}
 \wT_{ij}{}^\a_\beta(\t,\t') = \delta(\t-\t')\delta_{ij}\big(f^{\a\g}{}_I+ f^{\a\g}{}_a\braket{\OB^a_i(\t)}\big)K_{\g\beta} -\sum_{k,l}\int_0^\beta d\t'' \Big( 
 &f^{\a\delta}{}_a t_{il,\delta\e} \gT_{lk}{}^\e_\g(\t,\t'') \DV^a_i(\t)\wT_{kj}{}^\g_\beta(\t'',\t') \\
 &+f^{a\a}{}_\delta V_{il,ab} \gT_{ik}{}^\delta_\g(\t,\t'') \DV^b_l(\t)\wT_{kj}{}^{\g}_\beta(\t'',\t') \Big),
\end{split}
 \ee
 fixes the ratio between the two factors in Eq.~\eqref{Shansatz},
with the remainder satisfying 
\be\la{eq:gT}
\begin{split}
\sum_k \Big[ 
\delta_{ik}\Big(-\delta^\a_\g \pa_{\t} -f^{a\a}{}_\g \sou_{ia}(\t) 
 - \mu_a f^{a\a}{}_\g + \Ut f^{\Theta\a}{}_\g + \sum_l f^{a\a}{}_\g V_{il,ab} \big(\braket{\OB_l^b(\t)}
		 +  \DV_l^b(\t) \big)\Big)& \\
 ~~~~~~~~~~~~~~~~~~~~~~~~~~~ +  f^{\a\delta}{}_I t_{ik,\delta\g}
	+ f^{\a\delta}{}_a t_{ik,\delta\g} \big( \braket{\OB_i^a(\t)} 
	 + \DV_i^a(\t) \big)&\Big] \gT_{kj}{}^\g_\beta(\t,\t')	= \delta(\t-\t')\delta_{ij}\delta^\a_\beta.
\end{split}
\ee
These two coupled equations are an exact rewriting of the equation of motion Eq.~\eqref{GFeqn}.
We call $\gT$ the canonised Green's function and $\wT$ the spectral weight.

We proceed by introducing two functionals $\Sg[\gT,\wT]$ and $ \SW[\gT,\wT]$ of the full $\gT$ and $\wT$ as follows \footnote{Refs.~\cite{Shastry_2011,Shastry_2013} treats these as functionals of $\gT$ only, corresponding to a perturbative expansion of $\wT$.}. We 
define the self-energy $\Sg$ through
\be\la{gTI}
\gT^{-1}_{ij}{}^\a_\beta(\t,\t') = \gT_{0,ij}^{-1}{}^\a_\beta(\t,\t') - \Sg_{ij}{}^\a_\beta(\t,\t'),
\ee
where $\gT_0$ satisfies
\be
 \Big[ 
\delta_{ik}\big(-\delta^\a_\g \pa_{\t} -f^{a\a}{}_\g \sou_{ia}(\t) 
 - \mu_a f^{a\a}{}_\g + \Ut f^{\Theta\a}{}_\g\big) + f^{\a\delta}{}_I t_{ik,\delta\g}
 		\Big] \gT_{0,kj}{}^\g_\beta(\t,\t')\\
	= \delta(\t-\t')\delta_{ij}\delta^\a_\beta,
\ee
and the adaptive spectral weight $\SW$ through
\be\la{wT}
\wT_{ij}{}^\a_\beta(\t,\t') = \wT_{0,ij}{}^\a_\beta(\t,\t') + \SW_{ij}{}^\a_\beta(\t,\t'),
\ee
with
\be
\wT_{0,ij}{}^\a_\beta(\t,\t')= \delta(\t-\t')\delta_{ij}f^{\a\g}{}_I K_{\g\beta}.
\ee
We obtain a closed equation  for $\Sg$ by convolving Eq.~\eqref{eq:gT} on the right with $\gT^{-1}$, which gives
\be\la{Sg0}
\begin{split}
\Sg_{ij}{}^\a_\beta(\t,\t')  =& - \delta(\t-\t')\Big(
	f^{\a\g}{}_a t_{ij,\g\beta}  \braket{\OB_i^a(\t)} +
	\delta_{ij} \sum_l f^{a\a}{}_\beta V_{il,ab} \braket{\OB_l^b(\t)}
	\Big)\\
&- \delta(\t-\t')\Big(
	\delta_{ij}\sum_l f^{\a\delta}{}_a  t_{il,\delta\e} \gT_{li}{}^\e_\g(\t,\t^+)f^{a\g}{}_\beta
 	+f^{a\a}{}_\delta V_{ij,ab} \gT_{ij}{}^\delta_\g(\t,\t^+)f^{b\g}{}_\beta 
	\Big)\\
&-\sum_{k,l}\int_0^\beta d\t'' \Big(
	f^{\a\delta}{}_a t_{il,\delta\e} \gT_{lk}{}^\e_\g(\t,\t'') \DV_i^a(\t)\Sg_{kj}{}^{\g}_\beta(\t'',\t')
	+f^{a\a}{}_\delta V_{il,ab} \gT_{ik}{}^\delta_\g(\t,\t'') \DV_l^b(\t)\Sg_{kj}{}^{\g}_\beta(\t'',\t')
	\Big),
\end{split}
\ee
upon using $(\DV \gT)\gT^{-1}=-\gT\DV\gT^{-1}= -\gT\DV\gT_0^{-1} + \gT\DV\Sg$, with
\be
\DV_l^a(\t'')\gT_{0,ij}^{-1}{}^\a_\beta(\t,\t')=-\delta(\t-\t')\delta(\t-\t''-0^+)\delta_{ij}\delta_{il}f^{a\a}{}_\beta.
\ee
A closed equation for $\SW$ follows directly from Eq.~\eqref{eq:cW},
\be\la{SW0}
\begin{split}
 \SW_{ij}{}^\a_\beta(\t,\t') = \delta(\t-\t')\delta_{ij} f^{\a\g}{}_aK_{\g\beta} \braket{\OB^a_i(\t)} -\sum_{k,l}\int_0^\beta d\t'' \Big( 
 &f^{\a\delta}{}_a t_{il,\delta\e} \gT_{lk}{}^\e_\g(\t,\t'') \DV^a_i(\t)\SW_{kj}{}^\g_\beta(\t'',\t') \\
 &+f^{a\a}{}_\delta V_{il,ab} \gT_{ik}{}^\delta_\g(\t,\t'') \DV^b_l(\t)\SW_{kj}{}^\g_\beta(\t'',\t')\Big).
\end{split}
 \ee

Equations \eqref{Sg0} and \eqref{SW0} are exact. We now obtain successive approximate solutions with a perturbative expansion in $\k$. We introduce rescaled parameters $
\ft^{\a\beta}{}_a= f^{\a\beta}{}_a/\k$, $\Vt_{ij,ab} = V_{ij,ab}/\k$,
so that $t_{ij,\a\beta}$, $\Vt_{ij,ab}$, $f^{a\a}{}_\beta$ and $\ft^{\a\beta}{}_a$ are all independent of $\k$, and write $\Sg = \sum_{s=0}^\infty \k^s [\Sg]_s$ and $\SW = \sum_{s=0}^\infty \k^s [\SW]_s$. The
leading contributions are
\be\la{SW1}
\begin{split}
\lbrack \Sg_{ij}{}^\a_\beta(\t,\t') \rbrack_1= &-\delta(\t-\t')\sum_{k}\int_0^\beta d\t'' \Big(
	\ft^{\a\g}{}_a t_{ij,\g\beta} \vG^a{}^\r_\s \gT_{ik}{}^\s_\lam(\t,\t'') \wT_{ki}{}^\lam_\r(\t'',\t^+)\\
	&~~~~~~~~~~~~~~~~~~~~~~~~~~~~~ +\delta_{ij} \sum_l f^{a\a}{}_\beta \Vt_{il,ab} 
		\vG^b{}^\r_\s \gT_{lk}{}^\s_\lam(\t,\t'') \wT_{kl}{}^\lam_\r(\t'',\t^+)
	\Big),\\
&- \delta(\t-\t')\Big(
	\delta_{ij}\sum_l \ft^{\a\delta}{}_a  t_{il,\delta\e} \gT_{li}{}^\e_\g(\t,\t^+)f^{a\g}{}_\beta
 	+f^{a\a}{}_\delta \Vt_{ij,ab} \gT_{ij}{}^\delta_\g(\t,\t^+)f^{b\g}{}_\beta 
	\Big)\\
[ \SW_{ij}{}^\a_\beta(\t,\t') ]_1=&  \delta(\t-\t')\delta_{ij}\sum_{k}\int_0^\beta d\t'' 
	 \ft^{\a\g}{}_a K_{\g\beta} \vG^a{}^\r_\s \gT_{ik}{}^\s_\lam(\t,\t'') \wT_{ki}{}^\lam_\r(\t'',\t^+).
\end{split}
\ee
Higher order terms are then obtained recursively through
\be\la{SgWrec}
\begin{split}
\lbrack\Sg_{ij}{}^\a_\beta(\t,\t')\rbrack_{s+1} & = -\sum_{k,l}\int_0^\beta d\t'' \Big(
	\ft^{\a\delta}{}_a t_{il,\delta\e} \gT_{lk}{}^\e_\g(\t,\t'') \DV_i^a(\t)[\Sg_{kj}{}^{\g}_\beta(\t'',\t')]_s
	+f^{a\a}{}_\delta \Vt_{il,ab} \gT_{ik}{}^\delta_\g(\t,\t'') \DV_l^b(\t)[\Sg_{kj}{}^{\g}_\beta(\t'',\t')]_s
	\Big),\\
[\SW_{ij}{}^\a_\beta(\t,\t')]_{s+1} &= -\sum_{k,l}\int_0^\beta d\t'' \Big( 
	\ft^{\a\delta}{}_a t_{il,\delta\e} \gT_{lk}{}^\e_\g(\t,\t'') \DV^a_i(\t)[\SW_{kj}{}^\g_\beta(\t'',\t')]_s 	
	+f^{a\a}{}_\delta \Vt_{il,ab} \gT_{ik}{}^\delta_\g(\t,\t'') \DV^b_l(\t)[\SW_{kj}{}^\g_\beta(\t'',\t')]_s
	\Big).
\end{split}
\ee
These depend on the sources only through $\gT$ and $\wT$, and at each order we need use only the leading contributions from
\be
\DV_l^a(\t'') \gT_{ij}{}^\a_\beta(\t,\t') = \gT_{il}{}^\a_\g(\t,\t'') f^{a\g}{}_\delta \gT_{lj}{}^\delta_\beta(\t'',\t')+\cO(\k),\quad \DV_l^a(\t'') \wT_{ij}{}^\a_\beta(\t,\t') = 0+\cO(\k),
\ee

\newpage
\end{widetext} \noindent
where here we have suppressed  the infinitesimal regulator. In this way we can systematically construct the functionals $\Sg[\gT,\wT]$ and $ \SW[\gT,\wT]$ to any desired order. We provide the second order contributions explicitly in Appendix~\ref{app:2nd}.

We have thus succeeded in our goal. We have obtained a  series of successive approximations for the Green's function, mirroring the self-energy expansion of the canonical case. Let us summarise. Upon expanding $\Sg[\gT,\wT]$ and $\SW[\gT,\wT]$ to some desired order, the zero source limit is straightforwardly taken as $\sou$ enters only through $\gT_0$.  Equations~\eqref{gTI} and \eqref{wT}  then provide a set of coupled self-consistent equations for  $\gT$ and $\wT$. The solutions can be combined to give $\GT$, and  the electronic Green's function is  in turn obtained from Eqs.~\eqref{GeGQ}.

The simplest approximation is to take $\GT=\gT_0 \wT_0$. We will examine this in the following section, and find that it captures an essential feature of $\sucex$ degrees of freedom: a splitting of the electronic dispersion. The next approximation is to take just the first order contributions to the self-energy and adaptive spectral weight from Eqs.~\eqref{SW1}. This is the analogue of the Hartree-Fock approximation for the canonical case, see Eq.~\eqref{HF}, and likewise captures static correlations. The effects of collisions can be examined by including the second order contributions of Eqs.~\eqref{SW2}.

\section{A controlled approximation} \la{sec:approx}

In the previous section we have derived a systematic framework for characterising interacting electrons with $\sucex$ degrees of freedom. We now take the simplest approximation,  $\GT=\gT_0 \wT_0$, and investigate the resulting electronic Green's function. The unexpanded $\gT_0$ and $\wT_0$ contain explicit dependence on $\k$ through the structure constants $f^{\a\beta}{}_I$ and $f^{\Theta\a}{}_\beta$, expressed in Appendix~\ref{app:params}. That is, we are not setting $\k=0$, but rather are truncating the expansions of $\Sg$ and $\SW$ at the zeroth order.
The full dependence on the Hubbard interaction, as well as some of the hopping correlations, enter already here. The affect of the approximation is to suppress all spin and charge correlations. In particular, the Heisenberg spin-exchange interaction does not contribute at this order.

First we obtain the matrix Green's function of the $\OQ$. Setting the sources to zero, and recombining $\gT_0$ and $\wT_0$, the equation of motion
in this approximation becomes
\begin{widetext}
\be
\begin{split}
\sum_k \Big[ 
\delta_{ik}\big(-\delta^\a_\g \pa_{\t}  
 - \mu_a f^{a\a}{}_\g + \Ut f^{\Theta\a}{}_\g \big) 
+ f^{\a\delta}{}_I t_{ik,\delta\g}
	\Big] \GT_{kj}{}^\g_\beta(\t,\t')
= \delta(\t-\t')\delta_{ij} f^{\a\g}{}_I K_{\g\beta}.
\end{split}
\ee
\end{widetext}
\noindent
It is sufficient to restrict the greek indices to run over $\{1,2,3,4\}$.
Fourier transforming, performing matrix inversion, and analytically continuing to all non-real $\om$, we obtain
\be
{G_p}^\a_\beta(\om)=  
     \left(\begin{array}{cccc}
    \sfg^- & \sfh & 0 & 0 \\
    \sfh & \sfg^+ & 0 & 0 \\
    0 & 0 & \sfg^- & -\sfh \\
    0 & 0 & -\sfh & \sfg^+ \\
    \end{array}\right),
\ee
where
\be
\begin{split}
\sfg^\pm&=\frac{(1+\k^2)(\omega+\mu) -\frac{2\k ^2  \ve_p}{1+\k^2} \pm\k^2 (\Ut+\lamt \ve_p)}{4\big(\omega+\mu  -\frac{\ve_p}{1+\k^2}\big) \big(\omega+\mu -\frac{\k ^2\ve_p}{1+\k^2} \big)-\k^2(\Ut+\lamt \ve_p)^2},\\
\sfh&=\frac{(1-\k ^2) (\omega+\mu  )}{4\big(\omega+\mu  - \frac{\ve_p}{1+\k^2} \big) \big(\omega+\mu - \frac{\k^2 \ve_p}{1+\k^2} \big)-\k^2 (\Ut+\lamt \ve_p)^2},
\end{split}
\ee
with non-interacting dispersion  $\ve_p =- \frac{t}{\vol} \sum_{\braket{i,j}} e^{\ii p(i-j)}$, and $\lamt=\lam/\k$.

The electronic Green's function can now be immediately obtained via Eqs.~\eqref{GeGQ}, yielding
\be\la{eqG}
G_{p\s}(\om) = \frac{1}{\om +\mu - \frac{\ve_p}{1+\k^2} - \frac{\k^2}{4} \frac{(\Ut+\lamt \ve_p)^2}{\om +\mu -\frac{\k^2 \ve_p}{1+\k^2}}}.
\ee
We choose $t$, $\k$, $\Ut$ and $\lamt$ to parametrise the model, and ascribe the following roles: $t$ controls the strength of dispersion, $\k$ controls the strength of correlations, $\Ut$ controls the band splitting, and $\lamt$ controls asymmetry. They are related to the original parameters of the model by 
\be\la{chparams}
\k= \sqrt{\frac{t-t_\pm}{t+t_\pm}},~~\Ut=\frac{U}{\k},~~\lamt=\frac{\lam}{\k}.
\ee
While it may be tempting to view the term with prefactor $\frac{\k^2}{4}$ in the denominator as a self-energy, we suggest this would be a misinterpretation of the degrees of freedom. This is clarified by rewriting Eq.~\eqref{eqG} as  
\be\la{eqGS}
G_{p\s}(\om)  = \frac{a_{p\qe} }{\om+\mu - \om_{p\qe}} + \frac{a_{p\qd}}{\om+\mu - \om_{p\qd}},
\ee
which makes manifest the splitting of Eq.~\eqref{inv_rels}. 
There are now two dispersive bands, which we label with $\nu=\bqe,\bqd$ as follows
\be
\om_{p\nu} = \frac{\ve_p}{2} +\frac{\nu}{2} \sqrt{\Big(\frac{1-\k^2}{1+\k^2}\Big)^2\ve_p^2+\k^2 \big(\Ut+\lamt \ve_p\big)^2},
\ee
and the electronic spectral weight is split between them
\be
a_{p\nu} = \frac{1}{2}+\frac{\nu}{2}\frac{\frac{1-\k^2}{1+\k^2}\ve_p}{\sqrt{\big(\frac{1-\k^2}{1+\k^2}\big)^2\ve_p^2+\k^2 \big(\Ut+\lamt \ve_p\big)^2}},
\ee
with $a_{p\qe}+a_{p\qd}=1$.
This is in sharp contrast with the canonical perspective, i.e.~conventional band theory, where the entire electronic spectral weight is locked together in a single band.

\begin{figure}[!]
\centering
\includegraphics[width=0.99\columnwidth]{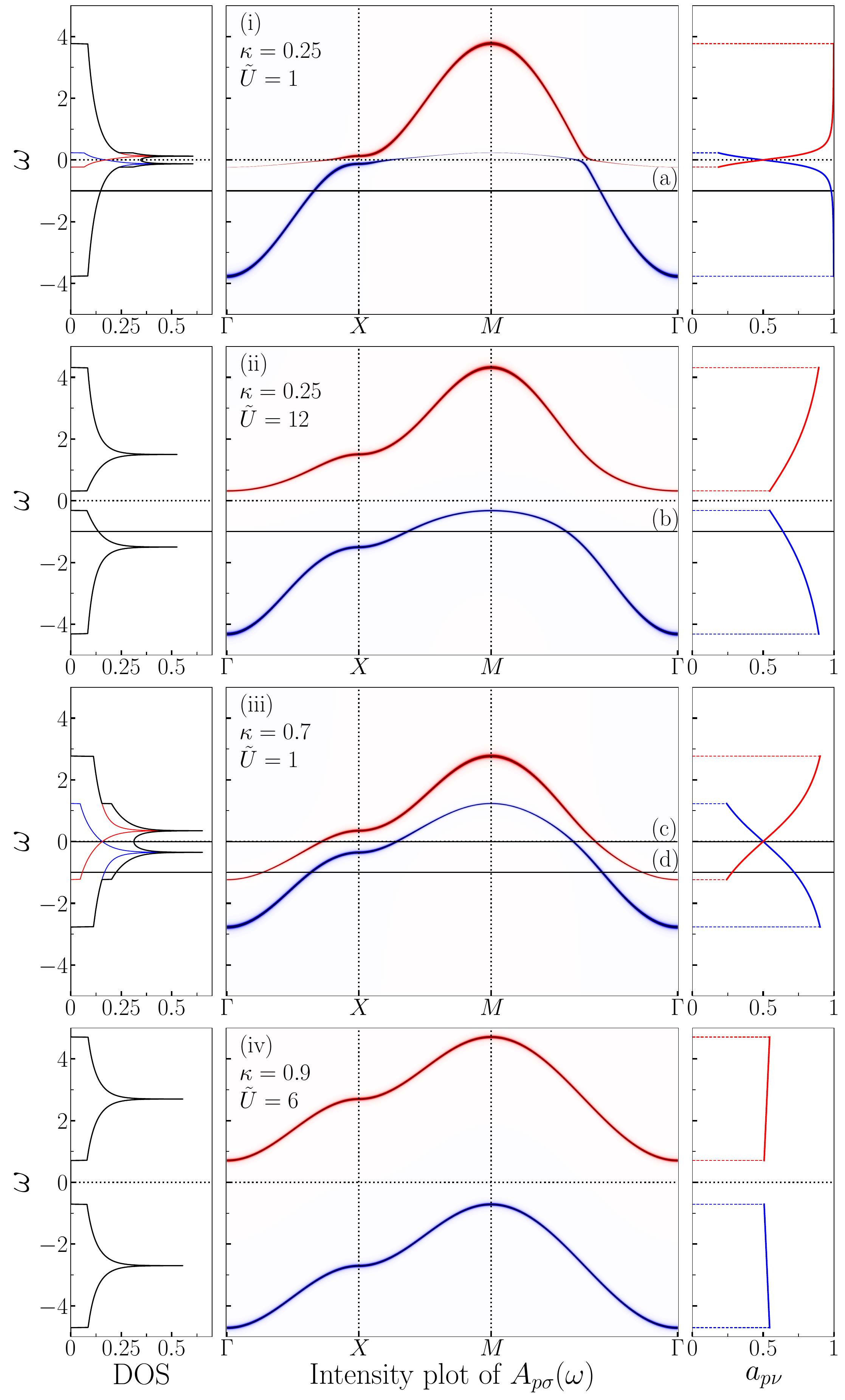}
\caption{\label{plot_BS}
The splitting of the electronic dispersion, exemplified on a square lattice.  
We focus on the symmetric case $\lamt=0$, and $J$ does not contribute at this order of approximation.
The left panel shows the electronic density of states (DOS), $\sum_\s\int_{BZ}\frac{d^2p}{(2\pi)^2} A_{p\s}(\om)$, with the contributions of the $\bqe$ (blue, lower) and $\bqd$ (red, upper) bands distinguished. 
The central panel shows the band structure, an intensity plot of the electronic spectral function (with Lorentzian broadening of $10^{-3}$), along the $\Gamma$-$X$-$M$-$\Gamma$ high-symmetry path in the Brillouin zone.  The right panel shows the spectral weights $a_{p\qe}$ and $a_{p\qd}$, which are momentum independent at this order of approximation. 
The horizontal lines (a)-(d) indicate slices along which the spectral function is plotted in Fig.~\ref{plot_FS}. Four examples (i)-(iv) of couplings $\k$ and $\Ut$ are presented: (i) on leaving the non-interacting point the band structure splits with the introduction of  weak and flat dispersion  near $\om=0$,  and the 2d Van Hove singularity of the DOS also splits in two. (ii) when $\Ut$ is increased above $\Ut_M=\frac{8}{1+\k^2}$ the two bands separate and a Mott gap opens. (iii) as the strength of correlated hopping is amplified the two bands overlap significantly for $\Ut<\Ut_M$. (iv) as $\k$ approaches 1 the two bands decouple, each with half the weight of an electron, and $\Ut$ behaves as an additional chemical potential that shifts the bands oppositely in $\om$.
}
\end{figure}

A Mott metal-insulator transition takes place when a gap opens between the two bands, i.e.~when $ \max_p \om_{p\qe} = \min_p \om_{p\qd}$.  For $\lamt=0$ and $W=2\max_p \ve_p=-2\min_p \ve_p$,  the transition occurs at $\Ut_M= \frac{W}{1+\k^2}$. The nature of the transition bears a close resemblance to a band insulator transition, but we emphasise the essential role of electronic correlations is reflected in the splitting of the electronic spectral weight  across the gap. 
This differs from the Brinkmann--Rice description of the Mott transition as the spectral weight  $a_{p\nu}$ does not go continuously to zero as the gap is opened  \cite{PhysRevB.2.4302}, and from the doublon-holon binding description \cite{doublon-holon} as there is no rearrangement of the degrees of freedom coincident with the Mott transition. 
The splitting of the electronic band is reminiscent of the foundational work of Hubbard \cite{Hubbard1,Hubbard3}, though our approach is very different from the large-$U$ perspective taken there.

It is illustrative to plot the electronic spectral function
\be
\begin{split}
A_{p\s}(\om)  &= - \frac{1}{\pi} \im G^\ret_{p\s}(\om)\\
&=a_{p\qe} \delta(\om - \om_{p\qe}) + a_{p\qd}\delta(\om - \om_{p\qd}).
\end{split}
\ee
 We focus on the example of nearest-neighbour hopping on the square lattice, with  dispersion relation $\ve_p=-2 \cos p_x -2\cos p_y$, setting $t=1$. In Fig.~\ref{plot_BS}  we plot the frequency dependence of the spectral function along the $\Gamma$-$X$-$M$-$\Gamma$ high-symmetry path in the Brillouin zone   for a choice of values of $\k$ and $\Ut$, with $\lamt=0$.  We also set $\mu=0$, but as $\mu$ enters Eq.~\eqref{eqGS} solely as a shift of $\omega$, the results for non-zero chemical potential correspond to translating the plots vertically in $\om$. The figure helps to visualise how the two bands emerge from a single band in the non-interacting limit, via hybridisation with an additional band carrying vanishing spectral weight.  
As the interactions are increased the two bands separate, and for $\Ut>\Ut_M$ a Mott gap is observed, with the vanishing of the carrier density at the transition evident through the density of states. 

\begin{figure}[tb]
\centering
\includegraphics[width=0.7\columnwidth]{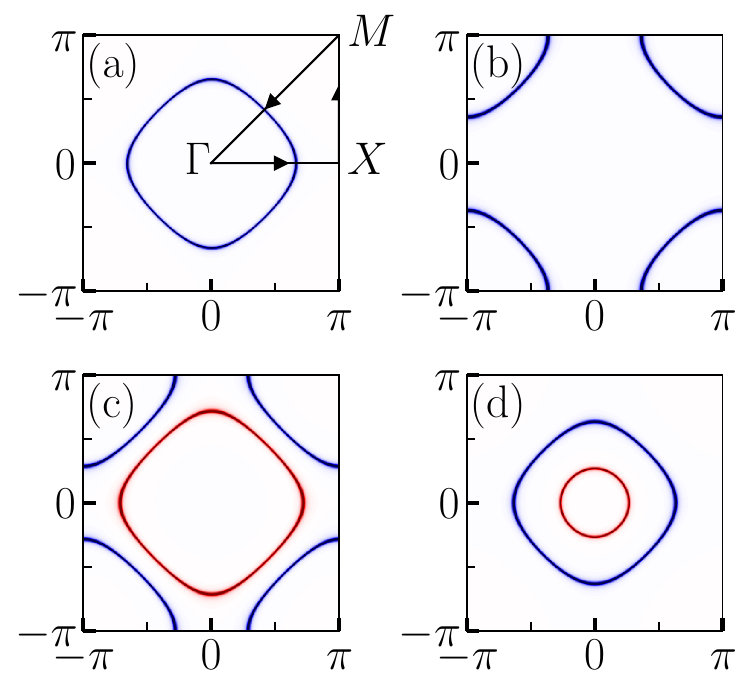}
\caption{\label{plot_FS}
Plots of the electronic spectral function throughout the 2d Brillouin zone on the slices (a)-(d) indicated in Fig.~\ref{plot_BS}. The violation of the Luttinger sum rule can be seen by contrasting between (a) and (b), both of which correspond to below half-filling: while less than half the Brillouin zone is enclosed in (a), this is clearly not the case in (b). In (c) and (d) the appearance of two surfaces is in sharp contrast to conventional band theory.
}
\end{figure}

Figure~\ref{plot_FS} displays cross sections of Fig.~\ref{plot_BS}, 
showing the spectral function throughout the 2d Brillouin zone for a choice of $\mu$, $\k$ and $\Ut$. 
This reveals surfaces which violate the Luttinger sum rule \footnote{The generalised sense of Luttinger's theorem argued in Ref.~\cite{Dzyaloshinskii_2003} is also violated, with the exception of the particle-hole symmetric case (i.e. here when $\lamt=\mu=0$) for which it has been proven to be true \cite{PhysRevB.75.104503,PhysRevB.96.085124}.}, clearly evidenced by contrasting Figs.~\ref{plot_FS}.(a) and \ref{plot_FS}.(b).
This is indeed reasonable as the sum rule relies on the existence of the Luttinger-Ward functional, which is tied to canonical characterisation of interactions \cite{LuttingerWard_1960,Luttinger_1960}.
The violation can be understood as a consequence of the non-canonical nature of the $\sucex$ degrees of freedom, for which the non-trivial spectral weight unties the link between electron density and Luttinger volume.

In summary, we have found that  $\sucex$ degrees of freedom govern a regime of behaviour which is fundamentally distinct from a Fermi liquid.

\section{Discussion}  \la{sec:disc}

The standard way to characterise the behaviour of interacting electrons is through perturbation theory from the non-interacting limit, built upon canonical degrees of freedom \cite{Abrikosov}. This logic is supported both by Landau's arguments on the robustness of the Fermi liquid \cite{landau1957,landau1959}, and Shankar's renormalisation group analysis \cite{Shankar_1994}. The approach has had great success, it underlies our understanding of a wide variety of materials.

Here we have identified a distinct way to characterise the electronic degree of freedom, and have demonstrated that it permits a description of a regime of behaviour different from the Fermi liquid. 
We have cast the electronic problem through the generators of the graded Lie algebra $\sucex$, and have shown how this provides a way to systematically organise the effects of electronic correlations. We have focused on the leading contribution, which reveals a splitting in two of the electronic band, see Fig.~\ref{plot_BS}. 
The Luttinger sum rule is violated, and a carrier-number-vanishing Mott metal-insulator transition is exhibited.
This reveals a scenario beyond Shankar's analysis, as that is formulated with canonical fermion coherent states which lack the freedom to capture the splitting of Eq.~\eqref{inv_rels}.

\begin{figure}[tb]
\centering
\includegraphics[width=0.85\columnwidth]{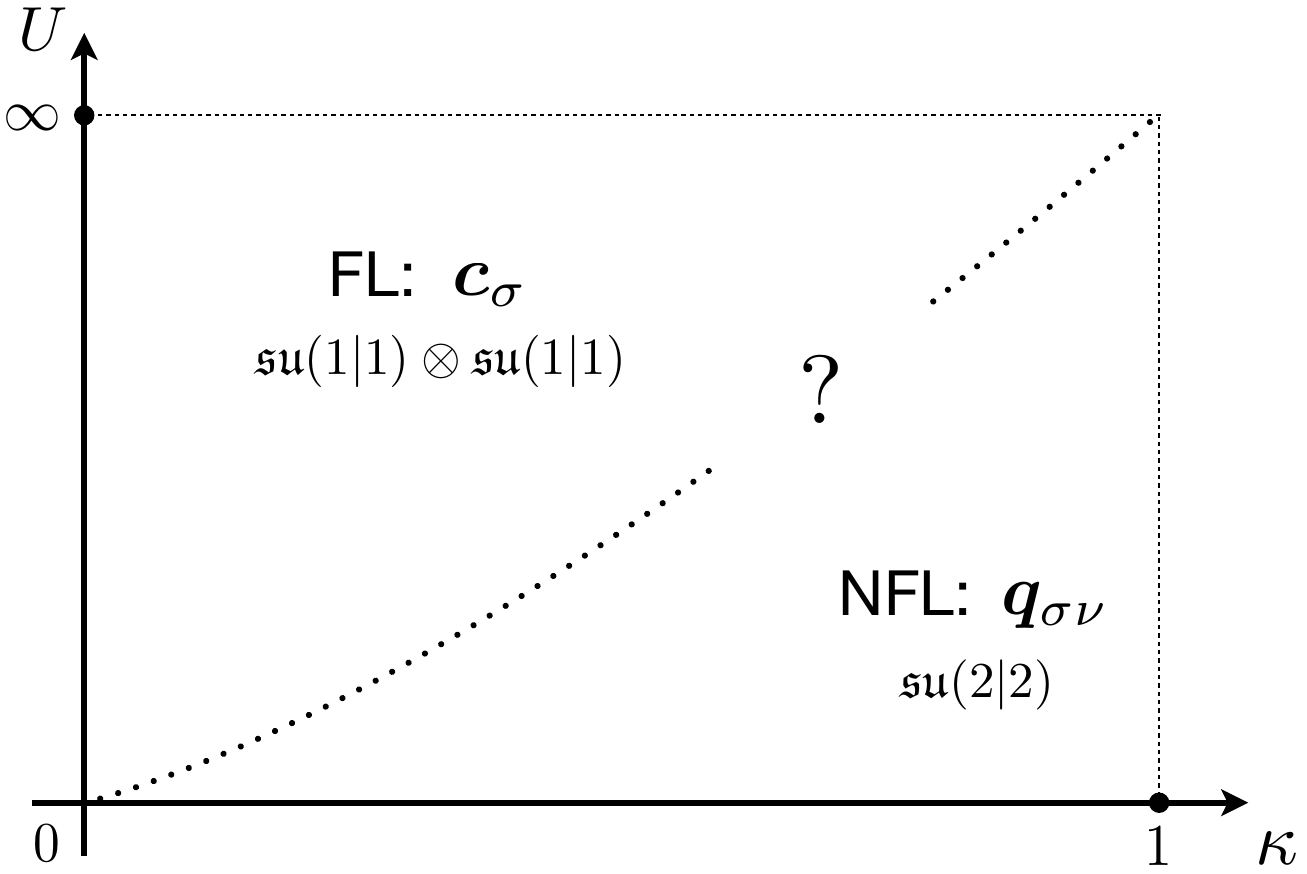}
\caption{\label{fig_kU}
A schematic depiction of how metallic  behaviour may be governed by either canonical or $\sucex$ degrees of freedom in different regions of parameter space.
Here correlated hopping, controlled by $\k$, and onsite  repulsion, controlled by $U$, compete to organise the electronic degree of freedom in distinct ways. While the $\sucex$ regime is restricted to small $U$ for small $\k$, it may extend to large $U$ when $\k$ is $\cO(1)$. We speculate on the nature of the `transition' between the two regimes towards the end of the Discussion.}
\end{figure}

The canonical description is expected to capture metallic behaviour for $\k\ll U$, i.e.~when any correlations in hopping are weak. We expect that $\sucex$ degrees of freedom may govern behaviour when the parameters  $\Ut$, $\lamt$ of Eq.~\eqref{chparams} are $\cO(1)$, in particular for $U\sim \k$ at small $\k$. We represent this  schematically in Fig.~\ref{fig_kU}. There is an argument to be made that the two regimes extend to either side of the point $\k=1$, $U=\infty$. On the one hand, one can consider departing from the degenerate atomic limit through a continuous unitary transformation organised in powers of $t/U$ \cite{Stein_1997}, 
but this breaks down when $t_\pm\sim t/U$, i.e.~close to $\k=1$.
On the other, the parametrisation of  Eq.~\eqref{eq:CHparam} is 
discontinuous at $t=t_\pm=0$, and equivalence with \eqref{eq:CH} requires that $t$ is not  taken to zero first, i.e.~to approach the atomic limit keeping $\k\sim1$.
These singular behaviours can be attributed to the fact that the Hubbard interaction commutes with correlated hopping when $\k=1$. 
See Ref.~\cite{Hidden_structure} for a closely related discussion. Let us also comment here that the framework pursued by Shastry sets $U=\infty$ from the outset, and this plays an important role throughout his analysis \cite{Shastry_2011,Shastry_2013}.

An important question is whether there exist materials whose behaviour is governed by $\sucex$?
We consider the pseudogap regime found in the cuprates to be an ideal candidate, 
as it is a metallic state with a distinct non-Fermi liquid character  \cite{Timusk_1999,Hashimoto_2014,Fradkin_2015}.
The cuprates are charge-transfer insulators \cite{Fujimori_1984,Zaanen_1985}, and their electronic structure suggests they should admit an effective single-orbital description with the eliminated low-lying ligand $p$ orbital inducing significant correlations in the hopping amplitudes \cite{Zhang_1988,Micnas89,MarsiglioHirsch,Sim_n_1993}. 
Quantum oscillation experiments indicate a clear violation of the Luttinger sum rule as the pseudogap regime is entered, and furthermore that the carrier density vanishes as the Mott transition is approached,  see e.g.~Fig.~4.b of Ref.~\cite{Badoux_2016}.

A next step is to go beyond the leading approximation upon which we focused in Sec.~\ref{sec:approx}.  Indeed, this is important to fully characterise the $\sucex$ regime of behaviour. Incorporating the first order contributions to the self-energy and adaptive spectral weight from Eqs.~\eqref{SW1} will capture the leading contributions of static spin and charge correlations. This is the analogue of the Hartree-Fock approximation for the canonical case.  In the context of cuprate physics it would be interesting to investigate if the phenomenological Yang--Rice--Zhang ansatz \cite{YRZ,YRZ_rev}, which bears a similar form to Eq.~\eqref{eqG}, can be justified in this way.
Another direction is to establish the thermodynamic properties of the regime. 
We hope such studies will clarify the relevance of $\sucex$ degrees of freedom for characterising the behaviour of strongly correlated materials.

 The details of the underlying lattice have not played an important role in our analysis. In practice, it is good to have translational invariance as Eqs.~\eqref{SgWrec} generate a local expansion, which is most conveniently handled in momentum space.

A special case  is when the lattice is a one-dimensional chain. Here the low-energy degrees of freedom are generically spin-charge separated \cite{HaldaneLL,GiamarchiLL}. Thus we do not expect $\sucex$ degrees of freedom to govern behaviour there, just as canonical fermions do not govern behaviour away from the non-interacting limit \cite{DL74}. 
Instead, degrees of freedom in one dimension are truly interacting. They can be characterised by their behaviour at integrable limits, where  scattering becomes completely elastic but remains non-trivial \cite{ZAMOLODCHIKOV1979253}, 
allowing for a complete description of the energy spectrum in terms of stable particles \cite{Bethe31,TakBook}. The classification of such integrable models is understood within the framework of algebraic Bethe ansatz \cite{Faddeev_2016}. It is noteworthy that the primary integrable models relevant for interacting electrons \cite{Hbook,EKS,AlcarazBariev,HS1}  descend from an R-matrix governed by the exceptional central extension of $\sucex$ symmetry we use here \cite{Beisert07,HS1}, or a q-deformation thereof \cite{BeisertKoroteev}. Indeed, the present work was greatly motivated by a combined study of these models \cite{Hidden_structure}.

Another important case is that of infinite dimensions, i.e. when the coordination number of the lattice diverges. While the notion of local degrees of freedom disappears in this limit, dynamical correlations can survive. The frequency dependent electronic Green's function of the Hubbard model can be determined in an exact way here through dynamical mean-field theory \cite{Metzner_1989,DMFT}. There exist works which incorporate correlated hopping into the formalism \cite{PhysRevB.67.075101,StanescuKotliar,PEREPELITSKY2013283}, but unfortunately we have not found an explicit study of the effect of correlated hopping on the electronic spectral function. We hope that this may be achieved, as it will provide a complementary controlled perspective on our description of the Mott transition.

The splitting of the electron in Eq.~\eqref{inv_rels} admits an interpretation in terms of slave particles. Slave bosons $\Obs$ and fermions $\Ofs$ fractionalise the canonical fermion generators as 
\be\la{cansp}
\Ocd_{\s} =\Obd_{\s} \Of_{\qe} +  \e_{\s\s'} \Ofd_{\qd} \Ob_{\s'},
\ee
or alternatively by interchanging $\Obs \oto \Ofs$ \cite{Barnes_1976,Coleman_1984,Arovas_1988,Yoshioka_1989}.
They are often invoked to characterise strongly correlated electrons \cite{Senthil_2003,LNWrev}.   
The $\OQs$ of $\sucex$ can be viewed as a decoupling of the two contributions to Eq.~\eqref{cansp} as follows 
\be
\OQd_{\s\nu}=\frac{1+\k}{2}\Ofd_{\nu} \Ob_{\bs}+\frac{1-\k}{2}\e_{\bs\s'}\e_{\nu\nu'}  \Obd_{\s'} \Of_{\nu'}.
\ee
Descriptions of correlated matter where deconfined slave particles govern the behaviour require  emergent gauge fields \cite{BaskaranAnderson}. This is not the case with the $\OQs$ however, which can be viewed as binding the $\Obs$ and $\Ofs$ to gauge invariant degrees of freedom. Such a binding has been considered from a phenomenological perspective in the context of cuprate physics \cite{PhysRevLett.76.503,PhysRevB.71.172509}.

Finally we offer a more general perspective. We have argued that there exist two distinct regimes of electronic behaviour, governed either by canonical $\su(1|1)\otimes\su(1|1)$  or non-canonical $\sucex$ degrees of freedom, which are Fermi liquid and non-Fermi liquid respectively, see Fig.~\ref{fig_kU}. This raises the question: what happens in between?  A phase transition in a conventional sense does not seem possible, as there is no clear notion of order parameter. Instead, each regime may be characterised by a quasi-particle description, where correlations are controlled in perturbative manner by distinct sets of degrees of freedom. While it is possible to connect the two regimes in a controlled way through the non-interacting point, this is highly singular due to the enhanced symmetry there, see Fig.~\ref{plot_BS}.(i). Instead, we suggest that connecting the two regimes along a generic path requires the breakdown of a quasi-particle description in between. This mirrors a previous proposal in an identical setting in one dimension \cite{Hidden_structure}.

More specifically, the robustness of the Fermi liquid  owes to the fact that the  lifetimes of the electronic quasi-particles scale as $(\om-\ve_F)^{-2}$, guaranteeing their stability in the vicinity of the Fermi surface. A `transition' may however occur if correlations shrink the domain over which this scaling is valid to zero. That is, the Fermi liquid may be destroyed by `coherence closing', while the spectrum remains gapless. Such a quantum chaotic  regime would permit a rearrangement of the spectrum, allowing in turn for a rearrangement of the electronic degree of freedom.

Again, the cuprates offer a prime candidate for identifying such behaviour in a material setting. They exhibit a `strange metal' regime, lying between the pseudogap and Fermi liquid regimes in their phase diagram, where the featureless linear in temperature resistivity has defied a quasi-particle interpretation \cite{Martin_1990,Chien_1991,Hussey_2011}. 
Our description of `coherence closing' is consistent with the phenomenological marginal Fermi liquid description of this regime \cite{marginalFL}. 
Establishing the necessity for a breakdown of a quasi-particle description in this way would provide a fresh starting point for understanding the anomalous behaviour there. 

From the perspective of either set of degrees of freedom,  the intermediate regime is where correlations grow out of control. Characterising such behaviour requires an alternative framework, not built upon 
underlying degrees of freedom. An intriguing possibility is holographic duality, which offers a controlled description through the semi-classical regime of a dual gravity theory \cite{zaanen2015holographic,hartnoll2016holographic}.

\section{Conclusion}  \la{sec:conc}

Characterising the behaviour of interacting electrons is an outstanding challenge, despite many decades of effort. 
Here we have offered a novel approach, based around characterising the electronic degree of freedom.

We have argued that strong electronic correlations are governed by the graded Lie algebra $\sucex$, as opposed to the canonical fermion algebra which underlies the Fermi liquid. 
We have derived a controlled description by obtaining a series of successive approximations for the electronic Green's function, mirroring the self-energy expansion of the canonical case. 
Focusing on the leading approximation, we found a splitting in two of the electronic band, a violation of the Luttinger sum rule, and a Mott transition when the split bands separate.

Much work is required to further characterise this non-Fermi liquid regime.
Ultimately, we hope this will lead to efficient techniques for understanding materials whose behaviour is driven by strong electronic correlations.


\section*{Acknowledgments}
We thank Jean-S\'ebastien Caux, Philippe Corboz, Sergey Frolov, Mark Golden, Enej Ilievski, Jasper van Wezel and Jan Zaanen for useful discussions. Support from the Foundation for Fundamental Research on Matter (FOM) and the Netherlands Organization for Scientific Research (NWO) is gratefully acknowledged.

\appendix


\section{The graded Lie algebra $\sucex$}\la{app:su22}

The graded Lie algebra $\sucex$ has fifteen generators 
\be
8\times\OQ,~~3\times\Os,~~3\times\Oet,~~ \bf{1},\ee 
with the $\OQ$ fermionic, and the remainder bosonic.
We focus on the fundamental 4-dimensional representation relevant for the electronic degree of freedom. The generators can be expressed in terms of spinful canonical fermions through Eqs.~\eqref{eq:spin}, \eqref{eq:charge} and \eqref{Q0s}.
The anti-commutation relations between the fermionic generators $\OQ$ are given by Eqs.~\eqref{eq:su22}. The commutation relations between the fermionic and bosonic generators are
\be
\begin{aligned}
&\lbrack\Os^z , \OQd_{\s\nu} \rbrack = \frac{\s}{2} \OQd_{\s\nu},\qquad 
&&	[\Os^z , \OQ_{\s\nu} ] = - \frac{\s}{2} \OQ_{\s\nu}, \\
&[\Os^+ , \OQd_{\ssd\nu} ] = - \OQd_{\ssu\nu},\qquad	
&&	[\Os^+ , \OQ_{\ssu\nu} ] =  \OQ_{\ssd\nu},\\
&[\Os^- , \OQd_{\ssu\nu} ] = - \OQd_{\ssd\nu},\qquad	
&&	[\Os^- , \OQ_{\ssd\nu} ] =  \OQ_{\ssu\nu},
\end{aligned}
\ee
and
\be
\begin{aligned}
&\lbrack\Oet^z , \OQd_{\s\nu} \rbrack = \frac{\nu}{2} \OQd_{\s\nu},\qquad 
&&	[\Oet^z , \OQ_{\s\nu} ] = - \frac{\nu}{2} \OQ_{\s\nu}, \\
&[\Oet^+ , \OQd_{\s\qe} ] = \OQd_{\s\qd},\qquad	
&&	[\Oet^+ , \OQ_{\s\qd} ] = - \OQ_{\s\qe},\\
&[\Oet^- , \OQd_{\s\qd} ] = \OQd_{\s\qe},\qquad	
&&	[\Oet^- , \OQ_{\s\qe} ] = - \OQ_{\s\qd}.
\end{aligned}
\ee
The commutation relations between the  bosonic generators are the spin and charge $\su(2)$ algebras given below Eqs.~\eqref{eq:spin} and \eqref{eq:charge} respectively. The algebra can be extended to $\ucex$ by adding $\OHu=\k \OV^H=\frac{\k}{3}(\vec{\Oet}\cdot\vec{\Oet}-\vec{\Os}\cdot\vec{\Os})$, with the extra relations given by Eqs.~\eqref{VHQ0}.

These relations can be neatly expressed through deformed Hubbard operators 
\be
\begin{split}
&\OX_\s^\s=\s \Os^z-\OHu +1/4,\quad   \OX_\ssd^\ssu = \Os^+,\quad   \OX_\ssu^\ssd = \Os^-,\\
&\OX_\nu^\nu=\nu \Oet^z+\OHu +1/4,\quad   \OX_\qe^\qd = \Oet^+,\quad   \OX_\qd^\qe = \Oet^-,\\
&\OX^\s_\nu = \OQ_{\bs\nu},\quad \OX^\nu_\s = \OQd_{\bs\nu},
\end{split}
\ee
which obey an  extended Hubbard algebra
\be
\begin{split}
\lbrack \OX^{\s_1}_{\s_2},\OX^{\s_3}_{\s_4}\rbrack & = 
	\delta^{\s_3}_{\s_2} \OX^{\s_1}_{\s_4}-\delta^{\s_1}_{\s_4} \OX^{\s_3}_{\s_2},\\ 
\lbrack \OX^{\nu_1}_{\nu_2},\OX^{\nu_3}_{\nu_4}\rbrack & = 
	\delta^{\nu_3}_{\nu_2} \OX^{\nu_1}_{\nu_4}-\delta^{\nu_1}_{\nu_4} \OX^{\nu_3}_{\nu_2},\\ 
\{ \OX^{\s}_{\nu},\OX^{\nu'}_{\s'} \} &= \frac{(1-\k)^2}{4} \delta^{\s}_{\s'} \delta^{\nu'}_{\nu} 
	+ \k \big( \delta^{\nu'}_{\nu} \OX^{\s}_{\s'} + \delta^{\s}_{\s'} \OX^{\nu'}_{\nu} \big),\\
\{ \OX^{\s}_{\nu},\OX^{\s'}_{\nu'} \} &= \frac{1-\k^2}{4} \e^{\s\s'} \e_{\nu\nu'},\\
\{ \OX_{\s}^{\nu},\OX_{\s'}^{\nu'} \} &= \frac{1-\k^2}{4} \e_{\s\s'} \e^{\nu\nu'},
\end{split}
\ee
Hubbard's operators and algebra are obtained at $\k=1$.

Formally, we are considering an exceptional central extension of $\su(2|2)$ \cite{Beisert07,Beisert08}. In Kac's classification of graded Lie algebras \cite{kac1977lie} the case corresponding to $\su(2|2)$ covers just $\k=1$ here, with the extension contained as a scaled limit of an exceptional algebra $\mathfrak{d}(2,1;\a)$.

\newpage
\begin{widetext}

\section{Constants and Parameters}\la{app:params}

In Sec.~\ref{sec:GF} we introduced compact notations, encoding the Green's function analysis in parameters $f$, $t$, $V$, $K$ and $\vG$. Here we present them explicitly.

\subsection{Structure constants: $f$}
\noindent
The structure constants $f^{\a\beta}{}_I$ and  $f^{\Theta\a}{}_\beta$ depend on $\k^2$ through $\sfa^\pm=\frac{1\pm \k^2}{4}$ as follows
\be
\begin{split}
f^{\a\beta}{}_I:\left(
\begin{array}{cccccccc}
 0 & 0 & 0 & 0 & \sfa^+ & \sfa^- & 0 & 0 \\
 0 & 0 & 0 & 0 & \sfa^- & \sfa^+ & 0 & 0 \\
 0 & 0 & 0 & 0 & 0 & 0 & \sfa^+ & -\sfa^- \\
 0 & 0 & 0 & 0 & 0 & 0 & -\sfa^- & \sfa^+ \\
 \sfa^+ & \sfa^- & 0 & 0 & 0 & 0 & 0 & 0 \\
 \sfa^- & \sfa^+ & 0 & 0 & 0 & 0 & 0 & 0 \\
 0 & 0 & \sfa^+ & -\sfa^- & 0 & 0 & 0 & 0 \\
 0 & 0 & -\sfa^- & \sfa^+ & 0 & 0 & 0 & 0 \\
\end{array}
\right),\quad
f^{\Theta\a}{}_\beta:\left(
\begin{array}{cccccccc}
 \sfa^+ & -\sfa^- & 0 & 0 & 0 & 0 & 0 & 0 \\
 \sfa^- & -\sfa^+ & 0 & 0 & 0 & 0 & 0 & 0 \\
 0 & 0 & \sfa^+ & \sfa^- & 0 & 0 & 0 & 0 \\
 0 & 0 & -\sfa^- & -\sfa^+ & 0 & 0 & 0 & 0 \\
 0 & 0 & 0 & 0 & -\sfa^+ & \sfa^- & 0 & 0 \\
 0 & 0 & 0 & 0 & -\sfa^- & \sfa^+ & 0 & 0 \\
 0 & 0 & 0 & 0 & 0 & 0 & -\sfa^+ & -\sfa^- \\
 0 & 0 & 0 & 0 & 0 & 0 & \sfa^- & \sfa^+ \\
\end{array}
\right).
\end{split}
\ee
The structure constants $f^{\a\beta}{}_a$ are proportional to $\k$ as follows
\be
\begin{split}
f^{\a\beta}{}_1:\left(
\begin{array}{cccccccc}
 0 & 0 & 0 & 0 & -\kappa  & 0 & 0 & 0 \\
 0 & 0 & 0 & 0 & 0 & \kappa  & 0 & 0 \\
 0 & 0 & 0 & 0 & 0 & 0 & \kappa  & 0 \\
 0 & 0 & 0 & 0 & 0 & 0 & 0 & -\kappa  \\
 -\kappa  & 0 & 0 & 0 & 0 & 0 & 0 & 0 \\
 0 & \kappa  & 0 & 0 & 0 & 0 & 0 & 0 \\
 0 & 0 & \kappa  & 0 & 0 & 0 & 0 & 0 \\
 0 & 0 & 0 & -\kappa  & 0 & 0 & 0 & 0 \\
\end{array}
\right),\quad 
f^{\a\beta}{}_2:\left(
\begin{array}{cccccccc}
 0 & 0 & 0 & 0 & 0 & 0 & 0 & 0 \\
 0 & 0 & 0 & 0 & 0 & 0 & 0 & 0 \\
 0 & 0 & 0 & 0 & \kappa  & 0 & 0 & 0 \\
 0 & 0 & 0 & 0 & 0 & \kappa  & 0 & 0 \\
 0 & 0 & \kappa  & 0 & 0 & 0 & 0 & 0 \\
 0 & 0 & 0 & \kappa  & 0 & 0 & 0 & 0 \\
 0 & 0 & 0 & 0 & 0 & 0 & 0 & 0 \\
 0 & 0 & 0 & 0 & 0 & 0 & 0 & 0 \\
\end{array}
\right),\quad 
f^{\a\beta}{}_3:\left(
\begin{array}{cccccccc}
 0 & 0 & 0 & 0 & 0 & 0 & \kappa  & 0 \\
 0 & 0 & 0 & 0 & 0 & 0 & 0 & \kappa  \\
 0 & 0 & 0 & 0 & 0 & 0 & 0 & 0 \\
 0 & 0 & 0 & 0 & 0 & 0 & 0 & 0 \\
 0 & 0 & 0 & 0 & 0 & 0 & 0 & 0 \\
 0 & 0 & 0 & 0 & 0 & 0 & 0 & 0 \\
 \kappa  & 0 & 0 & 0 & 0 & 0 & 0 & 0 \\
 0 & \kappa  & 0 & 0 & 0 & 0 & 0 & 0 \\
\end{array}
\right),\\
f^{\a\beta}{}_4:\left(
\begin{array}{cccccccc}
 0 & 0 & 0 & 0 & -\kappa  & 0 & 0 & 0 \\
 0 & 0 & 0 & 0 & 0 & \kappa  & 0 & 0 \\
 0 & 0 & 0 & 0 & 0 & 0 & -\kappa  & 0 \\
 0 & 0 & 0 & 0 & 0 & 0 & 0 & \kappa  \\
 -\kappa  & 0 & 0 & 0 & 0 & 0 & 0 & 0 \\
 0 & \kappa  & 0 & 0 & 0 & 0 & 0 & 0 \\
 0 & 0 & -\kappa  & 0 & 0 & 0 & 0 & 0 \\
 0 & 0 & 0 & \kappa  & 0 & 0 & 0 & 0 \\
\end{array}
\right),\quad
f^{\a\beta}{}_5:\left(
\begin{array}{cccccccc}
 0 & 0 & 0 & \kappa  & 0 & 0 & 0 & 0 \\
 0 & 0 & \kappa  & 0 & 0 & 0 & 0 & 0 \\
 0 & \kappa  & 0 & 0 & 0 & 0 & 0 & 0 \\
 \kappa  & 0 & 0 & 0 & 0 & 0 & 0 & 0 \\
 0 & 0 & 0 & 0 & 0 & 0 & 0 & 0 \\
 0 & 0 & 0 & 0 & 0 & 0 & 0 & 0 \\
 0 & 0 & 0 & 0 & 0 & 0 & 0 & 0 \\
 0 & 0 & 0 & 0 & 0 & 0 & 0 & 0 \\
\end{array}
\right),\quad
f^{\a\beta}{}_6:\left(
\begin{array}{cccccccc}
 0 & 0 & 0 & 0 & 0 & 0 & 0 & 0 \\
 0 & 0 & 0 & 0 & 0 & 0 & 0 & 0 \\
 0 & 0 & 0 & 0 & 0 & 0 & 0 & 0 \\
 0 & 0 & 0 & 0 & 0 & 0 & 0 & 0 \\
 0 & 0 & 0 & 0 & 0 & 0 & 0 & \kappa  \\
 0 & 0 & 0 & 0 & 0 & 0 & \kappa  & 0 \\
 0 & 0 & 0 & 0 & 0 & \kappa  & 0 & 0 \\
 0 & 0 & 0 & 0 & \kappa  & 0 & 0 & 0 \\
\end{array}
\right).
\end{split}
\ee
The structure constants $f^{a\a}{}_\beta$ are independent of $\k$ as follows
\be
\begin{split}
f^{1\a}{}_\beta:\left(
\begin{array}{cccccccc}
 \frac{1}{2} & 0 & 0 & 0 & 0 & 0 & 0 & 0 \\
 0 & \frac{1}{2} & 0 & 0 & 0 & 0 & 0 & 0 \\
 0 & 0 & -\frac{1}{2} & 0 & 0 & 0 & 0 & 0 \\
 0 & 0 & 0 & -\frac{1}{2} & 0 & 0 & 0 & 0 \\
 0 & 0 & 0 & 0 & -\frac{1}{2} & 0 & 0 & 0 \\
 0 & 0 & 0 & 0 & 0 & -\frac{1}{2} & 0 & 0 \\
 0 & 0 & 0 & 0 & 0 & 0 & \frac{1}{2} & 0 \\
 0 & 0 & 0 & 0 & 0 & 0 & 0 & \frac{1}{2} \\
\end{array}
\right),\quad 
f^{2\a}{}_\beta:\left(
\begin{array}{cccccccc}
 0 & 0 & -1 & 0 & 0 & 0 & 0 & 0 \\
 0 & 0 & 0 & 1 & 0 & 0 & 0 & 0 \\
 0 & 0 & 0 & 0 & 0 & 0 & 0 & 0 \\
 0 & 0 & 0 & 0 & 0 & 0 & 0 & 0 \\
 0 & 0 & 0 & 0 & 0 & 0 & 0 & 0 \\
 0 & 0 & 0 & 0 & 0 & 0 & 0 & 0 \\
 0 & 0 & 0 & 0 & 1 & 0 & 0 & 0 \\
 0 & 0 & 0 & 0 & 0 & -1 & 0 & 0 \\
\end{array}
\right),\quad 
f^{3\a}{}_\beta:\left(
\begin{array}{cccccccc}
 0 & 0 & 0 & 0 & 0 & 0 & 0 & 0 \\
 0 & 0 & 0 & 0 & 0 & 0 & 0 & 0 \\
 -1 & 0 & 0 & 0 & 0 & 0 & 0 & 0 \\
 0 & 1 & 0 & 0 & 0 & 0 & 0 & 0 \\
 0 & 0 & 0 & 0 & 0 & 0 & 1 & 0 \\
 0 & 0 & 0 & 0 & 0 & 0 & 0 & -1 \\
 0 & 0 & 0 & 0 & 0 & 0 & 0 & 0 \\
 0 & 0 & 0 & 0 & 0 & 0 & 0 & 0 \\
\end{array}
\right),\\
f^{4\a}{}_\beta:\left(
\begin{array}{cccccccc}
 -\frac{1}{2} & 0 & 0 & 0 & 0 & 0 & 0 & 0 \\
 0 & -\frac{1}{2} & 0 & 0 & 0 & 0 & 0 & 0 \\
 0 & 0 & -\frac{1}{2} & 0 & 0 & 0 & 0 & 0 \\
 0 & 0 & 0 & -\frac{1}{2} & 0 & 0 & 0 & 0 \\
 0 & 0 & 0 & 0 & \frac{1}{2} & 0 & 0 & 0 \\
 0 & 0 & 0 & 0 & 0 & \frac{1}{2} & 0 & 0 \\
 0 & 0 & 0 & 0 & 0 & 0 & \frac{1}{2} & 0 \\
 0 & 0 & 0 & 0 & 0 & 0 & 0 & \frac{1}{2} \\
\end{array}
\right),\quad 
f^{5\a}{}_\beta:\left(
\begin{array}{cccccccc}
 0 & 0 & 0 & 0 & 0 & 0 & 0 & 0 \\
 0 & 0 & 0 & 0 & 0 & 0 & 0 & 0 \\
 0 & 0 & 0 & 0 & 0 & 0 & 0 & 0 \\
 0 & 0 & 0 & 0 & 0 & 0 & 0 & 0 \\
 0 & 0 & 0 & -1 & 0 & 0 & 0 & 0 \\
 0 & 0 & 1 & 0 & 0 & 0 & 0 & 0 \\
 0 & -1 & 0 & 0 & 0 & 0 & 0 & 0 \\
 1 & 0 & 0 & 0 & 0 & 0 & 0 & 0 \\
\end{array}
\right),\quad 
f^{6\a}{}_\beta:\left(
\begin{array}{cccccccc}
 0 & 0 & 0 & 0 & 0 & 0 & 0 & 1 \\
 0 & 0 & 0 & 0 & 0 & 0 & -1 & 0 \\
 0 & 0 & 0 & 0 & 0 & 1 & 0 & 0 \\
 0 & 0 & 0 & 0 & -1 & 0 & 0 & 0 \\
 0 & 0 & 0 & 0 & 0 & 0 & 0 & 0 \\
 0 & 0 & 0 & 0 & 0 & 0 & 0 & 0 \\
 0 & 0 & 0 & 0 & 0 & 0 & 0 & 0 \\
 0 & 0 & 0 & 0 & 0 & 0 & 0 & 0 \\
\end{array}
\right).
\end{split}
\ee

\subsection{Hopping and interaction  parameters, and metric: $t$, $V$, $K$}
\noindent
The hopping and interaction parameters relating the general Hamiltonian of Eq.~\eqref{hamTTR} to the specific model considered in Sec.~\ref{sec:dof} are given as follows, alongside the metric defined by Eq.~\eqref{eq:metric},
\be
t_{\a\beta}:\left(
\begin{array}{cccccccc}
 0 & 0 & 0 & 0 & t_\qe & 0 & 0 & 0 \\
 0 & 0 & 0 & 0 & 0 & -t_\qd & 0 & 0 \\
 0 & 0 & 0 & 0 & 0 & 0 & t_\qe & 0 \\
 0 & 0 & 0 & 0 & 0 & 0 & 0 & -t_\qd \\
 -t_\qe & 0 & 0 & 0 & 0 & 0 & 0 & 0 \\
 0 & t_\qd & 0 & 0 & 0 & 0 & 0 & 0 \\
 0 & 0 & -t_\qe & 0 & 0 & 0 & 0 & 0 \\
 0 & 0 & 0 & t_\qd & 0 & 0 & 0 & 0 \\
\end{array}
\right),\quad 
V_{ab}:
\left(
\begin{array}{cccccc}
 J & 0 & 0 & 0 & 0 & 0 \\
 0 & 0 & \frac{J}{2} & 0 & 0 & 0 \\
 0 & \frac{J}{2} & 0 & 0 & 0 & 0 \\
 0 & 0 & 0 & 0 & 0 & 0 \\
 0 & 0 & 0 & 0 & 0 & 0 \\
 0 & 0 & 0 & 0 & 0 & 0 \\
\end{array}
\right),\quad
K_{\a\beta}:\left(
\begin{array}{cccccccc}
 0 & 0 & 0 & 0 & 1 & 0 & 0 & 0 \\
 0 & 0 & 0 & 0 & 0 & 1 & 0 & 0 \\
 0 & 0 & 0 & 0 & 0 & 0 & 1 & 0 \\
 0 & 0 & 0 & 0 & 0 & 0 & 0 & 1 \\
 1 & 0 & 0 & 0 & 0 & 0 & 0 & 0 \\
 0 & 1 & 0 & 0 & 0 & 0 & 0 & 0 \\
 0 & 0 & 1 & 0 & 0 & 0 & 0 & 0 \\
 0 & 0 & 0 & 1 & 0 & 0 & 0 & 0 \\
\end{array}
\right).
\ee

\subsection{Coefficients: $\vG$}
\noindent
The coefficients $\vG^a{}^\a_\beta$ of Eq.~\eqref{GtoT}, which relate $\OB^a = \vG^a{}^\a_\beta \OF_\a\OF^\beta$ in a $\k$ independent way, are 
\be
\begin{split}
\vG^1{}^\a_\beta:\left(
\begin{array}{cccccccc}
 0 & 0 & 0 & 0 & 0 & 0 & 0 & 0 \\
 0 & 0 & 0 & 0 & 0 & 0 & 0 & 0 \\
 0 & 0 & 0 & 0 & 0 & 0 & 0 & 0 \\
 0 & 0 & 0 & 0 & 0 & 0 & 0 & 0 \\
 0 & 0 & 0 & 0 & \frac{1}{2} & \frac{1}{2} & 0 & 0 \\
 0 & 0 & 0 & 0 & \frac{1}{2} & \frac{1}{2} & 0 & 0 \\
 0 & 0 & 0 & 0 & 0 & 0 & -\frac{1}{2} & \frac{1}{2} \\
 0 & 0 & 0 & 0 & 0 & 0 & \frac{1}{2} & -\frac{1}{2} \\
\end{array}
\right),\quad
&\vG^2{}^\a_\beta:\left(
\begin{array}{cccccccc}
 0 & 0 & 1 & -1 & 0 & 0 & 0 & 0 \\
 0 & 0 & 1 & -1 & 0 & 0 & 0 & 0 \\
 0 & 0 & 0 & 0 & 0 & 0 & 0 & 0 \\
 0 & 0 & 0 & 0 & 0 & 0 & 0 & 0 \\
 0 & 0 & 0 & 0 & 0 & 0 & 0 & 0 \\
 0 & 0 & 0 & 0 & 0 & 0 & 0 & 0 \\
 0 & 0 & 0 & 0 & 0 & 0 & 0 & 0 \\
 0 & 0 & 0 & 0 & 0 & 0 & 0 & 0 \\
\end{array}
\right),\quad
\vG^3{}^\a_\beta:\left(
\begin{array}{cccccccc}
 0 & 0 & 0 & 0 & 0 & 0 & 0 & 0 \\
 0 & 0 & 0 & 0 & 0 & 0 & 0 & 0 \\
 0 & 0 & 0 & 0 & 0 & 0 & 0 & 0 \\
 0 & 0 & 0 & 0 & 0 & 0 & 0 & 0 \\
 0 & 0 & 0 & 0 & 0 & 0 & -1 & 1 \\
 0 & 0 & 0 & 0 & 0 & 0 & -1 & 1 \\
 0 & 0 & 0 & 0 & 0 & 0 & 0 & 0 \\
 0 & 0 & 0 & 0 & 0 & 0 & 0 & 0 \\
\end{array}
\right),\\
\vG^4{}^\a_\beta:
\left(
\begin{array}{cccccccc}
 0 & 0 & 0 & 0 & 0 & 0 & 0 & 0 \\
 0 & 0 & 0 & 0 & 0 & 0 & 0 & 0 \\
 0 & 0 & \frac{1}{2} & -\frac{1}{2} & 0 & 0 & 0 & 0 \\
 0 & 0 & -\frac{1}{2} & \frac{1}{2} & 0 & 0 & 0 & 0 \\
 0 & 0 & 0 & 0 & -\frac{1}{2} & -\frac{1}{2} & 0 & 0 \\
 0 & 0 & 0 & 0 & -\frac{1}{2} & -\frac{1}{2} & 0 & 0 \\
 0 & 0 & 0 & 0 & 0 & 0 & 0 & 0 \\
 0 & 0 & 0 & 0 & 0 & 0 & 0 & 0 \\
\end{array}
\right),\quad
&\vG^5{}^\a_\beta:\left(
\begin{array}{cccccccc}
 0 & 0 & 0 & 0 & 0 & 0 & 0 & 0 \\
 0 & 0 & 0 & 0 & 0 & 0 & 0 & 0 \\
 0 & 0 & 0 & 0 & 0 & 0 & 0 & 0 \\
 0 & 0 & 0 & 0 & 0 & 0 & 0 & 0 \\
 0 & 0 & 0 & 0 & 0 & 0 & 0 & 0 \\
 0 & 0 & 0 & 0 & 0 & 0 & 0 & 0 \\
 1 & 1 & 0 & 0 & 0 & 0 & 0 & 0 \\
 -1 & -1 & 0 & 0 & 0 & 0 & 0 & 0 \\
\end{array}
\right),\quad
\vG^6{}^\a_\beta:\left(
\begin{array}{cccccccc}
 0 & 0 & 0 & 0 & 0 & 0 & 1 & -1 \\
 0 & 0 & 0 & 0 & 0 & 0 & 1 & -1 \\
 0 & 0 & 0 & 0 & 0 & 0 & 0 & 0 \\
 0 & 0 & 0 & 0 & 0 & 0 & 0 & 0 \\
 0 & 0 & 0 & 0 & 0 & 0 & 0 & 0 \\
 0 & 0 & 0 & 0 & 0 & 0 & 0 & 0 \\
 0 & 0 & 0 & 0 & 0 & 0 & 0 & 0 \\
 0 & 0 & 0 & 0 & 0 & 0 & 0 & 0 \\
\end{array}
\right).
\end{split}
\ee

\section{Review of the canonical Green's function analysis}\la{app:canGF}

We review the self-energy expansion for the canonical Green's function. We loosely follow Chap.~5 of the text~\cite{kadanoff1962quantum}, and do not invoke Wick's theorem.  We mirror the derivation with that of Sec.~\ref{sec:GF}. Differences in the signs of terms are primarily a consequence of  $[\On,\Oc]=-\Oc$ here, while $[\OB^a,\OF^\beta] = f^{a\beta}{}_\g \OF^\g$ there.

We focus on the case of spinless fermions, and consider the  Hamiltonian
\be
\OH=-\sum_{i,j}t_{ij}\Ocd_{i}\Oc_{j} + \frac{1}{2}\sum_{i,j} V_{ij} \On_i \On_j - \mu\sum_j\On_j,
\ee
with $t_{ii}=0$, $t_{ji}=t_{ij}$, and $V_{ii}=0$, $V_{ji}=V_{ij}$. 
The interaction term induces correlations, and to track these we
 introduce a source term to the thermal expectation value as follows 
 \be
 \braket{ \cO(\t_1,\t_2,\ldots)} = \frac{\Tr \Big( e^{-\beta  \OH} \cT \big[e^{\int_0^\beta d\t \cS(\t)} \cO(\t_1,\t_2,\ldots) \big]   \Big)}{\Tr \big( e^{-\beta \OH} \cT [e^{\int_0^\beta d\t \cS(\t)}]   \big)},
\ee
with $\cS(\t)  = \sum_i \sou_{i}(\t) \On_i(\t)$. 
Density correlations can then be reinterpreted as variations of the sources through
\be
\begin{split}
\DV_i(\t)\braket{\cO(\t_1,\t_2,\ldots,\t_n)}=& \braket{\On_i(\t)\cO(\t_1,\t_2,\ldots,\t_n)} -\braket{\On_i(\t)}\braket{\cO(\t_1,\t_2,\ldots,\t_n)},
\end{split}
\ee
with $\DV_i(\t) =  \frac{{\delta} }{ {\delta} \sou_{i}(\t^+)}$. 
The equation of motion for $\GT_{ij}(\t,\t') = - \braket{\Oc_{i}(\t) \Ocd_{j}(\t')}$ is
\be
\begin{split}
\pa_{\t} \GT_{ij}(\t,\t') =& - \delta(\t-\t')\braket{\{\Oc_{i}(\t),\Ocd_{j}(\t)\}}+\braket{ [\cS(\t),\Oc_{i}(\t)] \Ocd_{j}(\t')} -\braket{ [\OH,\Oc_{i}(\t)] \Ocd_{j}(\t')}.
\end{split}
\ee
Here $[\OH,\Oc_{i}] = \sum_l t_{il} \Oc_{l}+\mu\Oc_{i}-\sum_l V_{il} \On_{l}\Oc_{i}$, giving
\be
\braket{[\OH,\Oc_{i}(\t)]\Ocd_{j}(\t')} = -\sum_l t_{il}  \GT_{lj}(\t,\t')- \mu \GT_{ij}(\t,\t') + \sum_l V_{il}\big(\braket{\On_l(\t)} + \DV_l(\t)  \big)\GT_{ij}(\t,\t'),
\ee
and so the equation of motion can be recast as
\be\la{canGFeqn}
\begin{split}
\sum_k \Big[ \delta_{ik} \Big( &-\pa_\t + \mu+ \sou_i(\t)  -\sum_l V_{il}\big( \braket{\On_l(\t)} + \DV_l(\t)  \big)\Big)+ t_{ik} \Big] \GT_{kj}(\t,\t') = \delta(\t-\t')\delta_{ij}.
\end{split}
\ee
Unlike the non-canonical case, contrast Eq.~\eqref{GFeqn}, here the right-hand side is independent of $\GT$.

We proceed by introducing the self-energy as a functional $\S[\GT]$  of the full $\GT$ through
\be\la{eqGI}
 \GT^{-1}_{ij}(\t,\t') = \GT^{-1}_{0,ij}(\t,\t') - \S_{ij}(\t,\t')  ,
\ee
where $\GT_{0}$ satisfies
\be
\begin{split}
\sum_k \big[ \delta_{ik} \big( &-\pa_\t + \sou_i(\t)+ \mu \big)+ t_{ik} \big] \GT_{0,kj}(\t,\t') = \delta(\t-\t')\delta_{ij}.
\end{split}
\ee
We obtain a closed equation for $\S$ by convolving Eq.~\eqref{canGFeqn} on the right with $\GT^{-1}$, which gives
\be\la{Sexact}
\begin{split}
\S_{ij}(\t,\t') 
 &= \delta(\t-\t')\delta_{ij} \sum_l V_{il} \braket{\On_l(\t)} -\delta(\t-\t') V_{ij} \GT_{ij}(\t,\t^+)
	+\sum_{l,k} \int_0^\beta d\t'' V_{il} \GT_{ik}(\t,\t'') \DV_l(\t) \S_{kj}(\t'',\t'),
\end{split}
\ee
upon using $(\DV \GT)\GT^{-1}=-\GT\DV\GT^{-1}= -\GT\DV\GT_0^{-1} + \GT\DV\S$, with 
\be
\DV_l(\t'')\GT_{0,ij}^{-1}(\t,\t')= \delta(\t-\t')\delta(\t-\t''-0^+)\delta_{ij}\delta_{il}.
\ee

Equation~\eqref{Sexact} is exact. We now obtain successive approximate solutions by a perturbative expansion of $\S$ in the strength of interactions. We introduce rescaled parameters $\Vt_{ij} =  V_{ij}/\lam$, and write $\S = \sum_{s=0}^\infty \lam^s [\S]_s$. The leading contribution
\be\la{HF}
 [\S_{ij}(\t,\t')]_1 =  
	\delta(\t-\t')\delta_{ij} \sum_l \Vt_{il} \GT_{ll}(\t,\t^+) -\delta(\t-\t') \Vt_{ij} \GT_{ij}(\t,\t^+),
\ee
contains both the Hartree and Fock terms, and we have used $\braket{\On_i(\t)}=\GT_{ii}(\t,\t^+)$ to cast it explicitly in terms of $\GT$.
All higher contributions are then  obtained recursively through
\be
\lbrack \S_{ij}(\t,\t') \rbrack_{s+1}= \sum_{k,l}\int_0^\beta d\t'' 
  \Vt_{il} \GT_{ik}(\t,\t'') \DV_l(\t)  [\S_{kj}(\t'',\t')]_{s}.
\ee
Dependence on the source $J$ appears only through $\GT$, and  the functional derivative can be evaluated at each order using 
\be\la{DG:GG}
\DV_l(\t'')\GT_{ij}{}(\t,\t') = - \GT_{il}{}(\t,\t'') \GT_{lj}{}(\t'',\t')+\cO(\lam).
\ee 
In this way we systematically obtain the  functional $\S[\GT]$  to any desired order.
For example, the second order contribution to the self-energy is 
\be
\lbrack \S_{ij}(\t,\t') \rbrack_{2}= \sum_{k,l} \Vt_{il}\Vt_{jk} 
	\big(-\GT_{ij}(\t,\t')\GT_{kl}(\t',\t)\GT_{lk}(\t,\t')
		+ \GT_{ik}(\t,\t')\GT_{kl}(\t',\t)\GT_{lj}(\t,\t')\big),
\ee
which gives the Born collision approximation. 

\newpage

 \end{widetext}


\section{Schematic summary of Sec.~\ref{sec:GF}}\la{app:sch}

We consider a Hamiltonian of the form
\be
\begin{split}
\OH \sim& -\frac{1}{2}\sum_{i,j}  t_{ij}   \OF_i  \OF_{j} 
	+\frac{\k}{2}\sum_{i,j} \Vt_{ij}  \OB_i \OB_j\\
	 & \qquad -  \mu \sum_i \OB_i+\Ut\sum_i \OHu_i,
\end{split}
\ee
where $\OF$, $\OB$, $\OHu$ satisfy a non-canonical algebra \footnote{We suppress $\k^2$ terms in this schematic representation of the algebra. They enter Eq.~\eqref{eomshm} only through $A$, and play no role in the subsequent expansion.}
\be\la{algsch}
\begin{split}
\{\OF,\OF\}\sim 1 &+ \k\OB,\quad [\OB,\OF]\sim \OF,\quad [\OB,\OB]\sim \OB,\\
&[\OHu,\OF]\sim \OF,\quad [\OHu,\OB]\sim0.
\end{split}
\ee
We wish to obtain the Green's function
\be
\GT_{ij}(\t,\t')\sim-\braket{\OF_i(\t)\OF_j(\t')}.
\ee
We include sources for $\OB$ in the expectation value
\be
 \braket{ \cO(\t_1,\t_2,\ldots)} \sim \frac{\Tr \Big( e^{-\beta  \OH} \cT \big[e^{\int_0^\beta d\t \cS(\t)} \cO(\t_1,\t_2,\ldots) \big]   \Big)}{\Tr \big( e^{-\beta H} \cT [e^{\int_0^\beta d\t \cS(\t)}]   \big)},
\ee
with $\cS(\t)  \sim \sum_i \sou_{i}(\t) \OB_i(\t)$, so that we can 
organise correlations through 
\be
\begin{split}
\DV_i(\t)\braket{\cO(\t_1,\t_2,\ldots,\t_n)}\sim &\braket{\OB_i(\t)\cO(\t_1,\t_2,\ldots,\t_n)}\\
	& -\braket{\OB_i(\t)}\braket{\cO(\t_1,\t_2,\ldots,\t_n)},
\end{split}
\ee
with $\DV_i(\t) \sim  \frac{{\delta} }{ {\delta} \sou_{i}(\t^+)}$. 

The equation of motion for $\GT$ is 
\be
\begin{split}
\pa_{\t} \GT_{ij}(\t,\t') \sim  -\delta(\t-\t')\braket{\{\OF_{i}(\t), \OF_{j}(\t)\}}&\\
	 + \braket{[\cS(\t), \OF_i(\t)]\OF_{j}(\t')}&\\
	 - \braket{ [\OH,\OF_i(\t)]\OF_{j}(\t') }.&
\end{split}
\ee
Here we have
\be
\begin{split}
\braket{\{\OF_{i}(\t), \OF_{j}(\t)\}} &\sim \delta_{ij} \big(1+ \k\braket{\OB_i(\t)}\big),\\
\braket{[\cS(\t), \OF_i(\t)]\OF_{j}(\t')} &\sim - \sou_{i}(\t)  \GT_{ij}(\t,\t'),
\end{split}
\ee
and computing the commutator in the final term
\be
\begin{split}
\lbrack\OH,\OF_i\rbrack \sim&
	\pm\sum_{l}\big[  t_{il} \OF_j
	+ \k  t_{il} \OB_i \OF_j\big]
	+ \k \sum_{l} \Vt_{il}\OB_l \OF_i \\
	&\quad - \mu \OF_i + \Ut  \OF_i,
\end{split}
\ee
(here we use $\pm$ to denote a sign change coming from antisymmetry of $t_{\a\beta}$ not seen in this schematic analysis) 
gives
\be
\begin{split}
\braket{\lbrack\OH,\OF_i(\t)\rbrack \OF_{j}(\t')}\sim (\mu - \Ut)  \GT_{ij}(\t,\t')  
	- \sum_l t_{il}   \GT_{lj}(\t,\t')& \\ 
\qquad - \k\sum_l t_{il}  \big(
		\braket{\OB_i(\t)}+\DV_i(\t)\big) \GT_{lj}(\t,\t')&\\
\qquad - \k\sum_l \Vt_{il}  \big(\braket{\OB_l(\t)}
		+ \DV_l(\t) \big) \GT_{ij}(\t,\t').&
\end{split}
\ee 
The equation of motion then takes the form
\begin{widetext}
\be
\begin{split}
\sum_k \Big[ 
\delta_{ik}\Big(-\pa_{\t} -\sou_{i}(\t) 
 - \mu + \Ut +\k \sum_l  \Vt_{il} \big(\braket{\OB_l(\t)}
		 +  \DV_l(\t) \big)\Big)  
+   t_{ik} + \k t_{ik} \big( \braket{\OB_i(\t)} 
	 + \DV_i(\t) \big)\Big] \GT_{kj}(\t,\t')\quad&\\
\sim \delta(\t-\t')\delta_{ij}\big(1+ \k\braket{\OB_i(\t)}\big)&.
\end{split}
\ee
\end{widetext}

To demonstrate Shastry's Green's function factorisation it is useful to simplify notations further and cast the equation of motion as 
\be\la{eomshm}
[A + \k (\braket{\OB} + \DV)]\GT \sim 1 + \k \braket{\OB} ,
\ee
where $\DV A \sim - 1$ and $\braket{\OB}\sim\GT$. Factorising $\GT\sim\gT\wT$, we have $\DV\GT\sim(\DV \gT)\wT + \gT(\DV\wT)$, and the equation of motion becomes
\be
\big([A + \k (\gT\wT + \DV)]\gT\big)\wT \sim 1 + \k \gT \wT+\k\gT\DV\wT.
\ee
Setting 
\be
\wT \sim  1 + \k \gT \wT+\k\gT\DV\wT,
\ee
 gives the following equation for $\gT$
 \be
 [A + \k (\gT\wT + \DV)]\gT \sim 1.
 \ee
We now introduce two functionals of the full $\gT$ and $\wT$, the self-energy $\Sg[\gT,\wT]$ and the adaptive spectral weight $\SW[\gT,\wT]$, defined through 
 \be\la{gTcWscm}
 \begin{split}
 \gT^{-1} &\sim A-\Sg[\gT,\wT], \\
 \wT &\sim 1 + \SW[\gT,\wT].
 \end{split}
 \ee
 These obey the closed equations
 \be
 \begin{split}
\Sg &\sim -\k \gT \wT - \k \gT - \k \gT \DV \Sg,\\
\SW  &\sim  \k \gT \wT - \k \gT \DV \SW,
\end{split}
\ee
with the first obtained using $(\DV \gT)\gT^{-1} = -\gT \DV \gT^{-1} =  \gT +\gT\DV\Sg$.
We solve these with a perturbative expansion in $\k$. At leading order
\be
[\Sg]_1 \sim -\gT\wT - \gT,\quad [\SW]_1 \sim \gT\wT.
\ee
Higher terms are obtained recursively through 
\be
 \begin{split}
\lbrack\Sg\rbrack_{s+1} & \sim -  \gT \DV [\Sg]_s,\\
[\SW]_{s+1}  &\sim  - \gT \DV [\SW]_s,
\end{split}
\ee
where at each order we can use just the leading contributions from
\be
\DV \gT \sim  \gT \gT + \cO(\k),\quad \DV \wT \sim 0 + \cO(\k).
\ee
For example the second order contributions are
\be
[\Sg]_2 \sim \gT\gT\gT\wT + \gT\gT\gT,\quad [\SW]_2 \sim -\gT\gT\gT\wT.
\ee
This allows for successive approximate solutions to  Eq.~\eqref{eomshm}: expanding $\Sg[\gT,\wT]$ and $\SW[\gT,\wT]$ to some desired order, solving the coupled Eqs.~\eqref{gTcWscm} for $\gT$ and $\wT$ self-consistently, and thus obtaining 
\be
\GT\sim\gT\wT\sim \frac{1+\SW[\gT,\wT]}{A-\Sg[\gT,\wT]}.
\ee
In summary, a systematic expansion of the Green's function of the non-canonical degree of freedom has been achieved by obtaining control over the growth of correlations in both the numerator and the denominator, as opposed to just the denominator for the case of a canonical degree of freedom.

\begin{widetext}

\section{Second order contributions}\la{app:2nd}

\noindent
The second order contributions to the functionals $\Sg[\gT,\wT]$ and $\SW[\gT,\wT]$ of Sec.~\ref{sec:GF} are
\be\la{SW2}
\begin{split}
\lbrack \Sg_{ij}{}^\a_\beta(\t,\t') \rbrack_2= 
	&\sum_{k,l,m}\int_0^\beta d\t'' 
		\ft^{\a\delta}{}_a t_{il,\delta\e} \gT_{lk}{}^\e_\eta(\t,\t') \ft^{\eta\g}{}_b t_{kj,\g\beta} 
		\vG^b{}^\rho_\s \gT_{ki}{}^\s_\lam(\t',\t) f^{a\lam}{}_\mu 
		\gT_{im}{}^\mu_\nu(\t,\t'') \wT_{mk}{}^\nu_\rho(\t'',\t') \\
	&+\sum_{k,l,m}\int_0^\beta d\t''  
		f^{a\a}{}_\delta \Vt_{il,ab} \gT_{ik}{}^\delta_\e(\t,\t') \ft^{\e\g}{}_c t_{kj,\g\beta} 
		\vG^c{}^\rho_\s  \gT_{kl}{}^\s_\lam(\t',\t) 
		f^{b\lam}{}_\mu \gT_{lm}{}^\mu_\nu(\t,\t'') \wT_{mk}{}^\nu_\rho(\t'',\t') \\
	&+\sum_{k,l,m}\int_0^\beta d\t''  
		\ft^{\a\delta}{}_a t_{il,\delta\e} \gT_{lj}{}^\e_\g(\t,\t') f^{b\g}{}_\beta \Vt_{jk,bc} 
		\vG^c{}^\rho_\s \gT_{ki}{}^\s_\lam(\t',\t)
		 f^{a\lam}{}_\mu \gT_{im}{}^\mu_\nu(\t,\t'') \wT_{mk}{}^\nu_\rho(\t'',\t') \\
	&+\sum_{k,l,m}\int_0^\beta d\t''  
		f^{a\a}{}_\delta \Vt_{il,ab} \gT_{ij}{}^\delta_\g(\t,\t') f^{c\g}{}_\beta \Vt_{jk,cd} 
		\vG^d{}^\rho_\s \gT^\s_\lam{}_{kl}(\t',\t) 
		f^{b\lam}{}_\mu \gT_{lm}{}^\mu_\nu(\t,\t'') \wT_{mk}{}^\nu_\rho(\t'',\t')	\\
	&+\sum_{k,l} 
		\ft^{\a\delta}{}_a t_{il,\delta\e} \gT_{lj}{}^\e_{\eta}(\t,\t') \ft^{\eta\rho}{}_b t_{jk,\rho\s}  
		\gT_{ki}{}^\s_\lam(\t',\t) f^{a\lam}{}_\mu \gT_{ij}{}^\mu_\nu(\t,\t') f^{b\nu}{}_\beta \\
	&+\sum_{k,l} 
		f^{a\a}{}_\delta \Vt_{il,ab} \gT_{ij}{}^\delta_\e(\t,\t') \ft^{\e\rho}{}_c t_{jk,\rho\s}  
		\gT_{kl}{}^\s_\lam(\t',\t) f^{b\lam}{}_\mu \gT_{lj}{}^\mu_\nu(\t,\t') f^{c\nu}{}_\beta \\
	&+\sum_{k,l} 
		\ft^{\a\delta}{}_a t_{il,\delta\e} \gT_{lk}{}^\e_\rho(\t,\t') f^{c\rho}{}_\s \Vt_{kj,cd}  
		\gT_{ki}{}^\s_\lam(\t',\t) f^{a\lam}{}_\mu \gT_{ij}{}^\mu_\nu(\t,\t') f^{d\nu}{}_\beta \\
	&+\sum_{k,l} 
		f^{a\a}{}_\delta \Vt_{il,ab} \gT_{ik}{}^\delta_\rho(\t,\t') f^{c\rho}{}_\s \Vt_{kj,cd}  
		\gT_{kl}{}^\s_\lam(\t',\t) f^{b\lam}{}_\mu \gT_{lj}{}^\mu_\nu(\t,\t') f^{d\nu}{}_\beta, \\
[ \SW_{ij}{}^\a_\beta(\t,\t') ]_2=
	&-\sum_{k,l}\int_0^\beta d\t'' 
		\ft^{\a\delta}{}_a t_{il,\delta\e} \gT_{lj}{}^\e_\eta(\t,\t') \ft^{\eta\g}{}_b K_{\g\beta} 
		\vG^b{}^\rho_\s \gT_{ji}{}^\s_\lam(\t',\t) 
		f^{a\lam}{}_\mu \gT_{ik}{}^\mu_\nu(\t,\t'') \wT_{kj}{}^\nu_\rho(\t'',\t') \\
	&-\sum_{k,l}\int_0^\beta d\t'' 
		f^{a\a}{}_\delta \Vt_{il,ab} \gT_{ij}{}^\delta_\e(\t,\t') \ft^{\e\g}{}_c K_{\g\beta} 
		\vG^c{}^\rho_\s \gT_{jl}{}^\s_\lam(\t',\t) 
		f^{b\lam}{}_\mu \gT_{lk}{}^\mu_\nu(\t,\t'') \wT_{kj}{}^\nu_\rho(\t'',\t'). \\
\end{split}
\ee

\end{widetext}

\bibliographystyle{apsrev4-1}
\bibliography{CHbib}

\begin{thebibliography}{99}%
\makeatletter
\providecommand \@ifxundefined [1]{%
 \@ifx{#1\undefined}
}%
\providecommand \@ifnum [1]{%
 \ifnum #1\expandafter \@firstoftwo
 \else \expandafter \@secondoftwo
 \fi
}%
\providecommand \@ifx [1]{%
 \ifx #1\expandafter \@firstoftwo
 \else \expandafter \@secondoftwo
 \fi
}%
\providecommand \natexlab [1]{#1}%
\providecommand \enquote  [1]{``#1''}%
\providecommand \bibnamefont  [1]{#1}%
\providecommand \bibfnamefont [1]{#1}%
\providecommand \citenamefont [1]{#1}%
\providecommand \href@noop [0]{\@secondoftwo}%
\providecommand \href [0]{\begingroup \@sanitize@url \@href}%
\providecommand \@href[1]{\@@startlink{#1}\@@href}%
\providecommand \@@href[1]{\endgroup#1\@@endlink}%
\providecommand \@sanitize@url [0]{\catcode `\\12\catcode `\$12\catcode
  `\&12\catcode `\#12\catcode `\^12\catcode `\_12\catcode `\%12\relax}%
\providecommand \@@startlink[1]{}%
\providecommand \@@endlink[0]{}%
\providecommand \url  [0]{\begingroup\@sanitize@url \@url }%
\providecommand \@url [1]{\endgroup\@href {#1}{\urlprefix }}%
\providecommand \urlprefix  [0]{URL }%
\providecommand \Eprint [0]{\href }%
\providecommand \doibase [0]{http://dx.doi.org/}%
\providecommand \selectlanguage [0]{\@gobble}%
\providecommand \bibinfo  [0]{\@secondoftwo}%
\providecommand \bibfield  [0]{\@secondoftwo}%
\providecommand \translation [1]{[#1]}%
\providecommand \BibitemOpen [0]{}%
\providecommand \bibitemStop [0]{}%
\providecommand \bibitemNoStop [0]{.\EOS\space}%
\providecommand \EOS [0]{\spacefactor3000\relax}%
\providecommand \BibitemShut  [1]{\csname bibitem#1\endcsname}%
\let\auto@bib@innerbib\@empty
\bibitem [{\citenamefont {Landau}(1957)}]{landau1957}%
  \BibitemOpen
  \bibfield  {author} {\bibinfo {author} {\bibfnamefont {L.~D.}\ \bibnamefont
  {Landau}},\ }\href@noop {} {\bibfield  {journal} {\bibinfo  {journal} {Sov.
  Phys. JETP}\ }\textbf {\bibinfo {volume} {3}},\ \bibinfo {pages} {920}
  (\bibinfo {year} {1957})}\BibitemShut {NoStop}%
\bibitem [{\citenamefont {Landau}(1959)}]{landau1959}%
  \BibitemOpen
  \bibfield  {author} {\bibinfo {author} {\bibfnamefont {L.~D.}\ \bibnamefont
  {Landau}},\ }\href@noop {} {\bibfield  {journal} {\bibinfo  {journal} {Sov.
  Phys. JETP}\ }\textbf {\bibinfo {volume} {8}},\ \bibinfo {pages} {70}
  (\bibinfo {year} {1959})}\BibitemShut {NoStop}%
\bibitem [{\citenamefont {Gross}\ and\ \citenamefont
  {Dreizler}(2013)}]{gross2013density}%
  \BibitemOpen
  \bibfield  {author} {\bibinfo {author} {\bibfnamefont {E.~K.}\ \bibnamefont
  {Gross}}\ and\ \bibinfo {author} {\bibfnamefont {R.~M.}\ \bibnamefont
  {Dreizler}},\ }\href@noop {} {\emph {\bibinfo {title} {Density functional
  theory}}},\ Vol.\ \bibinfo {volume} {337}\ (\bibinfo  {publisher} {Springer
  Science \& Business Media},\ \bibinfo {year} {2013})\BibitemShut {NoStop}%
\bibitem [{\citenamefont {Bednorz}\ and\ \citenamefont
  {M\"uller}(1986)}]{BednorzMuller86}%
  \BibitemOpen
  \bibfield  {author} {\bibinfo {author} {\bibfnamefont {J.~G.}\ \bibnamefont
  {Bednorz}}\ and\ \bibinfo {author} {\bibfnamefont {K.~A.}\ \bibnamefont
  {M\"uller}},\ }\href {\doibase 10.1007/bf01303701} {\bibfield  {journal}
  {\bibinfo  {journal} {Z. Phys. B}\ }\textbf {\bibinfo {volume} {64}},\
  \bibinfo {pages} {189} (\bibinfo {year} {1986})}\BibitemShut {NoStop}%
\bibitem [{\citenamefont {Anderson}(1987)}]{ANDERSON_1987}%
  \BibitemOpen
  \bibfield  {author} {\bibinfo {author} {\bibfnamefont {P.~W.}\ \bibnamefont
  {Anderson}},\ }\href {\doibase 10.1126/science.235.4793.1196} {\bibfield
  {journal} {\bibinfo  {journal} {Science}\ }\textbf {\bibinfo {volume}
  {235}},\ \bibinfo {pages} {1196} (\bibinfo {year} {1987})}\BibitemShut
  {NoStop}%
\bibitem [{\citenamefont {Keimer}\ \emph {et~al.}(2015)\citenamefont {Keimer},
  \citenamefont {Kivelson}, \citenamefont {Norman}, \citenamefont {Uchida},\
  and\ \citenamefont {Zaanen}}]{Keimer_rev}%
  \BibitemOpen
  \bibfield  {author} {\bibinfo {author} {\bibfnamefont {B.}~\bibnamefont
  {Keimer}}, \bibinfo {author} {\bibfnamefont {S.~A.}\ \bibnamefont
  {Kivelson}}, \bibinfo {author} {\bibfnamefont {M.~R.}\ \bibnamefont
  {Norman}}, \bibinfo {author} {\bibfnamefont {S.}~\bibnamefont {Uchida}}, \
  and\ \bibinfo {author} {\bibfnamefont {J.}~\bibnamefont {Zaanen}},\ }\href
  {\doibase 10.1038/nature14165} {\bibfield  {journal} {\bibinfo  {journal}
  {Nature}\ }\textbf {\bibinfo {volume} {518}},\ \bibinfo {pages} {179}
  (\bibinfo {year} {2015})}\BibitemShut {NoStop}%
\bibitem [{\citenamefont {Fradkin}\ \emph {et~al.}(2015)\citenamefont
  {Fradkin}, \citenamefont {Kivelson},\ and\ \citenamefont
  {Tranquada}}]{Fradkin_2015}%
  \BibitemOpen
  \bibfield  {author} {\bibinfo {author} {\bibfnamefont {E.}~\bibnamefont
  {Fradkin}}, \bibinfo {author} {\bibfnamefont {S.~A.}\ \bibnamefont
  {Kivelson}}, \ and\ \bibinfo {author} {\bibfnamefont {J.~M.}\ \bibnamefont
  {Tranquada}},\ }\href {\doibase 10.1103/revmodphys.87.457} {\bibfield
  {journal} {\bibinfo  {journal} {Rev. Mod. Phys.}\ }\textbf {\bibinfo {volume}
  {87}},\ \bibinfo {pages} {457} (\bibinfo {year} {2015})}\BibitemShut
  {NoStop}%
\bibitem [{\citenamefont {Varma}\ \emph {et~al.}(1989)\citenamefont {Varma},
  \citenamefont {Littlewood}, \citenamefont {Schmitt-Rink}, \citenamefont
  {Abrahams},\ and\ \citenamefont {Ruckenstein}}]{marginalFL}%
  \BibitemOpen
  \bibfield  {author} {\bibinfo {author} {\bibfnamefont {C.~M.}\ \bibnamefont
  {Varma}}, \bibinfo {author} {\bibfnamefont {P.~B.}\ \bibnamefont
  {Littlewood}}, \bibinfo {author} {\bibfnamefont {S.}~\bibnamefont
  {Schmitt-Rink}}, \bibinfo {author} {\bibfnamefont {E.}~\bibnamefont
  {Abrahams}}, \ and\ \bibinfo {author} {\bibfnamefont {A.~E.}\ \bibnamefont
  {Ruckenstein}},\ }\href {\doibase 10.1103/physrevlett.63.1996} {\bibfield
  {journal} {\bibinfo  {journal} {Phys. Rev. Lett.}\ }\textbf {\bibinfo
  {volume} {63}},\ \bibinfo {pages} {1996} (\bibinfo {year}
  {1989})}\BibitemShut {NoStop}%
\bibitem [{\citenamefont {Si}\ \emph {et~al.}(2016)\citenamefont {Si},
  \citenamefont {Yu},\ and\ \citenamefont {Abrahams}}]{Si_2016}%
  \BibitemOpen
  \bibfield  {author} {\bibinfo {author} {\bibfnamefont {Q.}~\bibnamefont
  {Si}}, \bibinfo {author} {\bibfnamefont {R.}~\bibnamefont {Yu}}, \ and\
  \bibinfo {author} {\bibfnamefont {E.}~\bibnamefont {Abrahams}},\ }\href
  {\doibase 10.1038/natrevmats.2016.17} {\bibfield  {journal} {\bibinfo
  {journal} {Nat. Rev. Mats.}\ }\textbf {\bibinfo {volume} {1}},\ \bibinfo
  {pages} {16017} (\bibinfo {year} {2016})}\BibitemShut {NoStop}%
\bibitem [{\citenamefont {Gegenwart}\ \emph {et~al.}(2008)\citenamefont
  {Gegenwart}, \citenamefont {Si},\ and\ \citenamefont
  {Steglich}}]{Gegenwart_2008}%
  \BibitemOpen
  \bibfield  {author} {\bibinfo {author} {\bibfnamefont {P.}~\bibnamefont
  {Gegenwart}}, \bibinfo {author} {\bibfnamefont {Q.}~\bibnamefont {Si}}, \
  and\ \bibinfo {author} {\bibfnamefont {F.}~\bibnamefont {Steglich}},\ }\href
  {\doibase 10.1038/nphys892} {\bibfield  {journal} {\bibinfo  {journal} {Nat.
  Phys.}\ }\textbf {\bibinfo {volume} {4}},\ \bibinfo {pages} {186} (\bibinfo
  {year} {2008})}\BibitemShut {NoStop}%
\bibitem [{\citenamefont {Powell}\ and\ \citenamefont
  {McKenzie}(2011)}]{Powell_2011}%
  \BibitemOpen
  \bibfield  {author} {\bibinfo {author} {\bibfnamefont {B.~J.}\ \bibnamefont
  {Powell}}\ and\ \bibinfo {author} {\bibfnamefont {R.~H.}\ \bibnamefont
  {McKenzie}},\ }\href {\doibase 10.1088/0034-4885/74/5/056501} {\bibfield
  {journal} {\bibinfo  {journal} {Rep. Prog. Phys.}\ }\textbf {\bibinfo
  {volume} {74}},\ \bibinfo {pages} {056501} (\bibinfo {year}
  {2011})}\BibitemShut {NoStop}%
\bibitem [{\citenamefont {Faddeev}(2016)}]{Faddeev_2016}%
  \BibitemOpen
  \bibfield  {author} {\bibinfo {author} {\bibfnamefont {L.}~\bibnamefont
  {Faddeev}},\ }in\ \href {\doibase 10.1142/9789814340960_0031} {\emph
  {\bibinfo {booktitle} {Fifty Years of Mathematical Physics}}}\ (\bibinfo
  {publisher} {{World} {Scientific}},\ \bibinfo {year} {2016})\ pp.\ \bibinfo
  {pages} {370--439}\BibitemShut {NoStop}%
\bibitem [{\citenamefont {Essler}\ \emph {et~al.}(1992)\citenamefont {Essler},
  \citenamefont {Korepin},\ and\ \citenamefont {Schoutens}}]{EKS}%
  \BibitemOpen
  \bibfield  {author} {\bibinfo {author} {\bibfnamefont {F.~H.~L.}\
  \bibnamefont {Essler}}, \bibinfo {author} {\bibfnamefont {V.~E.}\
  \bibnamefont {Korepin}}, \ and\ \bibinfo {author} {\bibfnamefont
  {K.}~\bibnamefont {Schoutens}},\ }\href {\doibase
  10.1103/PhysRevLett.68.2960} {\bibfield  {journal} {\bibinfo  {journal}
  {Phys. Rev. Lett.}\ }\textbf {\bibinfo {volume} {68}},\ \bibinfo {pages}
  {2960} (\bibinfo {year} {1992})}\BibitemShut {NoStop}%
\bibitem [{\citenamefont {Essler}\ \emph {et~al.}(2005)\citenamefont {Essler},
  \citenamefont {Frahm}, \citenamefont {G{\"o}hmann}, \citenamefont
  {Kl{\"u}mper},\ and\ \citenamefont {Korepin}}]{Hbook}%
  \BibitemOpen
  \bibfield  {author} {\bibinfo {author} {\bibfnamefont {F.~H.}\ \bibnamefont
  {Essler}}, \bibinfo {author} {\bibfnamefont {H.}~\bibnamefont {Frahm}},
  \bibinfo {author} {\bibfnamefont {F.}~\bibnamefont {G{\"o}hmann}}, \bibinfo
  {author} {\bibfnamefont {A.}~\bibnamefont {Kl{\"u}mper}}, \ and\ \bibinfo
  {author} {\bibfnamefont {V.~E.}\ \bibnamefont {Korepin}},\ }\href {\doibase
  10.1017/CBO9780511534843} {\emph {\bibinfo {title} {{The one-dimensional
  Hubbard model}}}}\ (\bibinfo  {publisher} {Cambridge University Press},\
  \bibinfo {year} {2005})\BibitemShut {NoStop}%
\bibitem [{\citenamefont {Frolov}\ and\ \citenamefont {Quinn}(2012)}]{HS1}%
  \BibitemOpen
  \bibfield  {author} {\bibinfo {author} {\bibfnamefont {S.}~\bibnamefont
  {Frolov}}\ and\ \bibinfo {author} {\bibfnamefont {E.}~\bibnamefont {Quinn}},\
  }\href {\doibase 10.1088/1751-8113/45/9/095004} {\bibfield  {journal}
  {\bibinfo  {journal} {J. Phys. A}\ }\textbf {\bibinfo {volume} {45}},\
  \bibinfo {pages} {095004} (\bibinfo {year} {2012})}\BibitemShut {NoStop}%
\bibitem [{\citenamefont {Zamolodchikov}\ and\ \citenamefont
  {Zamolodchikov}(1979)}]{ZAMOLODCHIKOV1979253}%
  \BibitemOpen
  \bibfield  {author} {\bibinfo {author} {\bibfnamefont {A.~B.}\ \bibnamefont
  {Zamolodchikov}}\ and\ \bibinfo {author} {\bibfnamefont {A.~B.}\ \bibnamefont
  {Zamolodchikov}},\ }\href {\doibase
  http://dx.doi.org/10.1016/0003-4916(79)90391-9} {\bibfield  {journal}
  {\bibinfo  {journal} {Ann. Phys.}\ }\textbf {\bibinfo {volume} {120}},\
  \bibinfo {pages} {253 } (\bibinfo {year} {1979})}\BibitemShut {NoStop}%
\bibitem [{\citenamefont {Wiegmann}(1988)}]{Wiegmann_1988}%
  \BibitemOpen
  \bibfield  {author} {\bibinfo {author} {\bibfnamefont {P.~B.}\ \bibnamefont
  {Wiegmann}},\ }\href {\doibase 10.1103/physrevlett.60.2445.3} {\bibfield
  {journal} {\bibinfo  {journal} {Phys. Rev. Lett.}\ }\textbf {\bibinfo
  {volume} {60}},\ \bibinfo {pages} {2445} (\bibinfo {year}
  {1988})}\BibitemShut {NoStop}%
\bibitem [{\citenamefont {F\"orster}(1989)}]{Forster_1989}%
  \BibitemOpen
  \bibfield  {author} {\bibinfo {author} {\bibfnamefont {D.}~\bibnamefont
  {F\"orster}},\ }\href {\doibase 10.1103/physrevlett.63.2140} {\bibfield
  {journal} {\bibinfo  {journal} {Phys. Rev. Lett.}\ }\textbf {\bibinfo
  {volume} {63}},\ \bibinfo {pages} {2140} (\bibinfo {year}
  {1989})}\BibitemShut {NoStop}%
\bibitem [{\citenamefont {Chaichian}\ \emph {et~al.}(1991)\citenamefont
  {Chaichian}, \citenamefont {Ellinas},\ and\ \citenamefont
  {Pre{\v{s}}najder}}]{Chaichian_1991}%
  \BibitemOpen
  \bibfield  {author} {\bibinfo {author} {\bibfnamefont {M.}~\bibnamefont
  {Chaichian}}, \bibinfo {author} {\bibfnamefont {D.}~\bibnamefont {Ellinas}},
  \ and\ \bibinfo {author} {\bibfnamefont {P.}~\bibnamefont
  {Pre{\v{s}}najder}},\ }\href {\doibase 10.1063/1.529451} {\bibfield
  {journal} {\bibinfo  {journal} {J. Math. Phys.}\ }\textbf {\bibinfo {volume}
  {32}},\ \bibinfo {pages} {3381} (\bibinfo {year} {1991})}\BibitemShut
  {NoStop}%
\bibitem [{\citenamefont {Kochetov}(1996)}]{Kochetov_1996}%
  \BibitemOpen
  \bibfield  {author} {\bibinfo {author} {\bibfnamefont {E.}~\bibnamefont
  {Kochetov}},\ }\href {\doibase 10.1016/0375-9601(96)00320-9} {\bibfield
  {journal} {\bibinfo  {journal} {Phys. Lett. A}\ }\textbf {\bibinfo {volume}
  {217}},\ \bibinfo {pages} {65} (\bibinfo {year} {1996})}\BibitemShut
  {NoStop}%
\bibitem [{\citenamefont {Coleman}\ and\ \citenamefont
  {P{\'{e}}pin}(2002)}]{Coleman_2002}%
  \BibitemOpen
  \bibfield  {author} {\bibinfo {author} {\bibfnamefont {P.}~\bibnamefont
  {Coleman}}\ and\ \bibinfo {author} {\bibfnamefont {C.}~\bibnamefont
  {P{\'{e}}pin}},\ }\href {\doibase 10.1016/s0921-4526(01)01501-0} {\bibfield
  {journal} {\bibinfo  {journal} {Physica B Condens Matter.}\ }\textbf
  {\bibinfo {volume} {312-313}},\ \bibinfo {pages} {539} (\bibinfo {year}
  {2002})}\BibitemShut {NoStop}%
\bibitem [{\citenamefont {Anderson}(2008)}]{Anderson08}%
  \BibitemOpen
  \bibfield  {author} {\bibinfo {author} {\bibfnamefont {P.~W.}\ \bibnamefont
  {Anderson}},\ }\href {\doibase 10.1103/PhysRevB.78.174505} {\bibfield
  {journal} {\bibinfo  {journal} {Phys. Rev. B}\ }\textbf {\bibinfo {volume}
  {78}},\ \bibinfo {pages} {174505} (\bibinfo {year} {2008})}\BibitemShut
  {NoStop}%
\bibitem [{\citenamefont {Avella}\ and\ \citenamefont
  {Mancini}(2011)}]{Avella_2011}%
  \BibitemOpen
  \bibfield  {author} {\bibinfo {author} {\bibfnamefont {A.}~\bibnamefont
  {Avella}}\ and\ \bibinfo {author} {\bibfnamefont {F.}~\bibnamefont
  {Mancini}},\ }in\ \href {\doibase 10.1007/978-3-642-21831-6_4} {\emph
  {\bibinfo {booktitle} {Springer Series in Solid-State Sciences}}}\ (\bibinfo
  {publisher} {Springer Berlin Heidelberg},\ \bibinfo {year} {2011})\ pp.\
  \bibinfo {pages} {103--141}\BibitemShut {NoStop}%
\bibitem [{\citenamefont {Ramires}\ and\ \citenamefont
  {Coleman}(2016)}]{Ramires}%
  \BibitemOpen
  \bibfield  {author} {\bibinfo {author} {\bibfnamefont {A.}~\bibnamefont
  {Ramires}}\ and\ \bibinfo {author} {\bibfnamefont {P.}~\bibnamefont
  {Coleman}},\ }\href {\doibase 10.1103/PhysRevB.93.035120} {\bibfield
  {journal} {\bibinfo  {journal} {Phys. Rev. B}\ }\textbf {\bibinfo {volume}
  {93}},\ \bibinfo {pages} {035120} (\bibinfo {year} {2016})}\BibitemShut
  {NoStop}%
\bibitem [{\citenamefont {Hubbard}(1965)}]{Hubbard4}%
  \BibitemOpen
  \bibfield  {author} {\bibinfo {author} {\bibfnamefont {J.}~\bibnamefont
  {Hubbard}},\ }\href {\doibase 10.1098/rspa.1965.0124} {\bibfield  {journal}
  {\bibinfo  {journal} {Proc. R. Soc. A}\ }\textbf {\bibinfo {volume} {285}},\
  \bibinfo {pages} {542} (\bibinfo {year} {1965})}\BibitemShut {NoStop}%
\bibitem [{\citenamefont {Vedyaev}\ and\ \citenamefont
  {Nikolaev}(1984)}]{Vedyaev_1984}%
  \BibitemOpen
  \bibfield  {author} {\bibinfo {author} {\bibfnamefont {A.~V.}\ \bibnamefont
  {Vedyaev}}\ and\ \bibinfo {author} {\bibfnamefont {M.~Y.}\ \bibnamefont
  {Nikolaev}},\ }\href {\doibase 10.1007/bf01018187} {\bibfield  {journal}
  {\bibinfo  {journal} {Theor. Math. Phys.}\ }\textbf {\bibinfo {volume}
  {59}},\ \bibinfo {pages} {510} (\bibinfo {year} {1984})}\BibitemShut
  {NoStop}%
\bibitem [{\citenamefont {Ruckenstein}\ and\ \citenamefont
  {Schmitt-Rink}(1988)}]{RuckensteinSR}%
  \BibitemOpen
  \bibfield  {author} {\bibinfo {author} {\bibfnamefont {A.~E.}\ \bibnamefont
  {Ruckenstein}}\ and\ \bibinfo {author} {\bibfnamefont {S.}~\bibnamefont
  {Schmitt-Rink}},\ }\href {\doibase 10.1103/PhysRevB.38.7188} {\bibfield
  {journal} {\bibinfo  {journal} {Phys. Rev. B}\ }\textbf {\bibinfo {volume}
  {38}},\ \bibinfo {pages} {7188} (\bibinfo {year} {1988})}\BibitemShut
  {NoStop}%
\bibitem [{\citenamefont {Izyumov}\ and\ \citenamefont
  {Letfulov}(1990)}]{Izyumov_1990}%
  \BibitemOpen
  \bibfield  {author} {\bibinfo {author} {\bibfnamefont {Y.~A.}\ \bibnamefont
  {Izyumov}}\ and\ \bibinfo {author} {\bibfnamefont {B.~M.}\ \bibnamefont
  {Letfulov}},\ }\href {\doibase 10.1088/0953-8984/2/45/005} {\bibfield
  {journal} {\bibinfo  {journal} {J. Phys. Condens. Matter}\ }\textbf {\bibinfo
  {volume} {2}},\ \bibinfo {pages} {8905} (\bibinfo {year} {1990})}\BibitemShut
  {NoStop}%
\bibitem [{\citenamefont {Ovchinnikov}\ and\ \citenamefont
  {Val'kov}(2004)}]{Ovchinnikov_2004}%
  \BibitemOpen
  \bibfield  {author} {\bibinfo {author} {\bibfnamefont {S.~G.}\ \bibnamefont
  {Ovchinnikov}}\ and\ \bibinfo {author} {\bibfnamefont {V.~V.}\ \bibnamefont
  {Val'kov}},\ }\href {\doibase 10.1142/9781860945977} {\emph {\bibinfo {title}
  {Hubbard Operators in the Theory of Strongly Correlated Electrons}}}\
  (\bibinfo  {publisher} {Imperial College Press},\ \bibinfo {year}
  {2004})\BibitemShut {NoStop}%
\bibitem [{\citenamefont {Izyumov}\ \emph {et~al.}(2005)\citenamefont
  {Izyumov}, \citenamefont {Chaschin}, \citenamefont {Alexeev},\ and\
  \citenamefont {Mancini}}]{Izyumov_2005}%
  \BibitemOpen
  \bibfield  {author} {\bibinfo {author} {\bibfnamefont {Y.~A.}\ \bibnamefont
  {Izyumov}}, \bibinfo {author} {\bibfnamefont {N.~I.}\ \bibnamefont
  {Chaschin}}, \bibinfo {author} {\bibfnamefont {D.~S.}\ \bibnamefont
  {Alexeev}}, \ and\ \bibinfo {author} {\bibfnamefont {F.}~\bibnamefont
  {Mancini}},\ }\href {\doibase 10.1140/epjb/e2005-00166-7} {\bibfield
  {journal} {\bibinfo  {journal} {Eur. Phys. J B}\ }\textbf {\bibinfo {volume}
  {45}},\ \bibinfo {pages} {69} (\bibinfo {year} {2005})}\BibitemShut {NoStop}%
\bibitem [{\citenamefont {Stanescu}\ and\ \citenamefont
  {Kotliar}(2004)}]{StanescuKotliar}%
  \BibitemOpen
  \bibfield  {author} {\bibinfo {author} {\bibfnamefont {T.~D.}\ \bibnamefont
  {Stanescu}}\ and\ \bibinfo {author} {\bibfnamefont {G.}~\bibnamefont
  {Kotliar}},\ }\href {\doibase 10.1103/PhysRevB.70.205112} {\bibfield
  {journal} {\bibinfo  {journal} {Phys. Rev. B}\ }\textbf {\bibinfo {volume}
  {70}},\ \bibinfo {pages} {205112} (\bibinfo {year} {2004})}\BibitemShut
  {NoStop}%
\bibitem [{\citenamefont {Shastry}(2011{\natexlab{a}})}]{Shastry_2011}%
  \BibitemOpen
  \bibfield  {author} {\bibinfo {author} {\bibfnamefont {B.~S.}\ \bibnamefont
  {Shastry}},\ }\href {\doibase 10.1103/physrevlett.107.056403} {\bibfield
  {journal} {\bibinfo  {journal} {Phys. Rev. Lett.}\ }\textbf {\bibinfo
  {volume} {107}},\ \bibinfo {pages} {056403} (\bibinfo {year}
  {2011}{\natexlab{a}})}\BibitemShut {NoStop}%
\bibitem [{\citenamefont {Shastry}(2013)}]{Shastry_2013}%
  \BibitemOpen
  \bibfield  {author} {\bibinfo {author} {\bibfnamefont {B.~S.}\ \bibnamefont
  {Shastry}},\ }\href {\doibase 10.1103/physrevb.87.125124} {\bibfield
  {journal} {\bibinfo  {journal} {Phys. Rev. B}\ }\textbf {\bibinfo {volume}
  {87}},\ \bibinfo {pages} {125124} (\bibinfo {year} {2013})}\BibitemShut
  {NoStop}%
\bibitem [{\citenamefont {Beisert}(2007)}]{Beisert07}%
  \BibitemOpen
  \bibfield  {author} {\bibinfo {author} {\bibfnamefont {N.}~\bibnamefont
  {Beisert}},\ }\href {\doibase 10.1088/1742-5468/2007/01/p01017} {\bibfield
  {journal} {\bibinfo  {journal} {J. Stat. Mech.}\ }\textbf {\bibinfo {volume}
  {2007}},\ \bibinfo {pages} {P01017} (\bibinfo {year} {2007})}\BibitemShut
  {NoStop}%
\bibitem [{\citenamefont {Beisert}(2008)}]{Beisert08}%
  \BibitemOpen
  \bibfield  {author} {\bibinfo {author} {\bibfnamefont {N.}~\bibnamefont
  {Beisert}},\ }\href {\doibase 10.4310/atmp.2008.v12.n5.a1} {\bibfield
  {journal} {\bibinfo  {journal} {Adv. Theor. Math. Phys.}\ }\textbf {\bibinfo
  {volume} {12}},\ \bibinfo {pages} {948} (\bibinfo {year} {2008})}\BibitemShut
  {NoStop}%
\bibitem [{\citenamefont {Luttinger}(1960)}]{Luttinger_1960}%
  \BibitemOpen
  \bibfield  {author} {\bibinfo {author} {\bibfnamefont {J.~M.}\ \bibnamefont
  {Luttinger}},\ }\href {\doibase 10.1103/physrev.119.1153} {\bibfield
  {journal} {\bibinfo  {journal} {Phys. Rev.}\ }\textbf {\bibinfo {volume}
  {119}},\ \bibinfo {pages} {1153} (\bibinfo {year} {1960})}\BibitemShut
  {NoStop}%
\bibitem [{\citenamefont {Doiron-Leyraud}\ \emph {et~al.}(2007)\citenamefont
  {Doiron-Leyraud}, \citenamefont {Proust}, \citenamefont {LeBoeuf},
  \citenamefont {Levallois}, \citenamefont {Bonnemaison}, \citenamefont
  {Liang}, \citenamefont {Bonn}, \citenamefont {Hardy},\ and\ \citenamefont
  {Taillefer}}]{Doiron_Leyraud_2007}%
  \BibitemOpen
  \bibfield  {author} {\bibinfo {author} {\bibfnamefont {N.}~\bibnamefont
  {Doiron-Leyraud}}, \bibinfo {author} {\bibfnamefont {C.}~\bibnamefont
  {Proust}}, \bibinfo {author} {\bibfnamefont {D.}~\bibnamefont {LeBoeuf}},
  \bibinfo {author} {\bibfnamefont {J.}~\bibnamefont {Levallois}}, \bibinfo
  {author} {\bibfnamefont {J.-B.}\ \bibnamefont {Bonnemaison}}, \bibinfo
  {author} {\bibfnamefont {R.}~\bibnamefont {Liang}}, \bibinfo {author}
  {\bibfnamefont {D.~A.}\ \bibnamefont {Bonn}}, \bibinfo {author}
  {\bibfnamefont {W.~N.}\ \bibnamefont {Hardy}}, \ and\ \bibinfo {author}
  {\bibfnamefont {L.}~\bibnamefont {Taillefer}},\ }\href {\doibase
  10.1038/nature05872} {\bibfield  {journal} {\bibinfo  {journal} {Nature}\
  }\textbf {\bibinfo {volume} {447}},\ \bibinfo {pages} {565} (\bibinfo {year}
  {2007})}\BibitemShut {NoStop}%
\bibitem [{\citenamefont {Badoux}\ \emph {et~al.}(2016)\citenamefont {Badoux},
  \citenamefont {Tabis}, \citenamefont {Lalibert{\'{e}}}, \citenamefont
  {Grissonnanche}, \citenamefont {Vignolle}, \citenamefont {Vignolles},
  \citenamefont {B{\'{e}}ard}, \citenamefont {Bonn}, \citenamefont {Hardy},
  \citenamefont {Liang}, \citenamefont {Doiron-Leyraud}, \citenamefont
  {Taillefer},\ and\ \citenamefont {Proust}}]{Badoux_2016}%
  \BibitemOpen
  \bibfield  {author} {\bibinfo {author} {\bibfnamefont {S.}~\bibnamefont
  {Badoux}}, \bibinfo {author} {\bibfnamefont {W.}~\bibnamefont {Tabis}},
  \bibinfo {author} {\bibfnamefont {F.}~\bibnamefont {Lalibert{\'{e}}}},
  \bibinfo {author} {\bibfnamefont {G.}~\bibnamefont {Grissonnanche}}, \bibinfo
  {author} {\bibfnamefont {B.}~\bibnamefont {Vignolle}}, \bibinfo {author}
  {\bibfnamefont {D.}~\bibnamefont {Vignolles}}, \bibinfo {author}
  {\bibfnamefont {J.}~\bibnamefont {B{\'{e}}ard}}, \bibinfo {author}
  {\bibfnamefont {D.~A.}\ \bibnamefont {Bonn}}, \bibinfo {author}
  {\bibfnamefont {W.~N.}\ \bibnamefont {Hardy}}, \bibinfo {author}
  {\bibfnamefont {R.}~\bibnamefont {Liang}}, \bibinfo {author} {\bibfnamefont
  {N.}~\bibnamefont {Doiron-Leyraud}}, \bibinfo {author} {\bibfnamefont
  {L.}~\bibnamefont {Taillefer}}, \ and\ \bibinfo {author} {\bibfnamefont
  {C.}~\bibnamefont {Proust}},\ }\href {\doibase 10.1038/nature16983}
  {\bibfield  {journal} {\bibinfo  {journal} {Nature}\ }\textbf {\bibinfo
  {volume} {531}},\ \bibinfo {pages} {210} (\bibinfo {year}
  {2016})}\BibitemShut {NoStop}%
\bibitem [{\citenamefont {Imada}\ \emph {et~al.}(1998)\citenamefont {Imada},
  \citenamefont {Fujimori},\ and\ \citenamefont {Tokura}}]{Imada_1998}%
  \BibitemOpen
  \bibfield  {author} {\bibinfo {author} {\bibfnamefont {M.}~\bibnamefont
  {Imada}}, \bibinfo {author} {\bibfnamefont {A.}~\bibnamefont {Fujimori}}, \
  and\ \bibinfo {author} {\bibfnamefont {Y.}~\bibnamefont {Tokura}},\ }\href
  {\doibase 10.1103/revmodphys.70.1039} {\bibfield  {journal} {\bibinfo
  {journal} {Rev. Mod. Phys.}\ }\textbf {\bibinfo {volume} {70}},\ \bibinfo
  {pages} {1039} (\bibinfo {year} {1998})}\BibitemShut {NoStop}%
\bibitem [{\citenamefont {Lee}\ \emph {et~al.}(2006)\citenamefont {Lee},
  \citenamefont {Nagaosa},\ and\ \citenamefont {Wen}}]{LNWrev}%
  \BibitemOpen
  \bibfield  {author} {\bibinfo {author} {\bibfnamefont {P.~A.}\ \bibnamefont
  {Lee}}, \bibinfo {author} {\bibfnamefont {N.}~\bibnamefont {Nagaosa}}, \ and\
  \bibinfo {author} {\bibfnamefont {X.-G.}\ \bibnamefont {Wen}},\ }\href
  {\doibase 10.1103/revmodphys.78.17} {\bibfield  {journal} {\bibinfo
  {journal} {Rev. Mod. Phys.}\ }\textbf {\bibinfo {volume} {78}},\ \bibinfo
  {pages} {17} (\bibinfo {year} {2006})}\BibitemShut {NoStop}%
\bibitem [{\citenamefont {Metzner}\ and\ \citenamefont
  {Vollhardt}(1989)}]{Metzner_1989}%
  \BibitemOpen
  \bibfield  {author} {\bibinfo {author} {\bibfnamefont {W.}~\bibnamefont
  {Metzner}}\ and\ \bibinfo {author} {\bibfnamefont {D.}~\bibnamefont
  {Vollhardt}},\ }\href {\doibase 10.1103/physrevlett.62.1066.2} {\bibfield
  {journal} {\bibinfo  {journal} {Phys. Rev. Lett.}\ }\textbf {\bibinfo
  {volume} {62}},\ \bibinfo {pages} {1066} (\bibinfo {year}
  {1989})}\BibitemShut {NoStop}%
\bibitem [{\citenamefont {Georges}\ \emph {et~al.}(1996)\citenamefont
  {Georges}, \citenamefont {Kotliar}, \citenamefont {Krauth},\ and\
  \citenamefont {Rozenberg}}]{DMFT}%
  \BibitemOpen
  \bibfield  {author} {\bibinfo {author} {\bibfnamefont {A.}~\bibnamefont
  {Georges}}, \bibinfo {author} {\bibfnamefont {G.}~\bibnamefont {Kotliar}},
  \bibinfo {author} {\bibfnamefont {W.}~\bibnamefont {Krauth}}, \ and\ \bibinfo
  {author} {\bibfnamefont {M.~J.}\ \bibnamefont {Rozenberg}},\ }\href {\doibase
  10.1103/revmodphys.68.13} {\bibfield  {journal} {\bibinfo  {journal} {Rev.
  Mod. Phys.}\ }\textbf {\bibinfo {volume} {68}},\ \bibinfo {pages} {13}
  (\bibinfo {year} {1996})}\BibitemShut {NoStop}%
\bibitem [{\citenamefont {Brinkman}\ and\ \citenamefont
  {Rice}(1970)}]{PhysRevB.2.4302}%
  \BibitemOpen
  \bibfield  {author} {\bibinfo {author} {\bibfnamefont {W.~F.}\ \bibnamefont
  {Brinkman}}\ and\ \bibinfo {author} {\bibfnamefont {T.~M.}\ \bibnamefont
  {Rice}},\ }\href {\doibase 10.1103/PhysRevB.2.4302} {\bibfield  {journal}
  {\bibinfo  {journal} {Phys. Rev. B}\ }\textbf {\bibinfo {volume} {2}},\
  \bibinfo {pages} {4302} (\bibinfo {year} {1970})}\BibitemShut {NoStop}%
\bibitem [{\citenamefont {Ando}\ \emph {et~al.}(2004)\citenamefont {Ando},
  \citenamefont {Kurita}, \citenamefont {Komiya}, \citenamefont {Ono},\ and\
  \citenamefont {Segawa}}]{Ando1}%
  \BibitemOpen
  \bibfield  {author} {\bibinfo {author} {\bibfnamefont {Y.}~\bibnamefont
  {Ando}}, \bibinfo {author} {\bibfnamefont {Y.}~\bibnamefont {Kurita}},
  \bibinfo {author} {\bibfnamefont {S.}~\bibnamefont {Komiya}}, \bibinfo
  {author} {\bibfnamefont {S.}~\bibnamefont {Ono}}, \ and\ \bibinfo {author}
  {\bibfnamefont {K.}~\bibnamefont {Segawa}},\ }\href {\doibase
  10.1103/PhysRevLett.92.197001} {\bibfield  {journal} {\bibinfo  {journal}
  {Phys. Rev. Lett.}\ }\textbf {\bibinfo {volume} {92}},\ \bibinfo {pages}
  {197001} (\bibinfo {year} {2004})}\BibitemShut {NoStop}%
\bibitem [{\citenamefont {Segawa}\ and\ \citenamefont {Ando}(2004)}]{Ando2}%
  \BibitemOpen
  \bibfield  {author} {\bibinfo {author} {\bibfnamefont {K.}~\bibnamefont
  {Segawa}}\ and\ \bibinfo {author} {\bibfnamefont {Y.}~\bibnamefont {Ando}},\
  }\href {\doibase 10.1103/PhysRevB.69.104521} {\bibfield  {journal} {\bibinfo
  {journal} {Phys. Rev. B}\ }\textbf {\bibinfo {volume} {69}},\ \bibinfo
  {pages} {104521} (\bibinfo {year} {2004})}\BibitemShut {NoStop}%
\bibitem [{\citenamefont {Holstein}\ and\ \citenamefont
  {Primakoff}(1940)}]{Holstein_1940}%
  \BibitemOpen
  \bibfield  {author} {\bibinfo {author} {\bibfnamefont {T.}~\bibnamefont
  {Holstein}}\ and\ \bibinfo {author} {\bibfnamefont {H.}~\bibnamefont
  {Primakoff}},\ }\href {\doibase 10.1103/physrev.58.1098} {\bibfield
  {journal} {\bibinfo  {journal} {Phys. Rev.}\ }\textbf {\bibinfo {volume}
  {58}},\ \bibinfo {pages} {1098} (\bibinfo {year} {1940})}\BibitemShut
  {NoStop}%
\bibitem [{\citenamefont {Dyson}(1956)}]{Dyson_1956}%
  \BibitemOpen
  \bibfield  {author} {\bibinfo {author} {\bibfnamefont {F.~J.}\ \bibnamefont
  {Dyson}},\ }\href {\doibase 10.1103/physrev.102.1217} {\bibfield  {journal}
  {\bibinfo  {journal} {Phys. Rev.}\ }\textbf {\bibinfo {volume} {102}},\
  \bibinfo {pages} {1217} (\bibinfo {year} {1956})}\BibitemShut {NoStop}%
\bibitem [{\citenamefont {Fujimori}\ and\ \citenamefont
  {Minami}(1984)}]{Fujimori_1984}%
  \BibitemOpen
  \bibfield  {author} {\bibinfo {author} {\bibfnamefont {A.}~\bibnamefont
  {Fujimori}}\ and\ \bibinfo {author} {\bibfnamefont {F.}~\bibnamefont
  {Minami}},\ }\href {\doibase 10.1103/physrevb.30.957} {\bibfield  {journal}
  {\bibinfo  {journal} {Phys. Rev. B}\ }\textbf {\bibinfo {volume} {30}},\
  \bibinfo {pages} {957} (\bibinfo {year} {1984})}\BibitemShut {NoStop}%
\bibitem [{\citenamefont {Zaanen}\ \emph {et~al.}(1985)\citenamefont {Zaanen},
  \citenamefont {Sawatzky},\ and\ \citenamefont {Allen}}]{Zaanen_1985}%
  \BibitemOpen
  \bibfield  {author} {\bibinfo {author} {\bibfnamefont {J.}~\bibnamefont
  {Zaanen}}, \bibinfo {author} {\bibfnamefont {G.~A.}\ \bibnamefont
  {Sawatzky}}, \ and\ \bibinfo {author} {\bibfnamefont {J.~W.}\ \bibnamefont
  {Allen}},\ }\href {\doibase 10.1103/physrevlett.55.418} {\bibfield  {journal}
  {\bibinfo  {journal} {Phys. Rev. Lett.}\ }\textbf {\bibinfo {volume} {55}},\
  \bibinfo {pages} {418} (\bibinfo {year} {1985})}\BibitemShut {NoStop}%
\bibitem [{\citenamefont {Zhang}\ and\ \citenamefont
  {Rice}(1988)}]{Zhang_1988}%
  \BibitemOpen
  \bibfield  {author} {\bibinfo {author} {\bibfnamefont {F.~C.}\ \bibnamefont
  {Zhang}}\ and\ \bibinfo {author} {\bibfnamefont {T.~M.}\ \bibnamefont
  {Rice}},\ }\href {\doibase 10.1103/physrevb.37.3759} {\bibfield  {journal}
  {\bibinfo  {journal} {Phys. Rev. B.}\ }\textbf {\bibinfo {volume} {37}},\
  \bibinfo {pages} {3759} (\bibinfo {year} {1988})}\BibitemShut {NoStop}%
\bibitem [{\citenamefont {Micnas}\ \emph {et~al.}(1989)\citenamefont {Micnas},
  \citenamefont {Ranninger},\ and\ \citenamefont {Robaszkiewicz}}]{Micnas89}%
  \BibitemOpen
  \bibfield  {author} {\bibinfo {author} {\bibfnamefont {R.}~\bibnamefont
  {Micnas}}, \bibinfo {author} {\bibfnamefont {J.}~\bibnamefont {Ranninger}}, \
  and\ \bibinfo {author} {\bibfnamefont {S.}~\bibnamefont {Robaszkiewicz}},\
  }\href {\doibase 10.1103/PhysRevB.39.11653} {\bibfield  {journal} {\bibinfo
  {journal} {Phys. Rev. B}\ }\textbf {\bibinfo {volume} {39}},\ \bibinfo
  {pages} {11653} (\bibinfo {year} {1989})}\BibitemShut {NoStop}%
\bibitem [{\citenamefont {Marsiglio}\ and\ \citenamefont
  {Hirsch}(1990)}]{MarsiglioHirsch}%
  \BibitemOpen
  \bibfield  {author} {\bibinfo {author} {\bibfnamefont {F.}~\bibnamefont
  {Marsiglio}}\ and\ \bibinfo {author} {\bibfnamefont {J.~E.}\ \bibnamefont
  {Hirsch}},\ }\href {\doibase 10.1103/PhysRevB.41.6435} {\bibfield  {journal}
  {\bibinfo  {journal} {Phys. Rev. B}\ }\textbf {\bibinfo {volume} {41}},\
  \bibinfo {pages} {6435} (\bibinfo {year} {1990})}\BibitemShut {NoStop}%
\bibitem [{\citenamefont {Sim{\'{o}}n}\ \emph {et~al.}(1993)\citenamefont
  {Sim{\'{o}}n}, \citenamefont {Bali{\~{n}}a},\ and\ \citenamefont
  {Aligia}}]{Sim_n_1993}%
  \BibitemOpen
  \bibfield  {author} {\bibinfo {author} {\bibfnamefont {M.}~\bibnamefont
  {Sim{\'{o}}n}}, \bibinfo {author} {\bibfnamefont {M.}~\bibnamefont
  {Bali{\~{n}}a}}, \ and\ \bibinfo {author} {\bibfnamefont {A.}~\bibnamefont
  {Aligia}},\ }\href {\doibase 10.1016/0921-4534(93)90529-y} {\bibfield
  {journal} {\bibinfo  {journal} {Physica C}\ }\textbf {\bibinfo {volume}
  {206}},\ \bibinfo {pages} {297} (\bibinfo {year} {1993})}\BibitemShut
  {NoStop}%
\bibitem [{\citenamefont {Rapp}\ \emph {et~al.}(2012)\citenamefont {Rapp},
  \citenamefont {Deng},\ and\ \citenamefont {Santos}}]{Rapp12}%
  \BibitemOpen
  \bibfield  {author} {\bibinfo {author} {\bibfnamefont {A.}~\bibnamefont
  {Rapp}}, \bibinfo {author} {\bibfnamefont {X.}~\bibnamefont {Deng}}, \ and\
  \bibinfo {author} {\bibfnamefont {L.}~\bibnamefont {Santos}},\ }\href
  {\doibase 10.1103/PhysRevLett.109.203005} {\bibfield  {journal} {\bibinfo
  {journal} {Phys. Rev. Lett.}\ }\textbf {\bibinfo {volume} {109}},\ \bibinfo
  {pages} {203005} (\bibinfo {year} {2012})}\BibitemShut {NoStop}%
\bibitem [{\citenamefont {DiLiberto}\ \emph {et~al.}(2014)\citenamefont
  {DiLiberto}, \citenamefont {Creffield}, \citenamefont {Japaridze},\ and\
  \citenamefont {Morais~Smith}}]{Liberto2014}%
  \BibitemOpen
  \bibfield  {author} {\bibinfo {author} {\bibfnamefont {M.}~\bibnamefont
  {DiLiberto}}, \bibinfo {author} {\bibfnamefont {C.~E.}\ \bibnamefont
  {Creffield}}, \bibinfo {author} {\bibfnamefont {G.~I.}\ \bibnamefont
  {Japaridze}}, \ and\ \bibinfo {author} {\bibfnamefont {C.}~\bibnamefont
  {Morais~Smith}},\ }\href {\doibase 10.1103/PhysRevA.89.013624} {\bibfield
  {journal} {\bibinfo  {journal} {Phys. Rev. A}\ }\textbf {\bibinfo {volume}
  {89}},\ \bibinfo {pages} {013624} (\bibinfo {year} {2014})}\BibitemShut
  {NoStop}%
\bibitem [{\citenamefont {Kac}(1977)}]{kac1977lie}%
  \BibitemOpen
  \bibfield  {author} {\bibinfo {author} {\bibfnamefont {V.~G.}\ \bibnamefont
  {Kac}},\ }\href@noop {} {\bibfield  {journal} {\bibinfo  {journal} {Adv.
  Math.}\ }\textbf {\bibinfo {volume} {26}},\ \bibinfo {pages} {8} (\bibinfo
  {year} {1977})}\BibitemShut {NoStop}%
\bibitem [{\citenamefont {Abrikosov}\ \emph {et~al.}(1963)\citenamefont
  {Abrikosov}, \citenamefont {Gorkov},\ and\ \citenamefont
  {Dzyaloshinski}}]{Abrikosov}%
  \BibitemOpen
  \bibfield  {author} {\bibinfo {author} {\bibfnamefont {A.~A.}\ \bibnamefont
  {Abrikosov}}, \bibinfo {author} {\bibfnamefont {L.~P.}\ \bibnamefont
  {Gorkov}}, \ and\ \bibinfo {author} {\bibfnamefont {I.~E.}\ \bibnamefont
  {Dzyaloshinski}},\ }\href@noop {} {\emph {\bibinfo {title} {Methods of
  quantum field theory in statistical mechanics}}}\ (\bibinfo  {publisher}
  {Prentice-Hall, New Jersey},\ \bibinfo {year} {1963})\BibitemShut {NoStop}%
\bibitem [{\citenamefont {Kadanoff}\ and\ \citenamefont
  {Baym}(1962)}]{kadanoff1962quantum}%
  \BibitemOpen
  \bibfield  {author} {\bibinfo {author} {\bibfnamefont {L.~P.}\ \bibnamefont
  {Kadanoff}}\ and\ \bibinfo {author} {\bibfnamefont {G.~A.}\ \bibnamefont
  {Baym}},\ }\href@noop {} {\emph {\bibinfo {title} {Quantum statistical
  mechanics}}}\ (\bibinfo  {publisher} {Benjamin},\ \bibinfo {year}
  {1962})\BibitemShut {NoStop}%
\bibitem [{\citenamefont {Shastry}(2011{\natexlab{b}})}]{Shastry11_Anatomy}%
  \BibitemOpen
  \bibfield  {author} {\bibinfo {author} {\bibfnamefont {B.~S.}\ \bibnamefont
  {Shastry}},\ }\href {\doibase 10.1103/PhysRevB.84.165112} {\bibfield
  {journal} {\bibinfo  {journal} {Phys. Rev. B}\ }\textbf {\bibinfo {volume}
  {84}},\ \bibinfo {pages} {165112} (\bibinfo {year}
  {2011}{\natexlab{b}})}\BibitemShut {NoStop}%
\bibitem [{Note1()}]{Note1}%
  \BibitemOpen
  \bibinfo {note} {The asymmetry in this factorisation $\protect \mathcal
  G=\protect \mathfrak g\protect \mathcal W$ results from considering the
  equation of motion $\partial _\tau \protect \mathcal G(\tau ,\tau ')$.
  Alternatively we could consider $\partial _{\tau '} \protect \mathcal G(\tau
  ,\tau ')$, and then factorise the Green's function as $\protect \mathcal
  G=\protect \mathcal W\protect \mathfrak g$.}\BibitemShut {Stop}%
\bibitem [{Note2()}]{Note2}%
  \BibitemOpen
  \bibinfo {note} {Refs.~\cite {Shastry_2011,Shastry_2013} treats these as
  functionals of $\protect \mathfrak g$ only, corresponding to a perturbative
  expansion of $\protect \mathcal W$.}\BibitemShut {Stop}%
\bibitem [{\citenamefont {Yokoyama}\ \emph {et~al.}(2006)\citenamefont
  {Yokoyama}, \citenamefont {Ogata},\ and\ \citenamefont
  {Tanaka}}]{doublon-holon}%
  \BibitemOpen
  \bibfield  {author} {\bibinfo {author} {\bibfnamefont {H.}~\bibnamefont
  {Yokoyama}}, \bibinfo {author} {\bibfnamefont {M.}~\bibnamefont {Ogata}}, \
  and\ \bibinfo {author} {\bibfnamefont {Y.}~\bibnamefont {Tanaka}},\ }\href
  {\doibase 10.1143/JPSJ.75.114706} {\bibfield  {journal} {\bibinfo  {journal}
  {J. Phys. Soc. Jpn.}\ }\textbf {\bibinfo {volume} {75}},\ \bibinfo {pages}
  {114706} (\bibinfo {year} {2006})},\ \Eprint
  {http://arxiv.org/abs/https://doi.org/10.1143/JPSJ.75.114706}
  {https://doi.org/10.1143/JPSJ.75.114706} \BibitemShut {NoStop}%
\bibitem [{\citenamefont {Hubbard}(1963)}]{Hubbard1}%
  \BibitemOpen
  \bibfield  {author} {\bibinfo {author} {\bibfnamefont {J.}~\bibnamefont
  {Hubbard}},\ }\href {\doibase 10.1098/rspa.1963.0204} {\bibfield  {journal}
  {\bibinfo  {journal} {Proc. R. Soc. A}\ }\textbf {\bibinfo {volume} {276}},\
  \bibinfo {pages} {238} (\bibinfo {year} {1963})}\BibitemShut {NoStop}%
\bibitem [{\citenamefont {Hubbard}(1964)}]{Hubbard3}%
  \BibitemOpen
  \bibfield  {author} {\bibinfo {author} {\bibfnamefont {J.}~\bibnamefont
  {Hubbard}},\ }\href {\doibase 10.1098/rspa.1964.0190} {\bibfield  {journal}
  {\bibinfo  {journal} {Proc. R. Soc. A}\ }\textbf {\bibinfo {volume} {281}},\
  \bibinfo {pages} {401} (\bibinfo {year} {1964})}\BibitemShut {NoStop}%
\bibitem [{Note3()}]{Note3}%
  \BibitemOpen
  \bibinfo {note} {The generalised sense of Luttinger's theorem argued in
  Ref.~\cite {Dzyaloshinskii_2003} is also violated, with the exception of the
  particle-hole symmetric case (i.e. here when $\protect \tilde {\lambda }=\mu
  =0$) for which it has been proven to be true \cite
  {PhysRevB.75.104503,PhysRevB.96.085124}.}\BibitemShut {Stop}%
\bibitem [{\citenamefont {Luttinger}\ and\ \citenamefont
  {Ward}(1960)}]{LuttingerWard_1960}%
  \BibitemOpen
  \bibfield  {author} {\bibinfo {author} {\bibfnamefont {J.~M.}\ \bibnamefont
  {Luttinger}}\ and\ \bibinfo {author} {\bibfnamefont {J.~C.}\ \bibnamefont
  {Ward}},\ }\href {\doibase 10.1103/physrev.118.1417} {\bibfield  {journal}
  {\bibinfo  {journal} {Phys. Rev.}\ }\textbf {\bibinfo {volume} {118}},\
  \bibinfo {pages} {1417} (\bibinfo {year} {1960})}\BibitemShut {NoStop}%
\bibitem [{\citenamefont {Shankar}(1994)}]{Shankar_1994}%
  \BibitemOpen
  \bibfield  {author} {\bibinfo {author} {\bibfnamefont {R.}~\bibnamefont
  {Shankar}},\ }\href {\doibase 10.1103/revmodphys.66.129} {\bibfield
  {journal} {\bibinfo  {journal} {Rev. Mod. Phys.}\ }\textbf {\bibinfo {volume}
  {66}},\ \bibinfo {pages} {129} (\bibinfo {year} {1994})}\BibitemShut
  {NoStop}%
\bibitem [{\citenamefont {Stein}(1997)}]{Stein_1997}%
  \BibitemOpen
  \bibfield  {author} {\bibinfo {author} {\bibfnamefont {J.}~\bibnamefont
  {Stein}},\ }\href {\doibase 10.1007/bf02508481} {\bibfield  {journal}
  {\bibinfo  {journal} {J. Stat. Phys.}\ }\textbf {\bibinfo {volume} {88}},\
  \bibinfo {pages} {487} (\bibinfo {year} {1997})}\BibitemShut {NoStop}%
\bibitem [{\citenamefont {{Quinn}}(2015)}]{Hidden_structure}%
  \BibitemOpen
  \bibfield  {author} {\bibinfo {author} {\bibfnamefont {E.}~\bibnamefont
  {{Quinn}}},\ }\href@noop {} {\bibfield  {journal} {\bibinfo  {journal} {ArXiv
  e-prints}\ } (\bibinfo {year} {2015})},\ \Eprint
  {http://arxiv.org/abs/1512.00261} {arXiv:1512.00261 [cond-mat.str-el]}
  \BibitemShut {NoStop}%
\bibitem [{\citenamefont {Timusk}\ and\ \citenamefont
  {Statt}(1999)}]{Timusk_1999}%
  \BibitemOpen
  \bibfield  {author} {\bibinfo {author} {\bibfnamefont {T.}~\bibnamefont
  {Timusk}}\ and\ \bibinfo {author} {\bibfnamefont {B.}~\bibnamefont {Statt}},\
  }\href {\doibase 10.1088/0034-4885/62/1/002} {\bibfield  {journal} {\bibinfo
  {journal} {Rep. Prog. Phys.}\ }\textbf {\bibinfo {volume} {62}},\ \bibinfo
  {pages} {61} (\bibinfo {year} {1999})}\BibitemShut {NoStop}%
\bibitem [{\citenamefont {Hashimoto}\ \emph {et~al.}(2014)\citenamefont
  {Hashimoto}, \citenamefont {Vishik}, \citenamefont {He}, \citenamefont
  {Devereaux},\ and\ \citenamefont {Shen}}]{Hashimoto_2014}%
  \BibitemOpen
  \bibfield  {author} {\bibinfo {author} {\bibfnamefont {M.}~\bibnamefont
  {Hashimoto}}, \bibinfo {author} {\bibfnamefont {I.~M.}\ \bibnamefont
  {Vishik}}, \bibinfo {author} {\bibfnamefont {R.-H.}\ \bibnamefont {He}},
  \bibinfo {author} {\bibfnamefont {T.~P.}\ \bibnamefont {Devereaux}}, \ and\
  \bibinfo {author} {\bibfnamefont {Z.-X.}\ \bibnamefont {Shen}},\ }\href
  {\doibase 10.1038/nphys3009} {\bibfield  {journal} {\bibinfo  {journal} {Nat.
  Phys.}\ }\textbf {\bibinfo {volume} {10}},\ \bibinfo {pages} {483} (\bibinfo
  {year} {2014})}\BibitemShut {NoStop}%
\bibitem [{\citenamefont {Yang}\ \emph {et~al.}(2006)\citenamefont {Yang},
  \citenamefont {Rice},\ and\ \citenamefont {Zhang}}]{YRZ}%
  \BibitemOpen
  \bibfield  {author} {\bibinfo {author} {\bibfnamefont {K.-Y.}\ \bibnamefont
  {Yang}}, \bibinfo {author} {\bibfnamefont {T.~M.}\ \bibnamefont {Rice}}, \
  and\ \bibinfo {author} {\bibfnamefont {F.-C.}\ \bibnamefont {Zhang}},\ }\href
  {\doibase 10.1103/PhysRevB.73.174501} {\bibfield  {journal} {\bibinfo
  {journal} {Phys. Rev. B}\ }\textbf {\bibinfo {volume} {73}},\ \bibinfo
  {pages} {174501} (\bibinfo {year} {2006})}\BibitemShut {NoStop}%
\bibitem [{\citenamefont {Rice}\ \emph {et~al.}(2011)\citenamefont {Rice},
  \citenamefont {Yang},\ and\ \citenamefont {Zhang}}]{YRZ_rev}%
  \BibitemOpen
  \bibfield  {author} {\bibinfo {author} {\bibfnamefont {T.~M.}\ \bibnamefont
  {Rice}}, \bibinfo {author} {\bibfnamefont {K.-Y.}\ \bibnamefont {Yang}}, \
  and\ \bibinfo {author} {\bibfnamefont {F.~C.}\ \bibnamefont {Zhang}},\ }\href
  {\doibase 10.1088/0034-4885/75/1/016502} {\bibfield  {journal} {\bibinfo
  {journal} {Rep. Prog. Phys.}\ }\textbf {\bibinfo {volume} {75}},\ \bibinfo
  {pages} {016502} (\bibinfo {year} {2011})}\BibitemShut {NoStop}%
\bibitem [{\citenamefont {Haldane}(1981)}]{HaldaneLL}%
  \BibitemOpen
  \bibfield  {author} {\bibinfo {author} {\bibfnamefont {F.~D.~M.}\
  \bibnamefont {Haldane}},\ }\href
  {http://stacks.iop.org/0022-3719/14/i=19/a=010} {\bibfield  {journal}
  {\bibinfo  {journal} {J. Phys. C}\ }\textbf {\bibinfo {volume} {14}},\
  \bibinfo {pages} {2585} (\bibinfo {year} {1981})}\BibitemShut {NoStop}%
\bibitem [{\citenamefont {Giamarchi}(2003)}]{GiamarchiLL}%
  \BibitemOpen
  \bibfield  {author} {\bibinfo {author} {\bibfnamefont {T.}~\bibnamefont
  {Giamarchi}},\ }\href {\doibase 10.1093/acprof:oso/9780198525004.001.0001}
  {\emph {\bibinfo {title} {{Quantum Physics in One Dimension}}}}\ (\bibinfo
  {publisher} {Oxford University Press ({OUP})},\ \bibinfo {year}
  {2003})\BibitemShut {NoStop}%
\bibitem [{\citenamefont {Dzyaloshinskii}\ and\ \citenamefont
  {Larkin}(1974)}]{DL74}%
  \BibitemOpen
  \bibfield  {author} {\bibinfo {author} {\bibfnamefont {I.~E.}\ \bibnamefont
  {Dzyaloshinskii}}\ and\ \bibinfo {author} {\bibfnamefont {A.~I.}\
  \bibnamefont {Larkin}},\ }\href@noop {} {\bibfield  {journal} {\bibinfo
  {journal} {Sov. Phys. JETP}\ }\textbf {\bibinfo {volume} {38}},\ \bibinfo
  {pages} {202} (\bibinfo {year} {1974})}\BibitemShut {NoStop}%
\bibitem [{\citenamefont {Bethe}(1931)}]{Bethe31}%
  \BibitemOpen
  \bibfield  {author} {\bibinfo {author} {\bibfnamefont {H.}~\bibnamefont
  {Bethe}},\ }\href {\doibase 10.1007/bf01341708} {\bibfield  {journal}
  {\bibinfo  {journal} {{Z. Phys.}}\ }\textbf {\bibinfo {volume} {71}},\
  \bibinfo {pages} {205} (\bibinfo {year} {1931})}\BibitemShut {NoStop}%
\bibitem [{\citenamefont {Takahashi}(1999)}]{TakBook}%
  \BibitemOpen
  \bibfield  {author} {\bibinfo {author} {\bibfnamefont {M.}~\bibnamefont
  {Takahashi}},\ }\href {\doibase 10.1017/cbo9780511524332} {\emph {\bibinfo
  {title} {Thermodynamics of One-Dimensional Solvable Models}}}\ (\bibinfo
  {publisher} {Cambridge University Press ({CUP})},\ \bibinfo {year}
  {1999})\BibitemShut {NoStop}%
\bibitem [{\citenamefont {Alcaraz}\ and\ \citenamefont
  {Bariev}(1999)}]{AlcarazBariev}%
  \BibitemOpen
  \bibfield  {author} {\bibinfo {author} {\bibfnamefont {F.~C.}\ \bibnamefont
  {Alcaraz}}\ and\ \bibinfo {author} {\bibfnamefont {R.~Z.}\ \bibnamefont
  {Bariev}},\ }\href {http://stacks.iop.org/0305-4470/32/i=46/a=101} {\bibfield
   {journal} {\bibinfo  {journal} {J. Phys. A}\ }\textbf {\bibinfo {volume}
  {32}},\ \bibinfo {pages} {L483} (\bibinfo {year} {1999})}\BibitemShut
  {NoStop}%
\bibitem [{\citenamefont {Beisert}\ and\ \citenamefont
  {Koroteev}(2008)}]{BeisertKoroteev}%
  \BibitemOpen
  \bibfield  {author} {\bibinfo {author} {\bibfnamefont {N.}~\bibnamefont
  {Beisert}}\ and\ \bibinfo {author} {\bibfnamefont {P.}~\bibnamefont
  {Koroteev}},\ }\href {http://stacks.iop.org/1751-8121/41/i=25/a=255204}
  {\bibfield  {journal} {\bibinfo  {journal} {J. Phys. A}\ }\textbf {\bibinfo
  {volume} {41}},\ \bibinfo {pages} {255204} (\bibinfo {year}
  {2008})}\BibitemShut {NoStop}%
\bibitem [{\citenamefont {Shvaika}(2003)}]{PhysRevB.67.075101}%
  \BibitemOpen
  \bibfield  {author} {\bibinfo {author} {\bibfnamefont {A.~M.}\ \bibnamefont
  {Shvaika}},\ }\href {\doibase 10.1103/PhysRevB.67.075101} {\bibfield
  {journal} {\bibinfo  {journal} {Phys. Rev. B}\ }\textbf {\bibinfo {volume}
  {67}},\ \bibinfo {pages} {075101} (\bibinfo {year} {2003})}\BibitemShut
  {NoStop}%
\bibitem [{\citenamefont {Perepelitsky}\ and\ \citenamefont
  {Shastry}(2013)}]{PEREPELITSKY2013283}%
  \BibitemOpen
  \bibfield  {author} {\bibinfo {author} {\bibfnamefont {E.}~\bibnamefont
  {Perepelitsky}}\ and\ \bibinfo {author} {\bibfnamefont {B.~S.}\ \bibnamefont
  {Shastry}},\ }\href {\doibase https://doi.org/10.1016/j.aop.2013.09.010}
  {\bibfield  {journal} {\bibinfo  {journal} {Ann. Phys.}\ }\textbf {\bibinfo
  {volume} {338}},\ \bibinfo {pages} {283 } (\bibinfo {year}
  {2013})}\BibitemShut {NoStop}%
\bibitem [{\citenamefont {Barnes}(1976)}]{Barnes_1976}%
  \BibitemOpen
  \bibfield  {author} {\bibinfo {author} {\bibfnamefont {S.~E.}\ \bibnamefont
  {Barnes}},\ }\href {\doibase 10.1088/0305-4608/6/7/018} {\bibfield  {journal}
  {\bibinfo  {journal} {J. Phys. F}\ }\textbf {\bibinfo {volume} {6}},\
  \bibinfo {pages} {1375} (\bibinfo {year} {1976})}\BibitemShut {NoStop}%
\bibitem [{\citenamefont {Coleman}(1984)}]{Coleman_1984}%
  \BibitemOpen
  \bibfield  {author} {\bibinfo {author} {\bibfnamefont {P.}~\bibnamefont
  {Coleman}},\ }\href {\doibase 10.1103/physrevb.29.3035} {\bibfield  {journal}
  {\bibinfo  {journal} {Phys. Rev. B}\ }\textbf {\bibinfo {volume} {29}},\
  \bibinfo {pages} {3035} (\bibinfo {year} {1984})}\BibitemShut {NoStop}%
\bibitem [{\citenamefont {Arovas}\ and\ \citenamefont
  {Auerbach}(1988)}]{Arovas_1988}%
  \BibitemOpen
  \bibfield  {author} {\bibinfo {author} {\bibfnamefont {D.~P.}\ \bibnamefont
  {Arovas}}\ and\ \bibinfo {author} {\bibfnamefont {A.}~\bibnamefont
  {Auerbach}},\ }\href {\doibase 10.1103/physrevb.38.316} {\bibfield  {journal}
  {\bibinfo  {journal} {Phys. Rev. B}\ }\textbf {\bibinfo {volume} {38}},\
  \bibinfo {pages} {316} (\bibinfo {year} {1988})}\BibitemShut {NoStop}%
\bibitem [{\citenamefont {Yoshioka}(1989)}]{Yoshioka_1989}%
  \BibitemOpen
  \bibfield  {author} {\bibinfo {author} {\bibfnamefont {D.}~\bibnamefont
  {Yoshioka}},\ }\href {\doibase 10.1143/jpsj.58.1516} {\bibfield  {journal}
  {\bibinfo  {journal} {J. Phys. Soc. Jpn.}\ }\textbf {\bibinfo {volume}
  {58}},\ \bibinfo {pages} {1516} (\bibinfo {year} {1989})}\BibitemShut
  {NoStop}%
\bibitem [{\citenamefont {Senthil}\ \emph {et~al.}(2003)\citenamefont
  {Senthil}, \citenamefont {Sachdev},\ and\ \citenamefont
  {Vojta}}]{Senthil_2003}%
  \BibitemOpen
  \bibfield  {author} {\bibinfo {author} {\bibfnamefont {T.}~\bibnamefont
  {Senthil}}, \bibinfo {author} {\bibfnamefont {S.}~\bibnamefont {Sachdev}}, \
  and\ \bibinfo {author} {\bibfnamefont {M.}~\bibnamefont {Vojta}},\ }\href
  {\doibase 10.1103/physrevlett.90.216403} {\bibfield  {journal} {\bibinfo
  {journal} {Phys. Rev. Lett.}\ }\textbf {\bibinfo {volume} {90}},\ \bibinfo
  {pages} {216403} (\bibinfo {year} {2003})}\BibitemShut {NoStop}%
\bibitem [{\citenamefont {Baskaran}\ and\ \citenamefont
  {Anderson}(1988)}]{BaskaranAnderson}%
  \BibitemOpen
  \bibfield  {author} {\bibinfo {author} {\bibfnamefont {G.}~\bibnamefont
  {Baskaran}}\ and\ \bibinfo {author} {\bibfnamefont {P.~W.}\ \bibnamefont
  {Anderson}},\ }\href {\doibase 10.1103/PhysRevB.37.580} {\bibfield  {journal}
  {\bibinfo  {journal} {Phys. Rev. B}\ }\textbf {\bibinfo {volume} {37}},\
  \bibinfo {pages} {580} (\bibinfo {year} {1988})}\BibitemShut {NoStop}%
\bibitem [{\citenamefont {Wen}\ and\ \citenamefont
  {Lee}(1996)}]{PhysRevLett.76.503}%
  \BibitemOpen
  \bibfield  {author} {\bibinfo {author} {\bibfnamefont {X.-G.}\ \bibnamefont
  {Wen}}\ and\ \bibinfo {author} {\bibfnamefont {P.~A.}\ \bibnamefont {Lee}},\
  }\href {\doibase 10.1103/PhysRevLett.76.503} {\bibfield  {journal} {\bibinfo
  {journal} {Phys. Rev. Lett.}\ }\textbf {\bibinfo {volume} {76}},\ \bibinfo
  {pages} {503} (\bibinfo {year} {1996})}\BibitemShut {NoStop}%
\bibitem [{\citenamefont {Ng}(2005)}]{PhysRevB.71.172509}%
  \BibitemOpen
  \bibfield  {author} {\bibinfo {author} {\bibfnamefont {T.-K.}\ \bibnamefont
  {Ng}},\ }\href {\doibase 10.1103/PhysRevB.71.172509} {\bibfield  {journal}
  {\bibinfo  {journal} {Phys. Rev. B}\ }\textbf {\bibinfo {volume} {71}},\
  \bibinfo {pages} {172509} (\bibinfo {year} {2005})}\BibitemShut {NoStop}%
\bibitem [{\citenamefont {Martin}\ \emph {et~al.}(1990)\citenamefont {Martin},
  \citenamefont {Fiory}, \citenamefont {Fleming}, \citenamefont {Schneemeyer},\
  and\ \citenamefont {Waszczak}}]{Martin_1990}%
  \BibitemOpen
  \bibfield  {author} {\bibinfo {author} {\bibfnamefont {S.}~\bibnamefont
  {Martin}}, \bibinfo {author} {\bibfnamefont {A.~T.}\ \bibnamefont {Fiory}},
  \bibinfo {author} {\bibfnamefont {R.~M.}\ \bibnamefont {Fleming}}, \bibinfo
  {author} {\bibfnamefont {L.~F.}\ \bibnamefont {Schneemeyer}}, \ and\ \bibinfo
  {author} {\bibfnamefont {J.~V.}\ \bibnamefont {Waszczak}},\ }\href {\doibase
  10.1103/physrevb.41.846} {\bibfield  {journal} {\bibinfo  {journal} {Phys.
  Rev. B}\ }\textbf {\bibinfo {volume} {41}},\ \bibinfo {pages} {846} (\bibinfo
  {year} {1990})}\BibitemShut {NoStop}%
\bibitem [{\citenamefont {Chien}\ \emph {et~al.}(1991)\citenamefont {Chien},
  \citenamefont {Wang},\ and\ \citenamefont {Ong}}]{Chien_1991}%
  \BibitemOpen
  \bibfield  {author} {\bibinfo {author} {\bibfnamefont {T.~R.}\ \bibnamefont
  {Chien}}, \bibinfo {author} {\bibfnamefont {Z.~Z.}\ \bibnamefont {Wang}}, \
  and\ \bibinfo {author} {\bibfnamefont {N.~P.}\ \bibnamefont {Ong}},\ }\href
  {\doibase 10.1103/physrevlett.67.2088} {\bibfield  {journal} {\bibinfo
  {journal} {Phys. Rev. Lett.}\ }\textbf {\bibinfo {volume} {67}},\ \bibinfo
  {pages} {2088} (\bibinfo {year} {1991})}\BibitemShut {NoStop}%
\bibitem [{\citenamefont {Hussey}\ \emph {et~al.}(2011)\citenamefont {Hussey},
  \citenamefont {Cooper}, \citenamefont {Xu}, \citenamefont {Wang},
  \citenamefont {Mouzopoulou}, \citenamefont {Vignolle},\ and\ \citenamefont
  {Proust}}]{Hussey_2011}%
  \BibitemOpen
  \bibfield  {author} {\bibinfo {author} {\bibfnamefont {N.~E.}\ \bibnamefont
  {Hussey}}, \bibinfo {author} {\bibfnamefont {R.~A.}\ \bibnamefont {Cooper}},
  \bibinfo {author} {\bibfnamefont {X.}~\bibnamefont {Xu}}, \bibinfo {author}
  {\bibfnamefont {Y.}~\bibnamefont {Wang}}, \bibinfo {author} {\bibfnamefont
  {I.}~\bibnamefont {Mouzopoulou}}, \bibinfo {author} {\bibfnamefont
  {B.}~\bibnamefont {Vignolle}}, \ and\ \bibinfo {author} {\bibfnamefont
  {C.}~\bibnamefont {Proust}},\ }\href {\doibase 10.1098/rsta.2010.0196}
  {\bibfield  {journal} {\bibinfo  {journal} {Philos. T. Roy. Soc. A}\ }\textbf
  {\bibinfo {volume} {369}},\ \bibinfo {pages} {1626} (\bibinfo {year}
  {2011})}\BibitemShut {NoStop}%
\bibitem [{\citenamefont {Zaanen}\ \emph {et~al.}(2015)\citenamefont {Zaanen},
  \citenamefont {Liu}, \citenamefont {Sun},\ and\ \citenamefont
  {Schalm}}]{zaanen2015holographic}%
  \BibitemOpen
  \bibfield  {author} {\bibinfo {author} {\bibfnamefont {J.}~\bibnamefont
  {Zaanen}}, \bibinfo {author} {\bibfnamefont {Y.}~\bibnamefont {Liu}},
  \bibinfo {author} {\bibfnamefont {Y.-W.}\ \bibnamefont {Sun}}, \ and\
  \bibinfo {author} {\bibfnamefont {K.}~\bibnamefont {Schalm}},\ }\href@noop {}
  {\emph {\bibinfo {title} {Holographic duality in condensed matter physics}}}\
  (\bibinfo  {publisher} {Cambridge University Press},\ \bibinfo {year}
  {2015})\BibitemShut {NoStop}%
\bibitem [{\citenamefont {{Hartnoll}}\ \emph {et~al.}(2016)\citenamefont
  {{Hartnoll}}, \citenamefont {{Lucas}},\ and\ \citenamefont
  {{Sachdev}}}]{hartnoll2016holographic}%
  \BibitemOpen
  \bibfield  {author} {\bibinfo {author} {\bibfnamefont {S.~A.}\ \bibnamefont
  {{Hartnoll}}}, \bibinfo {author} {\bibfnamefont {A.}~\bibnamefont {{Lucas}}},
  \ and\ \bibinfo {author} {\bibfnamefont {S.}~\bibnamefont {{Sachdev}}},\
  }\href@noop {} {\bibfield  {journal} {\bibinfo  {journal} {ArXiv e-prints}\ }
  (\bibinfo {year} {2016})},\ \Eprint {http://arxiv.org/abs/1612.07324}
  {arXiv:1612.07324 [hep-th]} \BibitemShut {NoStop}%
\bibitem [{Note4()}]{Note4}%
  \BibitemOpen
  \bibinfo {note} {We suppress $\kappa ^2$ terms in this schematic
  representation of the algebra. They enter Eq.~\protect \textup {\hbox
  {\mathsurround \z@ \protect \normalfont (\ignorespaces \ref {eomshm}\unskip
  \@@italiccorr )}} only through $A$, and play no role in the subsequent
  expansion.}\BibitemShut {Stop}%
\bibitem [{\citenamefont {Dzyaloshinskii}(2003)}]{Dzyaloshinskii_2003}%
  \BibitemOpen
  \bibfield  {author} {\bibinfo {author} {\bibfnamefont {I.}~\bibnamefont
  {Dzyaloshinskii}},\ }\href {\doibase 10.1103/physrevb.68.085113} {\bibfield
  {journal} {\bibinfo  {journal} {Phys. Rev. B}\ }\textbf {\bibinfo {volume}
  {68}},\ \bibinfo {pages} {085113} (\bibinfo {year} {2003})}\BibitemShut
  {NoStop}%
\bibitem [{\citenamefont {Stanescu}\ \emph {et~al.}(2007)\citenamefont
  {Stanescu}, \citenamefont {Phillips},\ and\ \citenamefont
  {Choy}}]{PhysRevB.75.104503}%
  \BibitemOpen
  \bibfield  {author} {\bibinfo {author} {\bibfnamefont {T.~D.}\ \bibnamefont
  {Stanescu}}, \bibinfo {author} {\bibfnamefont {P.}~\bibnamefont {Phillips}},
  \ and\ \bibinfo {author} {\bibfnamefont {T.-P.}\ \bibnamefont {Choy}},\
  }\href {\doibase 10.1103/PhysRevB.75.104503} {\bibfield  {journal} {\bibinfo
  {journal} {Phys. Rev. B}\ }\textbf {\bibinfo {volume} {75}},\ \bibinfo
  {pages} {104503} (\bibinfo {year} {2007})}\BibitemShut {NoStop}%
\bibitem [{\citenamefont {Seki}\ and\ \citenamefont
  {Yunoki}(2017)}]{PhysRevB.96.085124}%
  \BibitemOpen
  \bibfield  {author} {\bibinfo {author} {\bibfnamefont {K.}~\bibnamefont
  {Seki}}\ and\ \bibinfo {author} {\bibfnamefont {S.}~\bibnamefont {Yunoki}},\
  }\href {\doibase 10.1103/PhysRevB.96.085124} {\bibfield  {journal} {\bibinfo
  {journal} {Phys. Rev. B}\ }\textbf {\bibinfo {volume} {96}},\ \bibinfo
  {pages} {085124} (\bibinfo {year} {2017})}\BibitemShut {NoStop}%
\end{thebibliography}%

\end{document}